\documentclass[10pt,fleqn,aps,prb,twocolumn,showpacs,floatfix]{revtex4-2}
\usepackage[a4paper,margin=1in]{geometry}
\usepackage{amsmath,amssymb,amsthm,mathtools,bm}

\usepackage{mathrsfs}
\usepackage{graphicx}
\usepackage{placeins}
\usepackage{url}
\usepackage[utf8]{inputenc}
\usepackage{subfigure}
\usepackage{slashed}
\usepackage{setspace}
\usepackage{wasysym}
\usepackage{makecell}
\usepackage{xcolor}
\usepackage{ulem}
\usepackage{hyperref}
\usepackage[shortlabels]{enumitem}
\usepackage[nameinlink,noabbrev]{cleveref}
\usepackage{booktabs,longtable,tabularx,array,multirow}
\usepackage{tikz}
\usetikzlibrary{arrows.meta,positioning,calc,fit,shapes.geometric}
\usepackage[most]{tcolorbox}
\usepackage{dsfont}
\newcommand{\id}{\mathds{1}}

\newcommand{\downtriangle}{%
	\mathord{\rotatebox[origin=c]{180}{$\triangle$}}%
}

\theoremstyle{definition}

\newcommand{\be}{\begin{equation}}
\newcommand{\ee}{\end{equation}}
\newcommand{\bea}{\begin{eqnarray}}
\newcommand{\eea}{\end{eqnarray}}

\hypersetup{
	colorlinks=true,
	linkcolor=blue!55!black,
	citecolor=blue!55!black,
	urlcolor=blue!55!black
}

\newcommand{\Tr}{\operatorname{Tr}}

\definecolor{defblue}{RGB}{235,244,252}
\definecolor{warnyellow}{RGB}{255,249,226}
\definecolor{examplegreen}{RGB}{238,249,240}

\newtcolorbox{definitionbox}[1][]{
	colback=defblue,
	colframe=blue!45!black,
	fonttitle=\bfseries,
	title=Definition,
	breakable,
	#1
}
\newtcolorbox{keybox}[1][]{
	colback=warnyellow,
	colframe=orange!55!black,
	fonttitle=\bfseries,
	title=Key point,
	breakable,
	#1
}
\newtcolorbox{examplebox}[1][]{
	colback=examplegreen,
	colframe=green!45!black,
	fonttitle=\bfseries,
	title=Worked example,
	breakable,
	#1
}

\makeatletter
\newcommand{\hidesubsections}{%
	\let\l@subsection\@gobbletwo
	\let\l@subsubsection\@gobbletwo
}
\makeatother
\begin{document}

\title{Dynamical splitting and a nodal Bose liquid in 2d chiral XYZ model}

\author{Tarun Grover}
\affiliation{Department of Physics, University of California at San Diego, La Jolla, California 92093, USA}

\begin{abstract}
We study a class of Hamiltonians with a structure that we call
``dynamical splitting'': the Hamiltonian terms can be divided into two
sets acting on the same degrees of freedom such that every term in one set commutes with every term in the other,
although terms within either set do not all commute.   This structure yields an algebraic duality to effective degrees of freedom on which the two parts of the Hamiltonian act disjointly, enabling exact
diagonalization on lattices with approximately twice as many spins as usually
accessible. We exploit it in the ``chiral XYZ model'', a geometrically
frustrated spin-$1/2$ model on the triangular lattice which was previously introduced as a special limit of a
Majorana--Hubbard model.  This model also possesses
anticommuting noncontractible line symmetries, which enforce an exact,
topology-dependent degeneracy between locally indistinguishable states.  We
first study a $\mathbb{Z}_N$ clock generalization and find, at large $N$, a gapless ground state with three subsystem-symmetry-protected nodal lines.  Exploiting dynamical splitting
and the subsystem symmetries, we carry out exact diagonalization of the $N=2$
model on lattices up to $9\times9$ spins.  The many-body gap and bipartite
entanglement provide strong evidence for a gapless state consistent with the
large-$N$ nodal structure: the entanglement scales as $L\log L$
and exhibits $1+1$-dimensional CFT-like chord scaling on cylinders.  Finally,
we study instabilities and proximate phases. In particular, we find evidence that a subsystem-symmetry-preserving deformation drives a finite coupling transition to a
gapped phase with $\mathbb Z_2 \times \mathbb Z_2$ topological order.
\end{abstract}

\maketitle
\tableofcontents

\section{Introduction}
\label{sec:introduction}

Symmetries  simplify quantum many-body problems by dividing the Hilbert
space into smaller sectors that do not mix.   Here we identify and explore the consequences of a related but distinct simplification.   In certain local Hamiltonians, the interaction terms can be divided into two or more groups such that every term in one group commutes with every term in the others, even though terms within  the same group need not commute and all the groups act on the same physical
spins. We call this structure ``dynamical splitting.'' Unlike a standard symmetry-based simplification, where the Hilbert space
decomposes into a direct sum of smaller Hilbert spaces, dynamical splitting
further decomposes each such space into a \textit{tensor product} of smaller spaces.  Each of the resulting factors generically depends on all the physical degrees of freedom, rather than on disjoint subsets of them. From a practical standpoint, the full quantum problem separates into smaller independent problems, thereby making unusually large systems accessible to exact diagonalization. Our main focus is a two-dimensional geometrically frustrated quantum spin system with this property. Combining a large-$N$ semiclassical analysis with exact diagonalization, we
provide evidence for a highly entangled, gapless nodal Bose liquid in this model, with momentum-space manifolds of soft modes sharing some features with previously studied Bose liquids~\cite{ParamekantiBalentsFisher2002}.  Notably, the gaplessness coexists with an exact, topology-dependent degeneracy reminiscent
of gapped topological order.

 The structure described above can be stated more precisely as follows: suppose that the Hamiltonian can be decomposed as
\begin{equation}
H=H_A+H_B,\,\,
H_A=\sum_i h_{iA},\,\,
H_B=\sum_j h_{jB}, \label{Eq:ABdecompose}
\end{equation}
such that 

\be [h_{iA},h_{jB}]=0 \ee 

 for every \(i,j\).  Let $\mathcal A$ and $\mathcal B$ denote the operator algebras generated by
$\{h_{iA}\}$ and $\{h_{jB}\}$, respectively. The condition $[h_{iA},h_{jB}]=0$ implies that the algebras
$\mathcal A$ and $\mathcal B$ commute.  In a finite-dimensional Hilbert space, this guarantees
a direct-sum decomposition into tensor-product sectors, although it need not
give a single tensor-product decomposition of the full Hilbert space because
the two algebras may be correlated through central quantum numbers \footnote{This is an instance of the finite-dimensional Artin--Wedderburn decomposition; see Ref.~\cite[Sec.~18.2]{DummitFoote} for the underlying theorem and	Ref.~\cite{ZanardiLidarLloyd2004} for the closely related
observable-induced tensor-product construction.}.  Concretely, let $\lambda$ label the simultaneous eigenspaces, or equivalently
the minimal central sectors, of the center of the jointly generated algebra
$\mathcal A\vee\mathcal B$.  The total Hilbert space then decomposes as
$\mathcal H=\bigoplus_\lambda\mathcal H_\lambda$, and within each sector
$\mathcal H_\lambda$ admits a tensor-product decomposition
\begin{equation}
\mathcal H_\lambda \simeq
\mathcal H_{A,\lambda}\otimes
\mathcal H_{B,\lambda}\otimes
\mathcal H_{0,\lambda},
\label{Eq:generalHilbertFactorization}
\end{equation}
such that every operator in $\mathcal A$ acts only on
$\mathcal H_{A,\lambda}$, every operator in $\mathcal B$ acts only on
$\mathcal H_{B,\lambda}$, and both act trivially on
the spectator, or multiplicity, space $\mathcal H_{0,\lambda}$ whose dimension gives
an exact degeneracy within the sector $\lambda$. Consequently, within each fixed sector the Hamiltonian
takes the form
\begin{equation}
\left.H\right|_{\mathcal H_\lambda}
=
H_A(\lambda)\otimes \mathds{1}_{B}\otimes\mathds{1}_{0}
+
\mathds{1}_{A}\otimes H_B(\lambda)\otimes\mathds{1}_{0}.
\label{Eq:generalHamiltonianFactorization}
\end{equation}
Crucially, the tensor-product form in
Eq.~\eqref{Eq:generalHamiltonianFactorization} does not arise from a
microscopic partition of the lattice degrees of freedom.  In the original
lattice description, $H_A$ and $H_B$ both act nontrivially on the same
physical qubits; the independent factors $\mathcal H_{A,\lambda}$ and
$\mathcal H_{B,\lambda}$, however, arise from an algebraic decomposition within each common central sector $\lambda$, with $H_A$ and $H_B$ acting disjointly on the corresponding effective degrees of freedom. Relatedly, although $H_A$ and $H_B$ may be geometrically local in the original lattice variables,
the sector Hamiltonians $H_A(\lambda)$ and $H_B(\lambda)$ need not inherit
this locality, since the factorization above is algebraic and need not arise from a local change of variables. 
The eigenvalues of $\left.H\right|_{\mathcal H_\lambda}$ are therefore sums of eigenvalues of $H_A(\lambda)$ and
$H_B(\lambda)$, with an additional degeneracy
$\dim\mathcal H_{0,\lambda}$. 

The factorization above is closely related to the  tensor-product
decompositions that arise in the bond-algebra/commutant framework~\cite{MoudgalyaMotrunichSymmetries}.  This framework has been used to organize conventional and
nonstandard symmetries \cite{MoudgalyaMotrunichSymmetries},
Hilbert-space fragmentation \cite{MoudgalyaMotrunichFragmentation}, and exact
quantum many-body scars \cite{MoudgalyaMotrunichScars}.  In the usual bond-algebra setting, the Hamiltonian belongs to the bond algebra, while the commutant of that algebra encodes its symmetries. Our focus, however, is different. Rather than using the bond-algebra decomposition  to classify symmetries, fragmented sectors, or scar states, we exploit a setting in which two mutually commuting but individually noncommutative local algebras both contribute to the Hamiltonian. After fixing their common central quantum numbers, the two contributions become
independent spectral problems. This dynamical splitting leads to reduction in computational effort to diagonalize the Hamiltonian and allows us to investigate the ground-state and low-energy physics of a two-dimensional
frustrated magnet of our interest at unusually large sizes.

From a practical standpoint, this reduction is most useful when the two active Hilbert-space factors $\mathcal H_{A,\lambda}$ and $\mathcal H_{B,\lambda}$ have comparable dimensions and the number of central
sectors that must be examined grows much more slowly than the original Hilbert-space dimension. Writing
$D_A=\dim\mathcal H_{A,\lambda}$ and
$D_B=\dim\mathcal H_{B,\lambda}$, diagonalization within a sector is reduced
from a problem of active dimension $D_A D_B$ to two separate problems of
dimensions $D_A$ and $D_B$.  When $D_A\sim D_B$, the largest matrix dimension
is thus approximately the square root of the original active dimension.
The spectator factor $\mathcal H_{0,\lambda}$ need not be small: because the
Hamiltonian acts trivially on it, it contributes only an exact degeneracy
and requires no additional diagonalization.

\begin{figure}[t]
	\centering
	\includegraphics[width=\columnwidth]{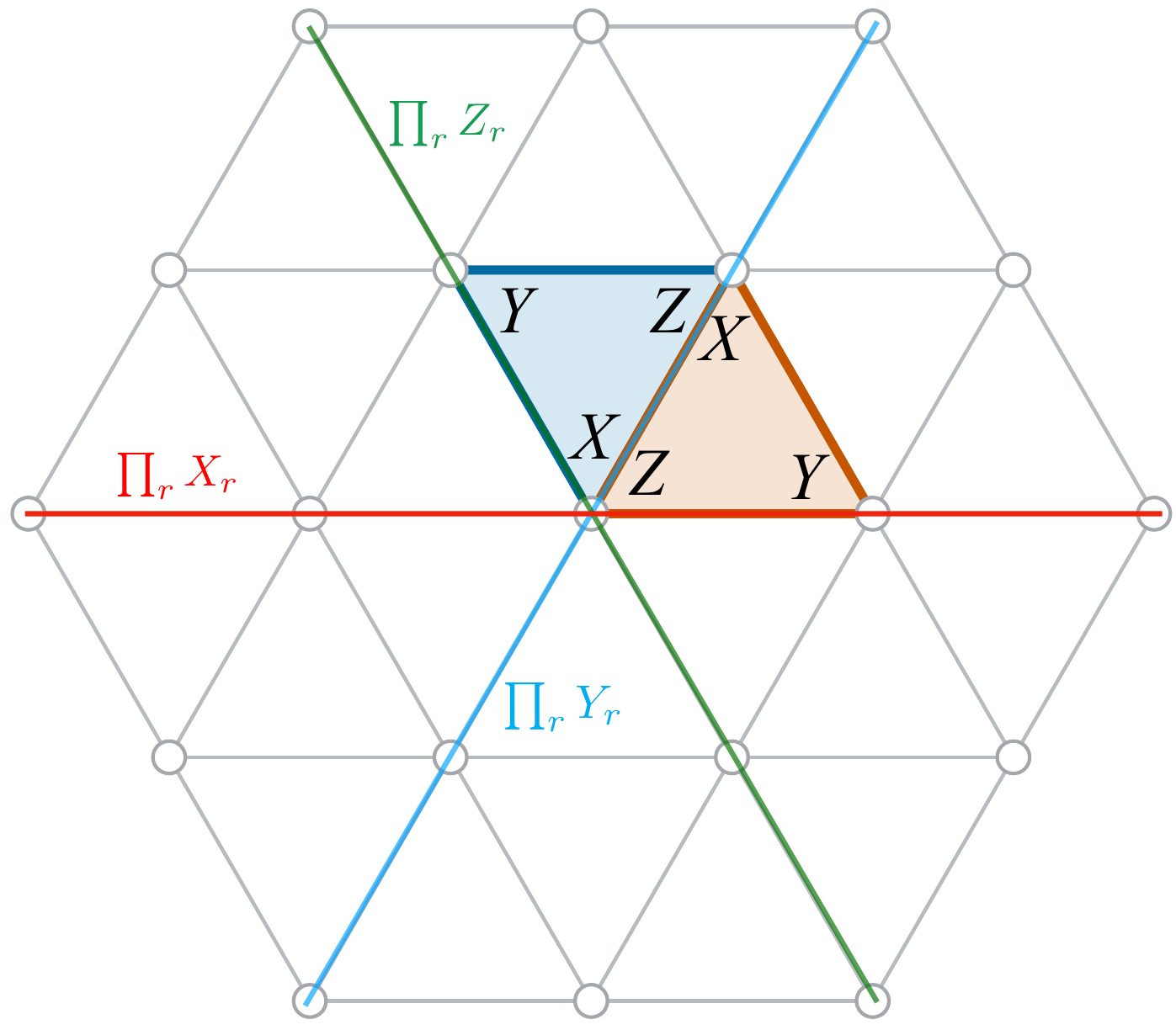}
	\caption{The chiral XYZ model on the triangular lattice.  A spin-$1/2$
	lives on every vertex.  The shaded upward- and downward-pointing triangles
	illustrate the two elementary interactions; the labels specify which Pauli
	operator acts at each vertex.  Their cyclic ordering is the same on the two
	triangle orientations.  The colored straight lines show representative
	row, column and diagonal subsystem symmetries, formed by products of \(X\), \(Y\),
	or \(Z\) operators along a noncontractible line on a torus.}
	\label{fig:modeldef}
\end{figure}

A simple one-dimensional example illustrates the idea. Consider a periodic chain of \(2N\) qubits with \(H=H_{XZ}+H_{XYZ}\), where \(H_{XZ}=J\sum_{i=1}^{2N}(X_iX_{i+1}+Z_iZ_{i+1})\) and \(H_{XYZ}=K\sum_{i=1}^{2N}X_iY_{i+1}Z_{i+2}\), with site labels understood modulo \(2N\). A brute-force treatment would require diagonalizing a \(\lvert\mathcal H\rvert\times\lvert\mathcal H\rvert\) matrix with \(\lvert\mathcal H\rvert=2^{2N}\). Separating the terms according to whether \(i\) is odd or even, however, gives \(H=H_A+H_B\), where \(H_A=J\sum_{i=1}^{N}(X_{2i-1}X_{2i}+Z_{2i}Z_{2i+1})+K\sum_{i=1}^{N}X_{2i-1}Y_{2i}Z_{2i+1}\) and \(H_B=J\sum_{i=1}^{N}(Z_{2i-1}Z_{2i}+X_{2i}X_{2i+1})+K\sum_{i=1}^{N}X_{2i}Y_{2i+1}Z_{2i+2}\). Every term in \(H_A\) commutes with every term in \(H_B\), and hence the Hamiltonian has the structure described above. The common center is generated by the global parity operators \(\prod_{i=1}^{2N}X_i\) and \(\prod_{i=1}^{2N}Z_i\); we denote their eigenvalues by \(\lambda=(\lambda_X,\lambda_Z)\), with \(\lambda_X,\lambda_Z=\pm1\). Within a fixed \(\lambda\) sector, defining \(\tau_i^x=X_{2i-1}X_{2i}\), \(\tau_i^z\tau_{i+1}^z=Z_{2i}Z_{2i+1}\), \(\sigma_i^x=Z_{2i-1}Z_{2i}\), and \(\sigma_i^z\sigma_{i+1}^z=X_{2i}X_{2i+1}\), one obtains \(H_\lambda=H_{A,\lambda}\otimes\mathds{1}+\mathds{1}\otimes H_{B,\lambda}\), with \(H_{A,\lambda}=\sum_{i=1}^{N}[J(\tau_i^x+\tau_i^z\tau_{i+1}^z)+K\tau_i^y\tau_{i+1}^z]\) and \(H_{B,\lambda}=\sum_{i=1}^{N}[J(\sigma_i^x+\sigma_i^z\sigma_{i+1}^z)-K\sigma_i^z\sigma_{i+1}^y]\). The boundary conditions of the two effective chains are fixed by \(\lambda_X\) and \(\lambda_Z\), and \(\mathcal H_\lambda\simeq\mathcal H_{A,\lambda}\otimes\mathcal H_{B,\lambda}\), with \(\lvert\mathcal H_{A,\lambda}\rvert=\lvert\mathcal H_{B,\lambda}\rvert=2^{N-1}\). Thus, in each of the four \(\lambda\) sectors, the spectrum of the original \(2^{2N}\)-dimensional Hamiltonian is obtained by diagonalizing two matrices of dimension \(2^{N-1}\) and adding their eigenvalues. This particular Hamiltonian is integrable, since it can be mapped to free
fermions by a Jordan--Wigner transformation, but the dynamical splitting does not
rely on integrability.  For example, one may add the interaction
\(\sum_i X_iX_{i+1}X_{i+2}X_{i+3}\) to $H$.  The terms beginning on odd and even sites
belong, respectively, to the two mutually commuting operator algebras, so the
decomposition into \(H_A\) and \(H_B\) survives, while the corresponding
effective Hamiltonians become interacting and are generically no longer
mappable to free fermions.

In the one-dimensional example above, the Hamiltonians
$H_{A/B,\lambda}$ remain local when expressed in terms of the original spin
variables.  A related factorization was obtained for an interacting Majorana
plaquette model through a duality to two decoupled local quantum compass
models \cite{KamiyaFurusakiTeoChern2018,RuckertRooszTimm2020}.  In both
constructions, the factor Hamiltonians can therefore be expressed locally in
appropriate variables.  More generally, however, the interactions between the effective qubits on which the factor Hamiltonians act (``factor qubits'' henceforth) need not be geometrically local. In such cases, efficient algebraic methods based on the Pauli group, familiar
from the binary symplectic formulation of stabilizer and subsystem
codes~\cite{GottesmanStabilizerThesis,PoulinSubsystemFormalism,
WildeLogicalOperators,HaahAlgebraicCodes}, turn out to be useful for
obtaining explicit forms of $H_{A/B,\lambda}$ even though the corresponding
Hamiltonians are not commuting-projector Hamiltonians.

The model of our primary interest is defined in Fig.~\ref{fig:modeldef}: for
each upward-pointing $\triangle$, there is a term $X_iY_jZ_k$, and a similar
term for each downward-pointing $\downtriangle$.  Both terms have the same
chirality, and we therefore call this the ``chiral XYZ model.''  This model
was introduced in Ref.~\cite{LiFranzMajoranaHubbard2018} as a special limit of Majorana-Hubbard model, and also discussed more recently
in the context of subsystem codes in
Ref.~\cite{BusseToikkaMajoranaXYZ2026}, although, to our knowledge, the precise nature of its
ground state is not known.
 
This model possesses several subsystem symmetries built out of string-like
operators $\prod_r X_r$, $\prod_r Y_r$, and $\prod_r Z_r$ that wrap around
noncontractible cycles, as shown in Fig.~\ref{fig:modeldef}. We discuss the
complete set of line operators in Sec.~\ref{subsec:subsystem-symmetries}, but
for now we simply note that one can choose a row and a column string that
intersect once and hence anticommute.  On a suitable lattice realization of
a closed surface of genus $g$, one obtains one such pair per handle.
Consequently, every energy level contains an exactly degenerate, locally
indistinguishable multiplet of dimension at least $2^g$.  This
symmetry-enforced degeneracy matches the topological ground-state degeneracy of the Kalmeyer--Laughlin chiral spin liquid, the lattice analog of the
$\nu=1/2$ bosonic Laughlin state~\cite{kalmeyer87}.  Indeed, one of our
motivations to study this model came from attempting to mimic chiral
Chern--Simons theory on the lattice while maintaining a tensor-product
Hilbert space and a finite local Hilbert space.  However, as we show below,
our numerical results indicate that the model is likely gapless and hence
does not correspond to a gapped chiral phase.  We also consider a
$\mathbb Z_N$ generalization, in which the line algebra enforces an at least
$N$-fold degeneracy of every level on the torus, reminiscent of
$SU(N)_1$ Chern--Simons theory.  Again, in the solvable limit $N\to\infty$,
we find that the model is gapless rather than a gapped topological state.

Since the chiral XYZ model does not appear to be exactly solvable, the tools
available to study its ground state are rather limited. Fortunately, all the
$\triangle$ terms commute with all the $\downtriangle$ terms, and therefore
this model possesses the advertised dynamical splitting; the two triangle families generate the commuting algebras $\mathcal A$ and $\mathcal B$. The center of $\mathcal A\vee\mathcal B$ is generated by the independent central
combinations of the subsystem line symmetries and their eigenvalues are the labels $\lambda$ in Eq.~\eqref{Eq:generalHilbertFactorization}.  The remaining
anticommuting row--column pair lies outside this center and supplies the
aforementioned logical-qubit and its exact twofold degeneracy on the torus. When supplemented
with subsystem symmetries, this dynamical splitting allows one to access its properties on
systems of up to $9\times9$ lattice sites with rather modest resources. To
illustrate the huge reduction associated with these symmetries, on a
$3\times3$ lattice the Hamiltonian in each symmetry sector takes the form in
Eq.~\eqref{Eq:generalHamiltonianFactorization}, where $H_A(\lambda)$ and
$H_B(\lambda)$ each act on a \textit{single} qubit:
$H_A(\lambda)=-\vec h_A\cdot\vec\tau_A$ and
$H_B(\lambda)=-\vec h_B\cdot\vec\tau_B$.  Here $\vec\tau_A$ and $\vec\tau_B$
are Pauli operators that act on the factor qubits. More generally, for a 
lattice with $L^2$ qubits, we reduce the diagonalization problem to two independent problems, each involving $(L-1)(L-2)/2$ qubits.  The reduction linear in $L$ originates from
the $O(L)$ subsystem symmetries, while the crucial factor of $1/2$ results
from the dynamical splitting.

We now summarize our main results:
\begin{enumerate}
	\item In Sec.~\ref{sec:model-symmetries}, we determine the exact
	subsystem-symmetry algebra of the chiral
	XYZ model.  The noncommuting line operators enforce a doubled on the torus and, more generally, a degeneracy of at least \(2^g\)
	on a closed surface of genus \(g\); the states within this 
	multiplet are locally indistinguishable.  After fixing the commuting
	center charges \(\lambda\), we establish the dynamical splitting
	\begin{align}
	\mathcal H_\lambda&\simeq
	\mathbb C^2_P\otimes\mathbb C^{2^q}_A\otimes\mathbb C^{2^q}_B,
	\nonumber\\
	H\big|_\lambda&=\mathds{1}_P\otimes\left[
	H_A(\lambda)\otimes\mathds{1}_B+
	\mathds{1}_A\otimes H_B(\lambda)\right].
	\label{eq:intro-split-result}
	\end{align}
	The factor Hamiltonians $H_A(\lambda), H_B(\lambda)$ remain interacting and are not known to be exactly solvable. The Hilbert space $\mathbb C^2_P$ corresponds to  the exact projective doublet.  For an
	$L\times L$ torus, $q=(L-1)(L-2)/2$.

  In Appendix~\ref{app:tensor-factorization}, we provide an explicit mapping from the triangle operators  defining the chiral XYZ Hamiltonian to Pauli operators acting on the factor qubits. This mapping may be viewed as a duality between the
  algebra of the triangle operators and the Pauli algebras corresponding to factor qubits. With open boundaries, the
   resulting factor Hamiltonians have bounded Pauli weight and are geometrically
   local.  On the torus, all but an $O(L)$ set of terms retain this local form;
   the remaining terms lie along convention-dependent cuts and can have
   $O(L)$ Pauli weight.
   
	\item In Sec.~\ref{sec:large-N}, we construct a $\mathbb{Z}_N$ clock
	generalization and solve its large-\(N\) limit.  We find a
	translationally invariant nodal Bose liquid with dispersion
	\begin{equation}
	\omega({\bm k})\sim\left|
	\sin\frac{k_x}{2}\sin\frac{k_y}{2}
	\sin\frac{k_x+k_y}{2}\right|,
	\label{eq:intro-largeN-dispersion}
	\end{equation}
	and hence three lines of zero-energy modes in the Brillouin zone.
	
	\item In Sec.~\ref{sec:ed-pure}, we use ED to study the ground state of the spin-$1/2$ chiral XYZ model.  The above tensor decomposition makes sizes up to \(9\times9\) accessible. We find strong evidence for a gapless state. Further, the half-torus entanglement is	consistent with $L\log L$ scaling, while entanglement on cylindrical/ladder geometries
	has the dependence expected from 1+1-D gapless channels whose number increases with the transverse size.
	Together, the gap and entanglement results provide evidence for a nodal Bose liquid at $N=2$.
	
	\item In Sec.~\ref{sec:perturbations}, we perturb the chiral XYZ model to probe the stability of the nodal liquid and access proximate phases:
	\begin{enumerate}[label=(\roman*),leftmargin=1.7em,itemsep=1pt,topsep=2pt]
		\item A deformation preserving the dynamical splitting structure produces a first-order
		transition at large \(N\) between uniform and a period-three nodal
		liquids.  At \(N=2\), exact diagonalization finds
		commensuration-dependent reconstruction of the center sector, consistent with a transition from a uniform state to an inhomogeneous state as a function of the deformation strength.
		\item The model admits a natural interpolation between the pure chiral XYZ Hamiltonian and two decoupled Wen--plaquette
		models~\cite{WenPlaquette2003}. This deformation preserves the subsystem symmetries but
		breaks the dynamical splitting.  The
		finite-size gap and Kitaev--Preskill topological entanglement entropy (TEE)~\cite{Kitaev06_1} are consistent with a finite-coupling
		transition from the nodal liquid to the gapped
		\((\mathrm{Wen\!-\!plaquette})^{\otimes2}\) topological phase.
		\item Finally, we consider perturbations that explicitly break the
		subsystem symmetries.  Within a controlled large-$N$ RG calculation, the leading subsystem-symmetry-breaking
		perturbations are relevant at the dynamically split point, while more generally, the relevance/irrelevance depends on a certain stiffness ratio, analogous to the ratio of the ring-exchange coupling to the charging energy in  exciton Bose liquid~\cite{ParamekantiBalentsFisher2002,Lake2022EBLRG}.  As a representative example of such a perturbation in the $N=2$ model, we study the chiral XYZ model supplemented with the uniform onsite field
		$\sum_{\bm r}(X_{\bm r}+Y_{\bm r}+Z_{\bm r})$.  ED on $3\times3$ and $4\times4$ tori finds a	strongly size-dependent reorganization of the low-energy spectrum that
		is suggestive of $\mathcal C_3$ breaking, a symmetry-breaking pattern also suggested by single-site mean-field.  However, the available sizes do not conclusively establish this symmetry breaking nor determine whether an arbitrarily weak field immediately opens a gap in the nodal liquid.  At larger
		fields, the ground state becomes nondegenerate and evolves toward the symmetric polarized state.
		
	\end{enumerate}
\end{enumerate}

We conclude with a discussion in Sec.~\ref{sec:discussion}.
  
\section{The Chiral XYZ model and its symmetries}
\label{sec:model-symmetries}

In this section we discuss in detail the symmetries of the chiral XYZ model
introduced above (Fig.~\ref{fig:modeldef}). This model has several distinctive
features. Its elementary terms do not commute, and hence the
ground state cannot be found by minimizing the terms independently.  At the
same time, these terms generate two mutually commuting operator algebras and
an $O(L)$ set of subsystem symmetries.  The latter guarantee an exact topology-dependent 
degeneracy for each distinct eigenvalue in its spectrum, while the former reduce the number of active qubits in exact
diagonalization by approximately a factor of two.

\subsection{The model}
\label{subsec:pure-model}
 The Hilbert space of our model consists of spin-$1/2$ spins on the sites of a triangular lattice (see Fig.~\ref{fig:modeldef}). We choose oblique primitive vectors
\begin{equation}
{\bm e}_x=(1,0),\qquad
{\bm e}_y=\left(-\frac12,\frac{\sqrt3}{2}\right),
\label{eq:triangular-primitive-vectors}
\end{equation}
and label the site
${\bm r}=x{\bm e}_x+y{\bm e}_y$ by the integer pair $(x,y)$.  We choose periodic boundary conditions unless otherwise mentioned, i.e.,
$(x,y)\sim(x+L_x,y)\sim(x,y+L_y)$ so that the topology of the lattice is a torus with $L_xL_y$ sites.  Notice that
${\bm e}_x$, ${\bm e}_y$, and $-({\bm e}_x+{\bm e}_y)$ are three
symmetry-related nearest-neighbor directions on the triangular lattice. When $L_x=L_y=L$, we refer to this geometry as an $L\times L$ torus.  

The primitive rhombus spanned by ${\bm e}_x$ and ${\bm e}_y$ contains two
elementary triangles $\triangle, \downtriangle$.  We define a three-spin operator on each of them:
\begin{align}
a_{x,y}&=X_{x,y}Y_{x,y+1}Z_{x+1,y+1},
\nonumber\\
b_{x,y}&=Z_{x,y}Y_{x+1,y}X_{x+1,y+1}.
\label{eq:model-triangles}
\end{align}
The Hamiltonian is given by
\begin{align}
H_{\rm XYZ}&=-J\sum_{x,y}\left(a_{x,y}+b_{x,y}\right)
\equiv H_A+H_B,
\nonumber\\
H_A&=-J\sum_{x,y}a_{x,y},\,\,\,
H_B=-J\sum_{x,y}b_{x,y},
\label{eq:model-HXYZ}
\end{align}
where we take $J>0$ \footnote{since the complex conjugate $H^{*}$ of $H$ satisfies $H^{*} = -H$, the ground state of $-H$ is the complex conjugate of the ground state of $H$ and is expected to be qualitatively similar}; we set $J=1$ in the numerical simulations.  Each elementary
triangle contains all three Pauli matrices once with the same cyclic orientation $X\rightarrow Y\rightarrow Z$, which motivates the name
``chiral XYZ model.''  This model was introduced in Ref.~\cite{LiFranzMajoranaHubbard2018} by
mapping the strong-coupling limit of the Majorana--Hubbard model on the
honeycomb lattice to spins on the triangular lattice through a Jordan--Wigner transformation.  It was further suggested that its ground state might be a quantum spin-liquid. Under the same transformation, however, the Majorana hopping term becomes a sum of
nonlocal spin strings and therefore does not correspond to a local perturbation of the spin-only model studied here. More recently, a subsystem code inspired by this model was studied in Ref.~\cite{BusseToikkaMajoranaXYZ2026}.

Neither $H_A$ nor $H_B$ is a commuting Hamiltonian separately: whenever two distinct triangles of the same orientation share a corner, the corresponding
operators anticommute.  However,
\begin{equation}
[a_{\bm r},b_{{\bm r}'}]=0
\qquad\text{for all }{\bm r},{\bm r}'.
\label{eq:model-cross-commute}
\end{equation}
This  distinctive property is stronger than $[H_A,H_B]=0$ and will play an essential role in our discussion.

As an aside, the Pauli algebra generated jointly by
$\{a_{\bm r},b_{\bm r}\}$ admits an embedding into the ``gauge algebra''
of the Bacon--Shor subsystem code~\cite{Bacon2006Subsystem}: each $XYZ$
triangle is, up to phase, the product of neighboring $XX$ and $ZZ$ gauge
bonds. The two algebras coincide only when \(\gcd(L_x,L_y)=1\), whereas the embedding is proper whenever \(\gcd(L_x,L_y)>1\), in
particular whenever $L_x = L_y > 1$.

\textbf{Motivation from Chern--Simons theory.}
Our original motivation for $H_{\rm XYZ}$ was to imitate the commutation relations in a Chern--Simons theory.  For this analogy, we regard the $\hat x$- and $\hat y$-directed bonds as forming a square lattice. On a square lattice, each lattice site can be associated with one horizontal and one vertical bond. After a choice of orientation, we regard $X_{\bm r}$ and $Z_{\bm r}$ as the
compact link variables
$U_x=e^{\mathrm{i}A_x}$ and $U_y=e^{\mathrm{i}A_y}$ assigned to these two
bonds.  Their Pauli algebra,
\begin{equation}
 U_yU_x=-U_xU_y,
\label{eq:model-CS-Weyl}
\end{equation}
can then be thought of as the exponentiated form of the commutation relation for the $U(1)_2$ Chern--Simons, 
$[A_x,A_y]=\mathrm{i}\pi$.

To make the relation with the two triangle operators $a_{\bm r}, b_{\bm r}$ in Eq.~\eqref{eq:model-triangles} explicit, let's define the link variables \begin{equation}
 U_x({\bm r}\rightarrow{\bm r}+\hat x)=X_{\bm r},
 \qquad
 U_y({\bm r}-\hat y\rightarrow{\bm r})=Z_{\bm r}.
\label{eq:model-CS-link-assignment}
\end{equation}
The magnetic flux through the plaquette whose lower-left corner is $\bm r$ is
\begin{align}
 W_p({\bm r})
 &=U_x({\bm r})U_y({\bm r}+\hat x+\hat y)
 \nonumber\\
 &\quad\times U_x({\bm r}+\hat y)^\dagger
 U_y({\bm r}+\hat y)^\dagger
 \nonumber\\
 &=X_{\bm r}Z_{{\bm r}+\hat x+\hat y}
 X_{{\bm r}+\hat y}Z_{{\bm r}+\hat y}
 \nonumber\\
 &=-\mathrm{i}a_{\bm r},
\label{eq:model-CS-plaquette}
\end{align}
Thus $a_{\bm r}=\mathrm{i}W_p({\bm r})$ is proportional to the flux operator.

To motivate the second triangle operator, let's temporarily introduce canonical
coordinates $A_i$ and their conjugate momenta $E_i$.  The time-derivative part of the level-$k$ Chern--Simons Lagrangian is $\frac{k}{4\pi}\left(A_y\dot A_x-A_x\dot A_y\right)$. For level $k=2$, one may therefore write
\begin{align}
 E_x\equiv\frac{\partial\mathcal L_{\rm CS}}{\partial\dot A_x}
 &=\frac{1}{2\pi}A_y,\,\,
 E_y\equiv\frac{\partial\mathcal L_{\rm CS}}{\partial\dot A_y}
 =-\frac{1}{2\pi}A_x,
 \nonumber\\
 \Rightarrow \nabla\!\cdot{\bm E}
 &=\frac{1}{2\pi}\left(\partial_xA_y-\partial_yA_x\right).
\label{eq:model-CS-electric-projection}
\end{align}
Thus, in the continuum, the electric divergence becomes proportional to the same curl represented
by $W_p$.  On the lattice, the momentum conjugate to a compact link variable
acts by shifting that variable.  At level two, $Z$ supplies the nontrivial
shift of an $X$ link, i.e., $ZXZ^\dagger=-X$ while $X$ similarly shifts a $Z$ link. The horizontal and vertical link shifts are therefore represented by
$Z_{\bm r}$ and $X_{\bm r}$, respectively.  Their product over the
four links incident on a vertex is the compact lattice divergence, and is given by
\begin{align}
G({\bm r} + \hat{x})
&=Z_{\bm r}Z_{{\bm r} + \hat{x}}X_{{\bm r} + \hat{x}}X_{{\bm r}+ \hat x + \hat y}
\nonumber\\
&=\mathrm{i}b_{{\bm r}},
\label{eq:model-CS-star}
\end{align}
Hence $b_{\bm r}$ is the
analogue of
$\nabla\!\cdot{\bm E}$.

This analogy is suggestive but it does not imply that
$H_{\rm XYZ}$ is a lattice realization of pure Chern--Simons theory.  In
particular, the model has no microscopic local gauge redundancy.  Moreover, in pure Chern--Simons theory,
$A_0$ is a Lagrange multiplier which imposes the local constraint $F_{xy}\equiv\partial_xA_y-\partial_yA_x=0$, whereas no analogous constraint is imposed in our
spin model.  Nevertheless, as discussed below in more detail, the model retains one  aspect of $SU(2)_1$
Chern--Simons theory: the noncontractible line operators on a torus include an
anticommuting pair that commutes with the Hamiltonian and enforces an exact
twofold degeneracy.

\subsection{Subsystem symmetries and  topology-dependent exact degeneracy}
\label{subsec:subsystem-symmetries}

We first consider periodic boundary conditions, so that the rectangular
lattice has the topology of a torus, and then briefly discuss an extension to higher-genus surfaces. The ED studies discussed in subsequent sections are also all done on a torus.

The most transparent symmetries of the model are Pauli strings that wrap
around the torus.  Let
\begin{equation}
n_D\equiv\gcd(L_x,L_y),\qquad
a\equiv\frac{L_x}{n_D},\qquad
b\equiv\frac{L_y}{n_D}.
\label{eq:model-number-diagonal-strings}
\end{equation}
Here $n_D$ is the number of distinct diagonal strings that wrap around the torus, while $a$ and $b$
are coprime winding integers along the two directions.  The model possesses several subsystem symmetries. To see this, let's define
\begin{align}
C_x&=\prod_{y=0}^{L_y-1}Z_{x,y},
\nonumber\\
R_y&=\prod_{x=0}^{L_x-1}X_{x,y},
\nonumber\\
D_s&=\prod_{x-y=s\ ({\rm mod}\ n_D)}Y_{x,y}.
\label{eq:model-line-operators}
\end{align}
Here $x=0,\ldots,L_x-1$, $y=0,\ldots,L_y-1$, and
$s=0,\ldots,n_D-1$.  The letters $C, R, D$ stand for column, row and diagonal respectively. The operators $D_s$ are the distinct closed strings in
the $(1,1)$ lattice direction.  Each elementary triangle of the lattice intersects any string in
Eq.~\eqref{eq:model-line-operators} in a way that produces an even number of
local anticommutations with any $a_{x,y}$ and $b_{x,y}$ term in the Hamiltonian.  Consequently, every $C_x$, $R_y$, and $D_s$
commutes with every term in the Hamiltonian.

Crucially, these symmetries cannot all be diagonalized simultaneously.  A
column and a row operator intersect once, whereas a column and a row intersect each diagonal string at $b$ and $a$ times, respectively.  Therefore,
\begin{align}
C_xR_y&=-R_yC_x,
\nonumber\\
C_xD_s&=(-1)^bD_sC_x,
\nonumber\\
R_yD_s&=(-1)^aD_sR_y.
\label{eq:model-line-Heisenberg}
\end{align}
More generally, due to the relation
$\prod_x C_x\prod_y R_y\prod_s D_s=i^{L_xL_y}$, the binary span of the line
operators has rank $L_x+L_y+n_D-1$. Of these,
$L_x+L_y+n_D-3$ independent combinations commute with every line operator.
The remaining noncommuting part consists of a single anticommuting pair,
which we may choose as
\begin{equation}
Z_P=C_0,\qquad X_P=R_0. \label{eq:logicalstrings}
\end{equation}
If a state is a simultaneous eigenstate of $H$ and $C_0$, acting with $R_0$
leaves its energy unchanged while reversing its $C_0$ eigenvalue.  Every
energy level is therefore at least twofold degenerate.
Although the operator algebra leading to this exact degeneracy is similar to
that typically associated with topologically ordered gapped ground states,
it does not rely on a spectral gap. Due to this resemblance, we refer to the
two-dimensional subspace associated with this anticommuting pair as a
``logical qubit.''

As mentioned above, the combinations of line operators that commute with
every line symmetry form a center of binary rank
$c=L_x+L_y+n_D-3$. Their eigenvalues provide convenient quantum numbers for
block-diagonalizing the Hamiltonian. A natural set of central operators is
$c_x=C_xC_0$, $r_y=R_yR_0$, and $d_s=D_sD_0$.  We use the same lowercase
symbols for their $\pm1$ eigenvalues when labeling a center sector; whether
an operator or its eigenvalue is meant will be clear from context.  We use
the same convention for the mixed generator $Q_{\rm mix}$ below.
These `relative strings' are independent unless both $L_x$ and $L_y$ are even. On an even-by-even lattice they obey one relation
$\prod_{x>0}c_x
\prod_{y>0}r_y
\prod_{s>0}d_s=\mathds{1}$, and therefore, one relative string must be omitted and replaced by an additional mixed central generator $Q_{\rm mix}$. The explicit form and sign convention for $Q_{\rm mix}$, needed for labeling the numerical symmetry sectors, are given in Appendix~\ref{app:mixed-center-generator}.

The chiral XYZ Hamiltonian has an additional antiunitary
symmetry whenever at least one of $L_x$ or $L_y$ is even.  For example, when $L_x$ is even,
\begin{equation}
\Theta_x=
\left(
\prod_{\substack{0\leq x<L_x\\ x\ {\rm even}}}
\prod_{y=0}^{L_y-1}X_{x,y}
\right)K,
\label{eq:model-checkerboard-antiunitary}
\end{equation}
commutes with $H_{\rm XYZ}$, where $K$ denotes complex conjugation in the
$Z$ basis; an analogous row/$Z$ symmetry exists when $L_y$ is even.  If
only one circumference is even, this symmetry pairs distinct center
sectors at the same energy, so together with the logical doublet every level
is at least fourfold degenerate. In Sec.~\ref{subsec:current-deformation-ed} we will consider a deformation that is odd under
$\Theta_x$, and which maintains all subsystem symmetries. Consequently, it will  lift the additional degeneracy associated with $\Theta_x$ while preserving the degeneracy associated with the logical qubit.

\subsubsection{Local indistinguishability and long-range entanglement in the logical sector }
\label{subsubsec:local-indistinguishability}
Here we show that the two states corresponding to the aforementioned logical qubit are locally indistinguishable. The argument is almost identical to that encountered in the context of gapped, topologically ordered states. Let
$O_A$ be any operator supported in a contractible region $A$ whose complement contains at  least one
complete row and one complete column. This condition is satisfied by any fixed bounded region on a sufficiently large torus. Let's choose such a column $C_x$ and a row $R_y$ lying entirely outside $A$.  Within a fixed center sector, they
differ from $C_0$ and $R_0$ only by fixed signs, and hence act as the same
logical $Z_P$ and $X_P$.  Since both strings avoid touching region $A$, the operator $O_A$
commutes with the full Pauli algebra of the logical qubit.  Thus, for fixed
values $\mu$ of all nonlogical labels, one obtains
\begin{equation}
\langle\psi_{s,\mu}|O_A|\psi_{s',\mu}\rangle
=\langle O_A\rangle_{\mu}\,\delta_{s,s'},
\label{eq:model-local-indistinguishability}
\end{equation}
where $\langle O_A\rangle_{\mu}$ is independent of the logical label $s$.
In particular, the two protected states have identical reduced density
matrices on $A$. Equation~\eqref{eq:model-local-indistinguishability} has the same form as the
standard local-indistinguishability condition for topologically ordered
ground states, or equivalently the quantum error-correction condition against
erasure of the region $A$
\cite{BravyiHastingsMichalakisTQO,KnillLaflammeQEC}.
In the present setting, however, it follows directly from the line symmetries
and does not by itself imply a gapped topologically ordered phase.

Local indistinguishability implies that  the ground state manifold is long-range entangled: any geometrically local circuit that prepares a state in this manifold must have depth $\Omega(L)$~\cite{BravyiHastingsVerstraete2006}.

\subsubsection{Extension to higher-genus surfaces}
\label{subsubsec:higher-genus}

The extension to a surface $\Sigma_g$ of genus $g$ requires some care.
Because the line operators are rigid subsystem symmetries rather than
deformable Wilson loops, their existence is not guaranteed on an arbitrary
triangulation of $\Sigma_g$.  For each fixed $g$, however, one can choose a
thermodynamic family of lattices, with linear size $L$, that agrees with the
regular triangular lattice except within $O(g)$ localized defect regions where the local coordination number differs from six.
The total number of sites in these regions is independent of $L$, and the
Hamiltonian is modified only on these isolated patches.

Precisely because the number of defects does not scale with system size, for sufficiently large $L$, each handle supports a noncontractible $Z$ string
$\mathcal Z_i$ following a complete lattice column and a noncontractible $X$
string $\mathcal X_i$ following a complete lattice row, both chosen to avoid
the defect regions.  The pair on the same handle intersects once, whereas
strings associated with different handles can be chosen disjoint.  The usual
local cancellation therefore gives
\begin{align}
\mathcal Z_i\mathcal X_j
&=(-1)^{\delta_{ij}}\mathcal X_j\mathcal Z_i,
\nonumber\\
[\mathcal Z_i,\mathcal Z_j]
&=[\mathcal X_i,\mathcal X_j]=0,
\nonumber\\
[H,\mathcal Z_i]&=[H,\mathcal X_i]=0,
\,\,\, i,j=1,\ldots,g.
\label{eq:model-genus-string-algebra}
\end{align}
Thus the symmetry algebra contains $g$ independent logical qubits. Therefore, every energy level
has multiplicity at least $2^g$. Since each handle contains many parallel defect-free rows and columns, the logical
strings can also be chosen to avoid any prescribed bounded contractible
region.  The resulting $2^g$-dimensional symmetry multiplets are therefore
locally indistinguishable by the same argument as on the torus.  Note this is only an existence statement for a family of lattice Hamiltonians constructed above, and not a claim that every triangulation of $\Sigma_g$ possesses these locally indistinguishable degenerate states.

\subsection{Dynamical splitting and associated tensor-product structure}
\label{subsec:tensor-product-structure}

Resolving the commuting line symmetries already divides the Hilbert space into
smaller sectors, but it does not account for the largest reduction used in our
exact diagonalization.  Recall that the Hamiltonian is
\begin{equation}
H_{\rm XYZ}=H_A+H_B,
\end{equation}
where $H_A$ and $H_B$ are sums of the $a$- and $b$-triangle terms,
respectively.  As already noted, every term in $H_A$ commutes with every term in $H_B$ and therefore $H$ has the structure discussed in the Introduction (Eq.~\eqref{Eq:ABdecompose}). As discussed there,
although the two sets of terms act on the same microscopic spins, this mutual
commutativity allows them to act independently once their common central quantum numbers are fixed. Let us discuss the associated structure in more detail.

Let $\mathfrak A$ and $\mathfrak B$ denote the operator algebras generated by
the $a$- and $b$-triangle terms, and let $V_A$ and $V_B$ be the corresponding
binary Pauli spaces. Writing $n=L_xL_y$ and $c=L_x+L_y+n_D-3$, exact binary row
reduction gives
\begin{gather}
\dim V_A=\dim V_B=n-1,
\nonumber\\
V_A\cap V_B=Z,\qquad \dim Z=c,
\nonumber\\
\dim(V_A/Z)=\dim(V_B/Z)=2q,
\nonumber\\
2q=n-L_x-L_y-n_D+2.
\label{eq:model-factor-ranks}
\end{gather}

Here $Z$ is precisely the span of the central line symmetries discussed
above.  After fixing their eigenvalues $\lambda$, the remaining generators in
each triangle family can be organized into $q$ independent anticommuting Pauli
pairs.  Thus, the $a$- and $b$-triangle algebras act on separate
$2^q$-dimensional factors.

A fixed center sector has dimension
\begin{equation}
\dim\mathcal H_\lambda=2^{n-c}=2^{2q+1}.
\end{equation}
The two triangle algebras account for two $q$-qubit factors, while the
remaining factor of two is the logical qubit associated with the
anticommuting row and column symmetries.  Therefore,
\begin{equation}
\boxed{
	\mathcal H_\lambda\simeq
	\mathbb C^2_P\otimes
	\mathbb C^{2^q}_A\otimes
	\mathbb C^{2^q}_B.}
\label{eq:model-factorization}
\end{equation}

In this decomposition, the two parts of the Hamiltonian act on different
factors,
\begin{equation}
H_{\rm XYZ}\big|_\lambda=
\mathds{1}_P\otimes
\left[
H_A(\lambda)\otimes\mathds{1}_B
+
\mathds{1}_A\otimes H_B(\lambda)
\right].
\label{eq:model-factor-H}
\end{equation}

It is important to note that Eq.~\eqref{eq:model-factorization} is an
algebraic, rather than spatial, tensor-product decomposition. The two
emergent sets of `factor qubits' are defined only after fixing the common
center sector $\lambda$ and can be encoded nonlocally across the entire
lattice; the precise mapping between the physical qubits and factor qubits is
not unique and depends on the details of the Gram--Schmidt procedure used to
define them. Accordingly, both $H_A(\lambda)$ and $H_B(\lambda)$, when
expressed in terms of factor qubits, need not be geometrically local.  Nevertheless, the factor-qubit mapping and the corresponding factor Hamiltonians can be constructed in time polynomial in $L$, since symplectic Gram--Schmidt reduces the
problem to linear algebra over $\mathbb F_2$. Of course, this efficient construction does not make the factor Hamiltonians exactly solvable; diagonalizing them still requires resources
exponential in the number of factor qubits. Recovering the wavefunction in the original spin basis requires applying the inverse Clifford
encoding. Appendix~\ref{app:tensor-factorization} gives details of this
encoding and its numerical implementation.  In particular, it gives a construction valid on every $L \times L$ torus with a geometrically local bulk and
$O(L)$ terms along certain convention-dependent cuts used to open the torus, together with explicit expressions for
$H_A(\lambda)$ and $H_B(\lambda)$ on several finite tori. The mapping may be viewed as a duality between each triangle algebra and the
Pauli algebra of $q=(L-1)(L-2)/2$ factor qubits, rather than as a site-by-site
mapping of the $L^2$ physical qubits.

If $e_i^A(\lambda)$ and $e_j^B(\lambda)$ are the eigenvalues of the two factor
Hamiltonians, the complete energies in this center sector are
\begin{equation}
E_{ij}(\lambda)=e_i^A(\lambda)+e_j^B(\lambda),
\label{eq:model-additive-spectrum}
\end{equation}
with an exact twofold multiplicity from the logical qubit.  The numerical consequence is dramatic.  A conventional calculation that
fixes the center and one projective charge acts on $2^{2q}$ states. Here, we instead need to diagonalize two
matrices of dimension $2^q$.

Despite dynamical splitting, it is important that the Hamiltonian contains both $H_A$ and $H_B$ terms, even though all terms in $H_A$ commute with all terms in $H_B$. If the Hamiltonian contained only $H_A$, then in a given sector $\lambda$,  the Hamiltonian would be just $\mathds{1}_P\otimes
H_A(\lambda)\otimes\mathds{1}_B$ leading to a   degeneracy of at least
\[
2\,\dim{\cal H}_{B,\lambda}=2^{q+1},
\]
where $q=\frac{(L-1)(L-2)}{2}$ for an $L \times L$ torus as discussed above. Therefore, if only one of the two terms, $H_A$ or $H_B$ is present then the system has an extensive zero-temperature entropy, contrary to the usual third-law expectation.  The low-energy
physics of the chiral XYZ model is therefore a property of the combined \(A\)- and \(B\)-triangle dynamics, not of either $H_A$ or $H_B$ separately.

\section{$\mathbb{Z}_N$ generalization and large-$N$ solution}
\label{sec:large-N}

The subsystem symmetry algebra admits a natural generalization from Pauli
spins to $\mathbb Z_N$ clock variables.  Apart from providing a controlled large-$N$
limit, this generalization gives a suggestive picture for the soft modes
of the $N=2$ model seen in the ED numerics.  A continuation away from
$N=2$ is not unique.  We require only that at $N=2$ the model reduces to the
one introduced in the previous section.

\subsection{$\mathbb{Z}_N$  chiral XYZ model}
\label{subsec:large-N-pure}

At each site $\bm r$, let $X_{\bm r}$ and $Z_{\bm r}$ obey the
$\mathbb Z_N$ clock algebra
\begin{equation}
Z_{\bm r}X_{\bm r}=\omega X_{\bm r}Z_{\bm r},
\qquad
\omega=e^{2\pi\mathrm{i}/N},
\qquad
X_{\bm r}^{N}=Z_{\bm r}^{N}=1 .
\label{eq:pure-largeN-Weyl}
\end{equation}
The unitary clock analogue of the Pauli $Y$ operator is
\begin{equation}
\mathcal Y_{\bm r}
=\eta X_{\bm r}^{\dagger}Z_{\bm r}^{\dagger},
\qquad
\eta=e^{\pi\mathrm{i}(N-1)/N}.
\label{eq:pure-largeN-Y}
\end{equation}
This convention gives $\mathcal Y^N=1$ and $\mathcal Y=Y$ at $N=2$.
We define the two clock-triangle operators (again on the same geometry as
Fig.~\ref{fig:modeldef}) as
\begin{align}
a_{\bm r}
&=-\omega X_{\bm r}\mathcal Y_{{\bm r}+\hat y}
Z_{{\bm r}+\hat x+\hat y},
\nonumber\\
b_{\bm r}
&=-\omega Z_{\bm r}\mathcal Y_{{\bm r}+\hat x}
X_{{\bm r}+\hat x+\hat y}.
\label{eq:pure-largeN-ab}
\end{align}
Analogous to the $N=2$ model, they obey
\begin{equation}
[a_{\bm r},b_{{\bm r}'}]
=[a_{\bm r},b_{{\bm r}'}^{\dagger}]
=0
\label{eq:pure-largeN-ab-commute}
\end{equation}
for all $\bm r,\bm r'$. Indeed, the vanishing of $[a_{\bm r},b_{{\bm r}'}]$ follows from cancellation
of the Weyl phases at every possible overlap, while the relations involving
adjoints then follow because the triangle operators $a_r, b_r$ are all unitary.
The Hamiltonian is
\begin{equation}
H_N=-\frac12\sum_{\bm r}
\left(a_{\bm r}+a_{\bm r}^{\dagger}
+b_{\bm r}+b_{\bm r}^{\dagger}\right)
\equiv H_A+H_B .
\label{eq:pure-largeN-H}
\end{equation}
Due to the commutation relations in Eq.~\eqref{eq:pure-largeN-ab-commute}, the Hamiltonian  possesses the dynamical splitting property analogous to the $N=2$ model, and at $N=2$,  it reduces  to
$H_{\rm XYZ}$ as desired.

\paragraph{Subsystem symmetries and exact degeneracy.}
On an $L_x\times L_y$ torus, define the column and row operators
\begin{equation}
C_x=\prod_y Z_{x,y},
\qquad
R_y=\prod_x X_{x,y}.
\label{eq:pure-largeN-row-column}
\end{equation}
One may verify that
\begin{equation}
[H_N,C_x]=[H_N,R_y]=0 .
\label{eq:pure-largeN-row-column-symmetry}
\end{equation}
There are also diagonal subsystem symmetries analogous to the $N=2$ case.  If
$d=\gcd(L_x,L_y)$, one convenient choice is
\begin{equation}
D_s=
\prod_{x-y=s\ ({\rm mod}\ d)}
\left(\eta X_{x,y}Z_{x,y}\right),
\,\,\, s=0,\ldots,d-1 ,
\label{eq:pure-largeN-diagonal}
\end{equation}
where the phase $\eta$ makes each on-site factor an order-$N$ unitary but
does not affect any commutation relation. Again, one may verify that
$[H_N,D_s]=0$.

The row and column symmetries have the standard projective commutation relation
\begin{equation}
C_xR_y
=\omega\,R_yC_x .
\label{eq:pure-largeN-line-Heisenberg}
\end{equation}
Choosing a simultaneous eigenstate of $H_N$ and $C_x$, with
$C_x|\psi\rangle=c|\psi\rangle$,
Eq.~\eqref{eq:pure-largeN-line-Heisenberg} gives
\begin{equation}
C_xR_y^m|\psi\rangle
=c\,\omega^mR_y^m|\psi\rangle,
\qquad m=0,\ldots,N-1 .
\label{eq:pure-largeN-Nplet}
\end{equation}
The $N$ eigenvalues $c\omega^m$ are distinct, while all the states in
Eq.~\eqref{eq:pure-largeN-Nplet} have the same energy.  Every energy level
is therefore at least $N$-fold degenerate on a torus.  The higher-genus
construction of Sec.~\ref{subsubsec:higher-genus} extends directly: there
are $g$ independent Heisenberg pairs of noncontractible subsystem strings,
giving an exact degeneracy of at least $N^g$. Note that the algebra of
$C_x,R_y$ is the same finite Heisenberg algebra obeyed, up to orientation conventions,
by noncontractible Wilson loops in $SU(N)_1$ Chern--Simons theory.  It therefore enforces the same $N$-dimensional multiplet structure.  This
algebraic analogy does not, however, imply that the present model realizes a gapped topologically ordered phase.

\subsection{Large-$N$ solution}
\label{subsec:large-N-solution}

In the $N\to\infty$ limit, $2\pi/N$ acts as an effective Planck
constant.  We expand semiclassically about the classical ground-state
configuration $A_x=A_y=0$, or any symmetry-related configuration satisfying
$\theta_A=\theta_B=0$ modulo $2\pi$.  In this semiclassical limit, the clock algebra may be represented as
$X_{\bm r}=e^{\mathrm{i}A_x(\bm r)}$ and
$Z_{\bm r}=e^{\mathrm{i}A_y(\bm r)}$, with
\begin{equation}
[A_x(\bm r),A_y(\bm r')]
=\frac{2\pi\mathrm{i}}{N}\delta_{\bm r,\bm r'},
\label{eq:pure-largeN-A-commutator}
\end{equation}
to leading order in $1/N$.
At leading order in the large-$N$ expansion,
\begin{equation}
a_{\bm r}\longrightarrow e^{\mathrm{i}\theta_A(\bm r)},
\qquad
b_{\bm r}\longrightarrow e^{\mathrm{i}\theta_B(\bm r)},
\label{eq:pure-largeN-ab-symbols}
\end{equation}
where $\theta_A(\bm r)
=A_x(\bm r)-A_x(\bm r+\hat y)
-A_y(\bm r+\hat y)+A_y(\bm r+\hat x+\hat y)$ and
$\theta_B(\bm r)
=A_y(\bm r)-A_y(\bm r+\hat x)
-A_x(\bm r+\hat x)+A_x(\bm r+\hat x+\hat y)$.  Up to an additive
constant, the resulting classical Hamiltonian is
\begin{equation}
H_N=-\sum_{\bm r}
\left[\cos\theta_A(\bm r)+\cos\theta_B(\bm r)\right].
\end{equation}
The harmonic approximation to $H_N$ is
\be 
H_N^{(2)}
=\frac12\sum_{\bm r}
\left[\theta_A(\bm r)^2+\theta_B(\bm r)^2\right]. \label{eq:thetaAthetaBquadratic}
\ee 

This harmonic approximation is controlled at large $N$: since
$[A_x,A_y]=O(N^{-1})$, fluctuations about the classical minimum satisfy
$\theta_{A,B}=O(N^{-1/2})$, so the quartic and higher terms in the cosine
produce relative corrections of order $1/N$ and smaller.

Within the harmonic approximation, straightforward algebra gives the
Heisenberg equations of motion for the Fourier-transformed fields:

\begin{equation}
\ddot A_i(\bm k)
=- \Omega(\bm k)^2 A_i(\bm k),
\label{eq:pure-largeN-second-order-eom}
\end{equation}

where 
\begin{equation}
\boxed{
	\Omega(\bm k)=\frac{16\pi}{N}
	\left|
	\sin\frac{k_x}{2}\sin\frac{k_y}{2}
	\sin\frac{k_x+k_y}{2}
	\right| .}
\label{eq:pure-largeN-lattice-dispersion}
\end{equation}

The dispersion vanishes on the three subsystem-symmetry lines \begin{equation}
k_x=0,\qquad k_y=0,\qquad k_x+k_y=0
\quad(\operatorname{mod}2\pi).
\label{eq:pure-largeN-nodal-lines}
\end{equation}
The first two factors in Eq.~\eqref{eq:pure-largeN-lattice-dispersion} are identical to the nodal structure of the exciton Bose liquid in Ref.~\cite{ParamekantiBalentsFisher2002}.  In the present model, the diagonal subsystem symmetry supplies the additional factor $\sin[(k_x+k_y)/2]$ and hence the third nodal line.

\paragraph{Continuum gradient expansion and subsystem symmetries.}

Now we discuss a continuum gradient expansion that also makes the
subsystem symmetries apparent. Within the Gaussian approximation, one finds

\begin{equation}
	\mathcal L_2=
	\frac{N}{4\pi}(A_y\dot A_x-A_x\dot A_y)
	-F^2-\frac{1}{4}C^2+\cdots ,
\label{eq:pure-largeN-continuum-L}
\end{equation}
where
\begin{equation}
F=\partial_xA_y-\partial_yA_x,
\qquad
C=\partial_x\partial_y(A_x+A_y).
\label{eq:pure-largeN-FC}
\end{equation}

At leading order in the gradient expansion, $F$ and $C$ are related to
$\theta_A$ and $\theta_B$ [Eq.~\eqref{eq:pure-largeN-ab-symbols}] through
$F\propto (\theta_A-\theta_B)/2$ and $C\propto\theta_A+\theta_B$.

The equation of motion for this Lagrangian gives
\begin{equation}
	\Omega(\bm k)
	=\frac{2\pi}{N}|k_xk_y(k_x+k_y)| .
\label{eq:pure-largeN-continuum-dispersion}
\end{equation}
This agrees with the small-momentum expansion of
Eq.~\eqref{eq:pure-largeN-lattice-dispersion}.

The continuous large-$N$ limits of the column, row, and diagonal
symmetries act through the time-independent shifts
\begin{align}
(i) \,\,  A_x(x,y)&\longrightarrow A_x(x,y)+u(x),
\nonumber\\
(ii) \,\, A_y(x,y)&\longrightarrow A_y(x,y)+v(y),
\nonumber\\
(iii) \,\, A_x(x,y)&\longrightarrow A_x(x,y)+h(x-y),\nonumber\\
A_y(x,y)&\longrightarrow A_y(x,y)-h(x-y).
\label{eq:pure-largeN-continuum-shifts}
\end{align}
Both $F$ and $C$ are invariant under these three distinct transformations.
The Hamiltonian is therefore invariant, while the first-order Berry term
changes only by a total time derivative.  Moreover, one may verify that $F$
and $C$, together with their derivatives, generate every local invariant linear in $A_x,A_y$ that is invariant under the subsystem symmetries. The most general
quadratic Hamiltonian consistent with the subsystem symmetries is therefore a
local differential-operator quadratic form in $(F,C)$:
\begin{equation}
\mathcal H_{\rm eff}^{(2)}
=\kappa F^2+2\mu FC+\lambda C^2+\cdots .
\label{eq:pure-largeN-Wilsonian}
\end{equation}
Stability requires
$\kappa>0$, $\lambda>0$, and $\kappa\lambda-\mu^2>0$. The pure clock model has the stronger dynamical splitting where there is no
$\theta_A\theta_B\propto C^2/4-F^2$ term.  The subsystem symmetries by themselves would allow such a term and its absence is therefore not symmetry enforced and is a special feature of the leading large-$N$ Hamiltonian. As long as subsystem symmetries exist, the low energy physics, however, remains the same whether one has dynamical splitting or not (i.e. one obtains three nodal lines). Therefore, from a computational perspective, dynamical splitting is a convenient fine-tuning to explore the physics of such a phase. On that note, we now turn to the ED study of the $N=2$ model. 

The relation between function-valued subsystem symmetries and nodal
manifolds is familiar from the XY-plaquette and exciton Bose liquid
theories~\cite{ParamekantiBalentsFisher2002,XuMoore2005,XuFisher2007,SeibergShao,GorantlaLamSeibergShao}.  In the present large-$N$ theory, the three
subsystem shifts similarly force the nodal lines $k_x=0$, $k_y=0$, and
$k_x+k_y=0$.  On an $L_x\times L_y$ torus, their union contains
\begin{equation}
N_{\rm node}=L_x+L_y+\gcd(L_x,L_y)-2
\end{equation}
distinct momenta.  This is also the maximum number of independent
mutually commuting line-symmetry operators in the qubit model.

\begin{table*}[t!]
	\centering
	\small
	\begin{tabular}{@{}c
			>{\raggedright\arraybackslash}p{0.28\textwidth}
			>{\raggedright\arraybackslash}p{0.47\textwidth}
			c@{}}
		\toprule
		\(L\) & Ground state center sector(s)
		& Center sector(s) of the first excited state
		& \((g_0,g_1)\)\\
		\midrule
		3
		& \((U_3,U_3,U_3)\)
		& \(\mathcal O[(+--),(+--),U_3]\) \quad (27 center sectors)
		& \((2,54)\)\\[2pt]
		4
		& \((U_4,U_4,U_4;Q_{\rm mix}=\pm1)\)
		& \(\mathcal O[(+-+-),U_4,U_4;Q_{\rm mix}=\pm1]\)
		(6 sectors), together with
		\(\mathcal O[(+--+),(+--+),U_4;Q_{\rm mix}=\pm1]\)
		(24 sectors)
		& \((4,60)\)\\[2pt]
		5
		& \((U_5,U_5,U_5)\)
		& \(\mathcal O[(+--++),(+--++),U_5]\) \quad (75 center sectors)
		& \((2,150)\)\\[2pt]
		6
		& \(\mathcal O[(+-+-+-),(+-+-+-),U_6;Q_{\rm mix}=\pm1]\)
		\quad (6 center sectors)
		& \((U_6,U_6,U_6;Q_{\rm mix}=+1)\) \quad (one center sector)
		& \((12,8)\)\\[2pt]
		7
		& \((U_7,U_7,U_7)\)
		& Internal excitation in the same sector \((U_7,U_7,U_7)\)
		& \((2,12)\)\\[2pt]
		8
		& \((U_8,U_8,U_8;Q_{\rm mix}=\pm1)\)
		& \(\mathcal O[(+--++--+),(+--++--+),U_8;
		Q_{\rm mix}=\pm1]\) \quad (24 center sectors)
		& \((4,48)\)\\
		\bottomrule
	\end{tabular}
	\caption{Center-sector content of the two lowest distinct global energy
		levels of the pure chiral XYZ model.  Here \(g_0\) is the full physical
		ground-state degeneracy and \(g_1\) is the full multiplicity of the first
		distinct excited level, including the projective doublet and any additional degeneracy within a fixed
		center sector.  Every entry for \(L=3,\ldots,8\) follows from an
		exhaustive comparison over all center sectors. $\mathcal O[c,r,d;Q_{\rm mix}]$ denotes the orbit of symmetry-equivalent center sectors generated from the displayed representative by translations and
		lattice symmetries; all sectors in this orbit are  isospectral.}
	\label{tab:pure-global-sector-table}
\end{table*}

\section{Exact diagonalization on the Chiral XYZ model}
\label{sec:ed-pure}

\subsection{Finite-size scaling of the many-body gap}
\label{subsec:pure-gap-scaling}

\begin{figure*}[t!]
 \centering
 \subfigure[Global many-body gap.]{
  \includegraphics[width=0.47\textwidth]
  {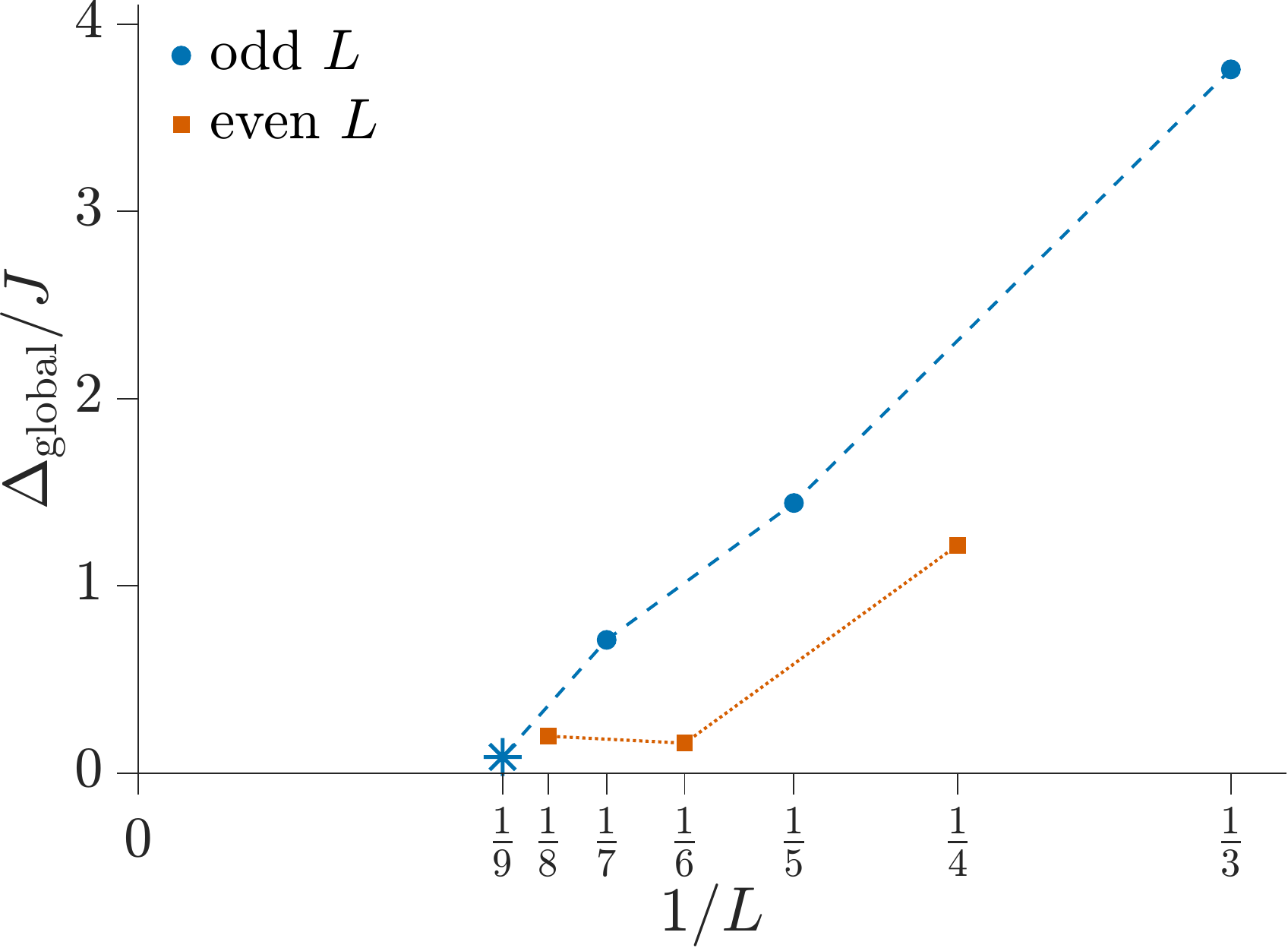}
  \label{fig:pure-global-gap}
 }
 \hfill
 \subfigure[Gap in the uniform center sector.]{
  \includegraphics[width=0.47\textwidth]
  {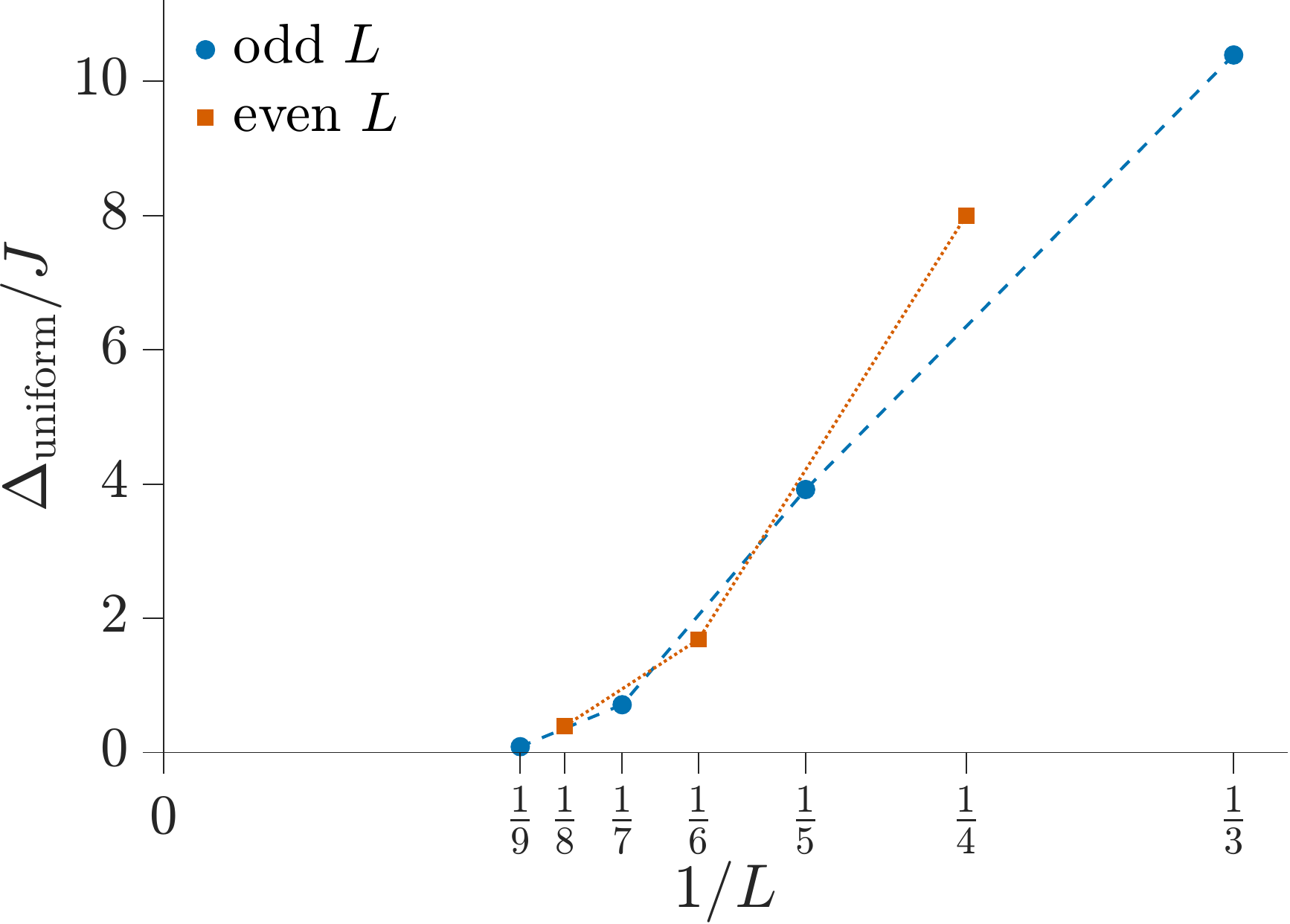}
  \label{fig:pure-uniform-gap}
 }
 \caption{Finite-size scaling of the many-body gap in the pure chiral XYZ
 model on $L \times L$ tori.  Panel (a) shows the global gap above the complete
 ground-state manifold.  The points for \(L=3,\ldots,8\) follow from an
 exhaustive comparison of all center sectors; the \(L=9\) point, marked by
 an asterisk, is a targeted estimate obtained from the uniform sector and
 several competing center-sector patterns.  Panel (b) shows the first distinct gap
 within the fixed uniform center sector.  In both panels, the odd and even
 sequences are connected by lines separately as guides to the eye.}
 \label{fig:pure-gap-comparison}
\end{figure*}

The large-$N$ analysis in Sec.~\ref{subsec:large-N-solution} finds a gapless nodal phase. Motivated by this result, we begin our ED study on the $N=2$ chiral XYZ model with the most direct diagnostic of gaplessness, namely, the scaling of its many-body gap with the system size.  Due to the exactly degenerate logical doublet (Sec.~\ref{subsec:subsystem-symmetries}), the relevant many-body gap is the \textit{first distinct energy} above the exactly degenerate ground-state manifold,
\begin{equation}
 \Delta_{\rm global}(L)=\min_{E_\alpha>E_0}(E_\alpha-E_0).
 \label{eq:pure-global-gap-definition}
\end{equation}
This definition includes both possibilities for the lowest excitation: an
 excitation in the same center sector as the ground state, or the ground state of
a different center sector. We carry out such an exhaustive minimization over all center sectors for $L \times L$ tori for $L=3,\ldots,8$.  At $L=9$,
where an exhaustive scan over the $2^{24}$ center characters is no longer
practical, we utilize the information gained from smaller sizes to focus on a few competitive sectors in which we compare the ground state in the uniform sector with a few other patterns favored at smaller odd sizes. The resulting point is therefore displayed with an asterisk in Fig.~\ref{fig:pure-global-gap}. Access to the unusually large sizes studied here relies crucially on dynamical splitting as explained in Sec.~\ref{subsec:tensor-product-structure}.

Table~\ref{tab:pure-global-sector-table} summarizes the center quantum
numbers of the ground state and the first excited level.  These quantum
numbers are the \(\pm1\) eigenvalues of the relative line operators
\(c_x=C_xC_0\), \(r_y=R_yR_0\), and
\(d_s=D_sD_0\).  For even \(L\), one must additionally specify the
mixed quantum number \(Q_{\rm mix}\); see
Appendix~\ref{app:mixed-center-generator}.  In the table,
\(U_L=(+\,+\,\cdots\,+)\) denotes a uniform relative-charge pattern, while
\(\mathcal O[c,r,d;Q_{\rm mix}]\) denotes the orbit of distinct center
characters generated from the displayed representative by translations and
permutations of the three line families.  The first sign in each pattern is
fixed to \(+\) by convention.  For example, the \(L=5\) ground-state entry
\((U_5,U_5,U_5)\) means \(c=r=d=U_5\).  For \(L=8\), a representative of
the first excited level is
\[
c=r=(+,-,-,+,+,-,-,+),\qquad d=U_8.
\]
Translations, line-family permutations, and the two values of \(Q_{\rm mix}\)
generate its orbit of 24 center sectors.

Figure~\ref{fig:pure-gap-comparison}(a) shows the gap between the global
ground state and the global first excited level.  We find even--odd and
commensuration effects, as also indicated by
Table~\ref{tab:pure-global-sector-table}.  Nevertheless, the gap falls
rapidly with size, strongly suggesting a gapless state.

The global gap mixes two effects: dynamics within a center sector and the
competition between different center characters.  We therefore also study the
gap within the uniform sector $c=r=d=U_L$, with a definite value of
$Q_{\rm mix}$ when $L$ is even. As shown in Fig.~\ref{fig:pure-gap-comparison}(b), this gap shows a more uniform decay with increasing system size, and drops by more than
two orders of magnitude between $L=3$ and $L=9$.  Separate log--log fits to
the odd and even sequences give approximate dynamical exponents $z\simeq4.2$ and $z\simeq4.3$, respectively.  These exponents drift  when the fit window is changed, so we do not identify either value with the actual dynamical exponent but overall data are indicative of a gapless state with $z > 1$.

\subsection{Entanglement scaling}
\label{subsec:pure-entanglement-scaling}

Entanglement scaling provides a
ground-state diagnostic for excitations.  We divide the torus by two
straight cuts, so that the system is divided into two cylinders $A, \overline{A}$ of size $\ell \times L_y$ and $(L_x - \ell) \times L_y$. For such a bipartition, we calculate the von Neumann entanglement
entropy
\begin{equation}
 S_1=-\Tr\rho_A\log\rho_A.
 \label{eq:pure-entropy-definitions}
\end{equation}

As discussed in the last section, within the large-$N$ limit, the dispersion of the excitations vanishes along three subsystem symmetry related nodal lines, $k_x = 0, k_y = 0, k_x + k_y = 0$ (Eq.~\eqref{eq:pure-largeN-lattice-dispersion}). For such a system, one expects~\cite{zhang_criticalee,LaiYangBonesteel2013,BlockShengMotrunichFisher2011,you2022fracton} entanglement scaling similar to that for the ground state of a Fermi gas~\cite{Wolf06,Gioev06,Swingle10}. For a generic fixed transverse momentum $k_y$,
the dispersion vanishes at a finite set of
$k_x$ values.  Expanding about any such zero gives a one-dimensional
linearly dispersing mode along the $x$ direction.  On a cylinder, $k_y$ is
quantized in units of $2\pi/L_y$, and hence the number of one-dimensional
critical channels intersecting the nodal lines grows in proportion to
$L_y$.  This implies that at the leading large-$N$ approximation, one expects 

\begin{equation}
S_{1}(\ell) \sim L_y
\log\!\left[
\frac{L_x}{\pi}
\sin\!\left(\frac{\pi\ell}{L_x}\right)\right]
\label{eq:pure-line-node-entropy-scaling}
\end{equation}
where we have again set the lattice constant $a = 1$. The points where nodal lines meet can modify the count by an
$O(1)$ number of channels, but not the overall scaling.
We will now test this scaling for the $N=2$ chiral XYZ ground state.

\begin{figure}[t]
 \centering
 \subfigure[Half-torus scaling for an $L \times L$ system.]{
  \includegraphics[width=0.94\columnwidth]
  {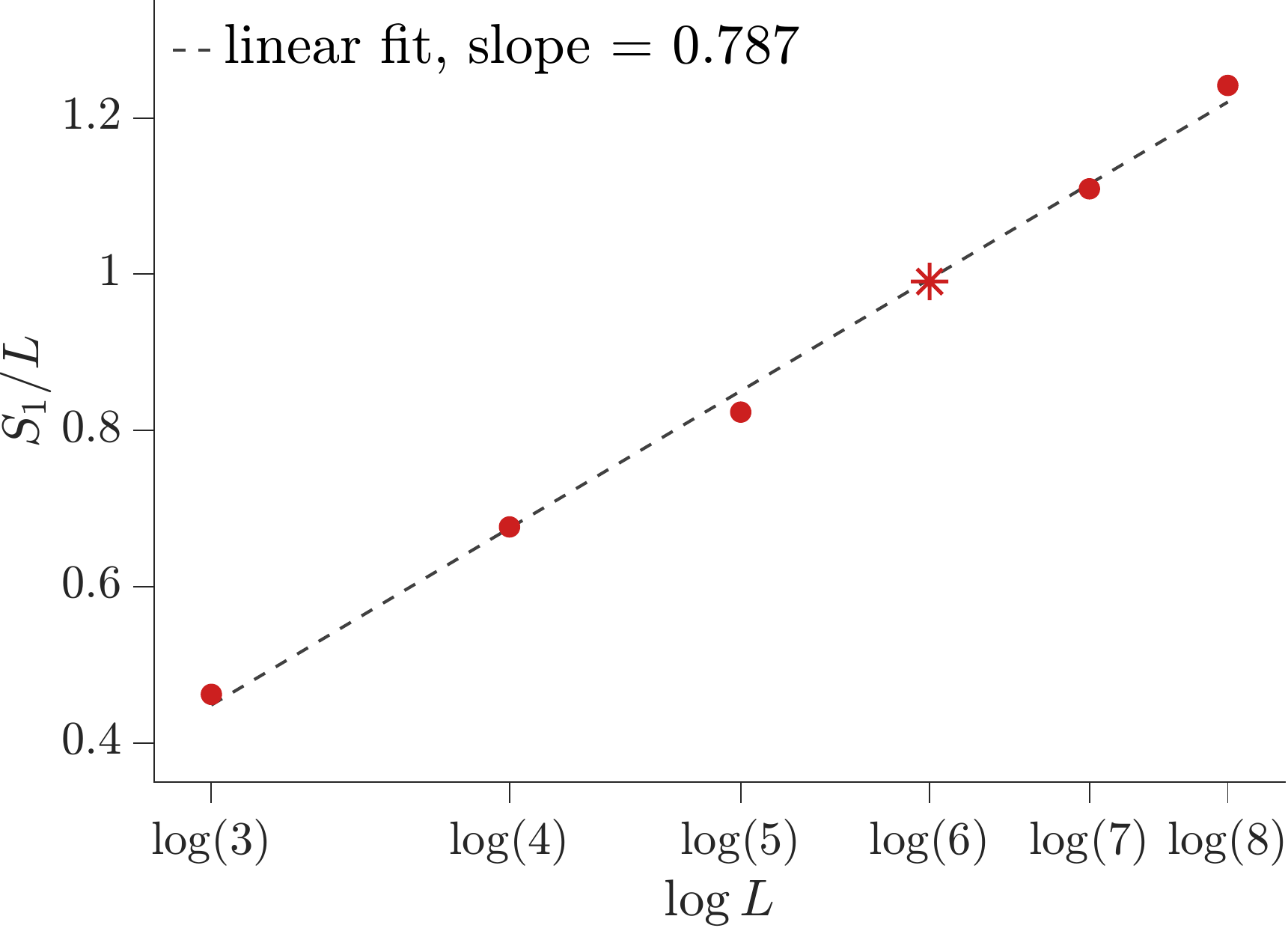}
 }
 \par\vspace{2mm}
 \subfigure[Chord scaling on the \(24\times3\) torus.]{
  \includegraphics[width=0.94\columnwidth]
  {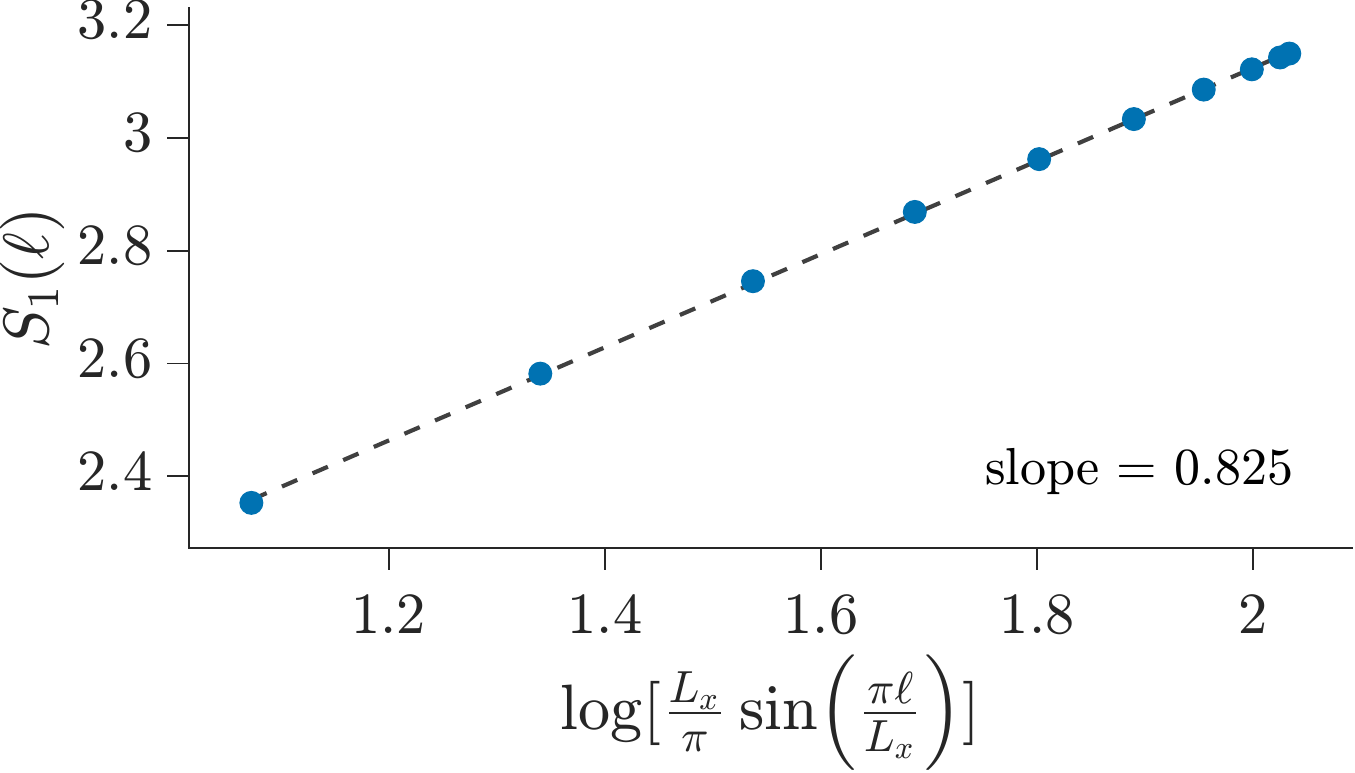}
 }
 \par\vspace{2mm}
 \subfigure[Chord scaling on the \(15\times5\) torus.]{
  \includegraphics[width=0.94\columnwidth]
  {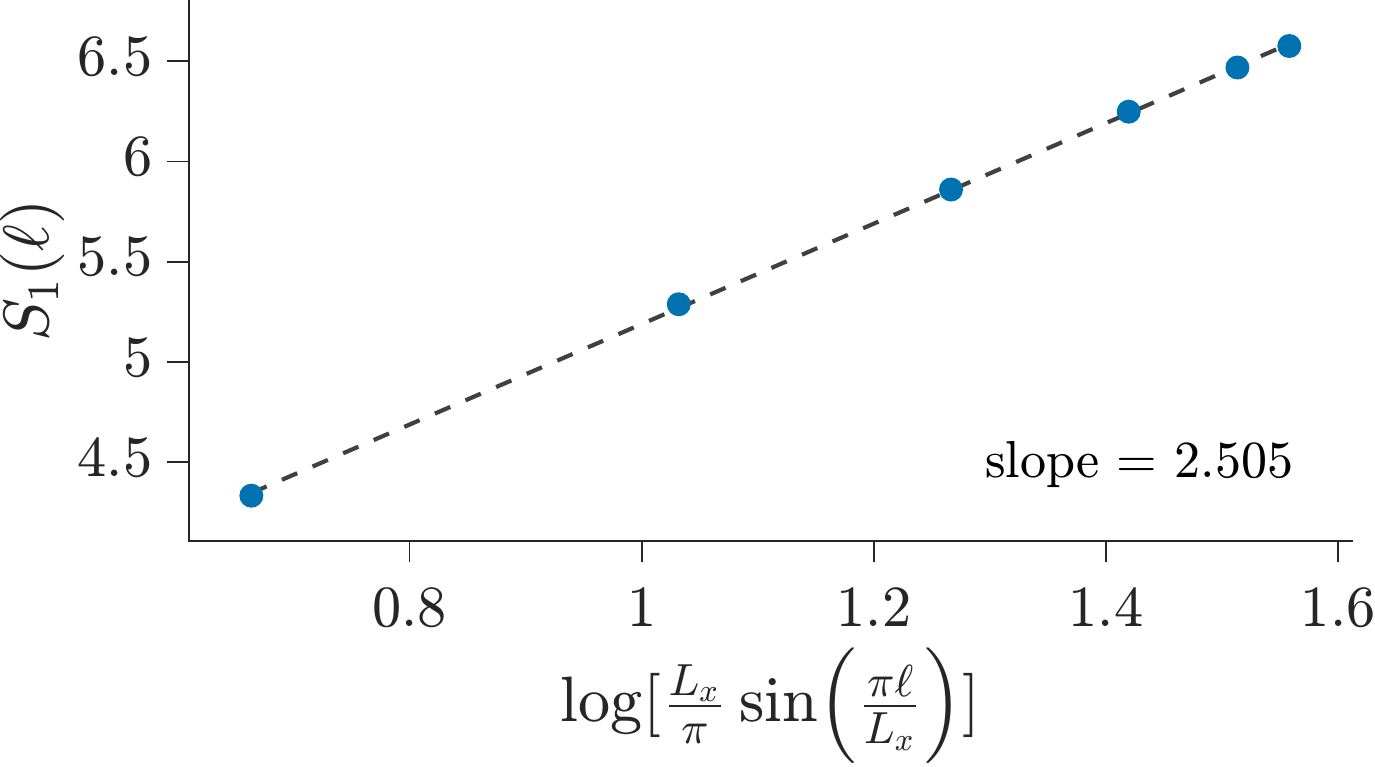}
 }
 \caption{Ground-state entanglement scaling in the pure chiral XYZ model. Panel (a) shows $S_1/L$ for the half-torus region $A=\lfloor L/2\rfloor\times L$.  For $L=3,4,5,7$, and $8$, the plotted
state is a global ground state, which lies in the uniform center sector.  At $L=6$, the lowest uniform-sector state lies $\Delta E\simeq0.160$ above the nonuniform global ground state; the asterisk denotes this uniform-sector state.  The dashed line is a least-squares guide through all six points. Panels (b) and (c) show the entropy obtained by moving the cylindrical cut
 within one fixed ground state. The fitted slopes are \(0.825\)
 and \(2.505\), respectively.}
 \label{fig:pure-entanglement-scaling}
\end{figure}

We first consider $L\times L$ tori with $L=3,\ldots,8$ and choose a state                  
with definite center charges and $C_0$.  For $L=3,4,5,7$, and $8$, the
global ground state lies in the uniform center sector.  For $L=6$, the               
global ground state instead lies in a nonuniform center sector, while the            
lowest uniform-sector state is the first excited level and lies only                 
$\Delta E\simeq0.160$ above it; see                                                  
Table~\ref{tab:pure-global-sector-table}.  To compare the same center-charge
pattern across sizes, we use this uniform-sector state for $L=6$ and mark
the corresponding point with an asterisk.  Figure~\ref{fig:pure-entanglement-scaling}(a)
shows $S_1/L$ versus $\log L$.  We find a nearly linear increase, consistent   
with Eq.~\eqref{eq:pure-line-node-entropy-scaling}.

We next keep $L_x, L_y$ and the ground state fixed and move only the entanglement cut.  This directly tests the $\ell$ dependence in
Eq.~\eqref{eq:pure-line-node-entropy-scaling}, which, for a fixed $L_y$ has the same form as entanglement in a 1+1-D CFT. The data in panels (b) and (c) of
Fig.~\ref{fig:pure-entanglement-scaling} show near-perfect agreement with Eq.~\eqref{eq:pure-line-node-entropy-scaling}. Interpreting the
slopes through Eq.~\eqref{eq:pure-line-node-entropy-scaling} gives
$c_{\rm eff}=2.48$ at $L_y=3$ and $7.51$ at $L_y=5$.  Their increase with
$L_y$ is consistent with the theoretical expectation for a nodal liquid although just two
widths do not permit a quantitative scaling analysis. 
Together, these results provide strong evidence for
a nodal Bose liquid ground state in the $N = 2$ chiral XYZ model.

\section{Proximate phases and instabilities of the chiral XYZ model}
\label{sec:perturbations}

As discussed above, the chiral XYZ Hamiltonian has several special properties such as dynamical splitting and subsystem symmetries. Therefore, it is helpful to consider various perturbations of this model to build understanding about which features of the ground state are tied to these special features. Perturbing the model may also allow us access to proximate phases and phase transitions. We will consider three such perturbations. The first preserves both the subsystem symmetries and the dynamical
splitting, the second preserves the subsystem symmetries but breaks the dynamical splitting, and the third breaks the subsystem symmetries and the dynamical splitting while retaining a global threefold rotation.

\subsection{Split-preserving deformation}
\label{subsec:current-deformation-ed}

We first consider a deformation that retains
the dynamical splitting structure. Any such deformation is guaranteed to also preserve the subsystem symmetries.  Let
$\mathcal D_\triangle=\{\hat x,\hat y,-\hat x-\hat y\}$ be the three
oriented bonds of the triangular lattice.  We define
\begin{align}
 J_A&=\mathrm{i}\sum_{\bm r}\sum_{{\bm\delta}\in\mathcal D_\triangle}
 a_{\bm r}a_{{\bm r}+{\bm\delta}},\nonumber\\
 J_B&=\mathrm{i}\sum_{\bm r}\sum_{{\bm\delta}\in\mathcal D_\triangle}
 b_{\bm r}b_{{\bm r}+{\bm\delta}},\nonumber\\
 H(x)&=xH_{\rm XYZ}-(1-x)(J_A-J_B).
 \label{eq:main-current-deformation}
\end{align}
Here $0\leq x\leq1$.
The relative sign between $J_A$ and $J_B$ is chosen to maintain the symmetries of the triangular lattice, as under twofold rotation $J_A \to - J_B, J_B \to - J_A$.  Since $J_A$ and
$J_B$ belong entirely to the $A$ and $B$ triangle algebras, respectively, $H(x)$ preserves every subsystem symmetry, the exact
logical doublet, and the dynamical splitting.  $H(x)$ does not preserve the antiunitary symmetry in Eq.~\eqref{eq:model-checkerboard-antiunitary} under which $J_A \leftrightarrow J_B$.

One can similarly write down a large-$N$ version of $H(x)$ and solve it exactly at $N = \infty$. See Appendix~\ref{app:split-preserving-current} for details. We find a first-order transition at $x_c=3/4$, with $x > x_c$ corresponding to a translationally invariant ground state (i.e. same symmetries as the pure chiral XYZ model) and $x < x_c$  corresponding to  an inhomogeneous phase with ordering wave vector ${\bm Q}=(2\pi/3,2\pi/3)$.  Both phases remain gapless and nodal however. For $x > x_c$, one finds the same dispersion as the pure chiral XYZ model, i.e., $\Omega(\bm k) \propto f(\bm k) =  \left|\sin\frac{k_x}{2}\sin\frac{k_y}{2}
\sin\frac{k_x+k_y}{2}\right|$.  In the inhomogeneous phase, the lowest excitation near the center of the reduced Brillouin zone retains the cubic form
$ \Omega(\bm k)\propto
 |k_xk_y(k_x+k_y)|$ while at $x=0$ it becomes quintic. See Appendix~\ref{app:split-preserving-current} for details.

\begin{figure}[t]
 \centering
 \begin{tikzpicture}[x=0.76cm,y=0.76cm]
  \draw[very thick] (0,0)--(7,0);
  \draw[-{Latex[length=2.1mm]},thick] (0,0)--(7.45,0)
       node[right=-1mm] {$x$};
  \fill[red] (3.615,0) circle[radius=2.3pt];
  \node[align=center] at (1.75,0.75)
       {inhomogeneous\\nodal liquid};
  \node[align=center] at (5.35,0.75)
       {uniform\\nodal liquid};
  \node[below=2pt] at (0,0) {$0$};
  \node[below=2pt] at (7,0) {$1$};
  \node[below=2pt] at (3.615,-0.23) {$x_c$};
 \end{tikzpicture}
 \caption{Schematic phase diagram of the split-preserving deformation in Eq.~\eqref{eq:main-current-deformation}. Large $N$ gives a first-order transition at $x_c=3/4$ from a nonuniform 	nodal liquid at small $x$ to a uniform nodal liquid at large $x$. At $N=2$, ED on commensurate finite tori similarly favors a nonuniform center sector at small $x$ and the uniform sector at larger $x$, although the reconstruction occurs near $x\simeq 0.30$ and depends strongly on the geometry.}
 \label{fig:current-deformation-schematic}
\end{figure}
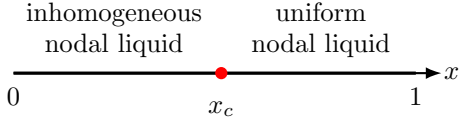

Exact diagonalization supports this qualitative competition between the uniform and non-uniform center sectors, consistent with translation breaking, while also
highlighting strong commensuration effects.  On the $6\times6$ torus, the ground state switches near $x\approx0.31$ from a nonuniform center character, containing both period-two and
period-three line-charge patterns, to the uniform center sector.  It remains uniform for $0.31\lesssim x\lesssim0.995$, but switches back to a nonuniform sector in a narrow interval near the pure chiral XYZ endpoint, $0.995\lesssim x\leq1$, including at $x=1$ as discussed in
Sec.~\ref{subsec:pure-gap-scaling}.  An exhaustive scan of all
$32768$ center sectors on $6\times9$ finds the corresponding reconstruction
between $x=0.275$ and $0.280$, with a narrow intervening center-sector
orbit.

The excitation and entanglement data further indicate that the uniform
branch remains gapless.  At $x=0.4,0.6$, and $0.8$, both the global gap
through $L=7$ and the fixed-uniform-sector gap through $L=9$ decrease with
system size.  Half-torus entropies retain a positive $L\log L$ contribution,
and single-torus cuts on $15\times3$ and $10\times5$ geometries continue to
follow the 1+1-D CFT-like form.  These
results corroborate the large-$N$ picture of a nodal uniform phase at sufficiently large $x$ but they do not fully establish the nature of the inhomogeneous state at small $x$. See Appendix~\ref{app:split-preserving-current} for the ED data.

\FloatBarrier

\subsection{Preserving subsystem symmetries while breaking dynamical splitting}
\label{subsec:split-breaking}

The deformation considered above preserves both the subsystem symmetries and
the dynamical splitting.  We now consider a deformation which breaks dynamical splitting while preserving all subsystem symmetries.  A simple way to achieve this is to multiply an $A$-triangle by a
$B$-triangle.  Specifically, we consider
\begin{align}
 R_{x,y}
 &=a_{x,y}b_{x+1,y+1}
 \nonumber\\
 &=X_{x,y}Y_{x,y+1}Y_{x+2,y+1}X_{x+2,y+2}.
 \label{eq:two-wp-plaquette}
\end{align}
The operators $R_{\bm r}$ mutually commute and preserve all column, row,
and diagonal subsystem symmetries. In fact, the Hamiltonian $H_{\mathrm{WP}^{\otimes 2}}=-\sum_{\bm r}R_{\bm r}$ precisely corresponds to two copies of the Wen-Plaquette Hamiltonian when $L_x$ is even. To see this, note that for even $L_x$, the
even and odd columns form two independent layers, and the operators in each layer are precisely the Wen--plaquette stabilizers on an effective square lattice that is sheared relative to the original triangular lattice.

We study the following Hamiltonian:
\begin{equation}
H(x)=(1-x)H_{\rm XYZ}
 +xH_{\mathrm{WP}^{\otimes 2}}
 \label{eq:two-wp-H}
\end{equation}
where  $0 \leq x\leq1$. The two end points correspond to the putative nodal liquid phase and the two copies of the Wen-plaquette model.
\begin{figure}[t]
 \centering
 \begin{tikzpicture}[x=0.72cm,y=0.72cm]
  \draw[very thick] (0,0)--(7,0);
  \draw[-{Latex[length=2.1mm]},thick] (0,0)--(7.45,0)
       node[right=-1mm] {$x$};
  \fill[red] (3.35,0) circle[radius=2.3pt];
  \node[align=center,font=\small] at (1.50,0.75)
       {gapless\\nodal liquid};
  \node[align=center,font=\small] at (5.25,0.82)
       {$(\mathrm{Wen\!-\!plaquette})^{\otimes 2}$\\topological phase};
  \node[below=2pt] at (0,0) {$0$};
  \node[below=2pt] at (7,0) {$1$};
  \node[below=2pt] at (3.35,0) {$x_c$};
 \end{tikzpicture}
 \caption{Schematic phase diagram for the Hamiltonian in
 Eq.~\eqref{eq:two-wp-H}.  The red point denotes the transition from the
 gapless nodal liquid to the gapped
 $(\mathrm{Wen\!-\!plaquette})^{\otimes 2}$ phase.  The finite-size results below place the transition
 near $x\simeq0.4$--$0.5$.}
 \label{fig:two-wp-schematic}
\end{figure}
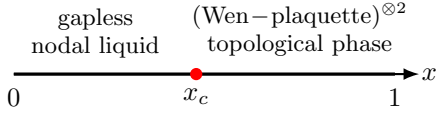
To determine how these two endpoints are connected, we diagonalize
Eq.~\eqref{eq:two-wp-H} on $L\times L$ tori with
$L=3,\ldots,6$ and $x=0,0.1,\ldots,1$.  We follow the uniform center
character
\begin{equation}
 c_x=r_y=d_s=+1,\qquad Q_{\rm mix}=+1\quad(L\ {\rm even}),
 \label{eq:two-wp-uniform-sector}
\end{equation}
which contains a ground state of $H_{\mathrm{WP}^{\otimes 2}}$ at $x=1$.
We define $\Delta_{\rm unif}$ as the first distinct excitation energy in
this sector, excluding the exact logical partner.  Thus
$\Delta_{\rm unif}$ is a fixed-sector gap rather than a minimization over
all center characters.

• \begin{figure}[t]
	\centering     
	\begin{minipage}{\linewidth}
		\centering   
		\includegraphics[width=\linewidth]{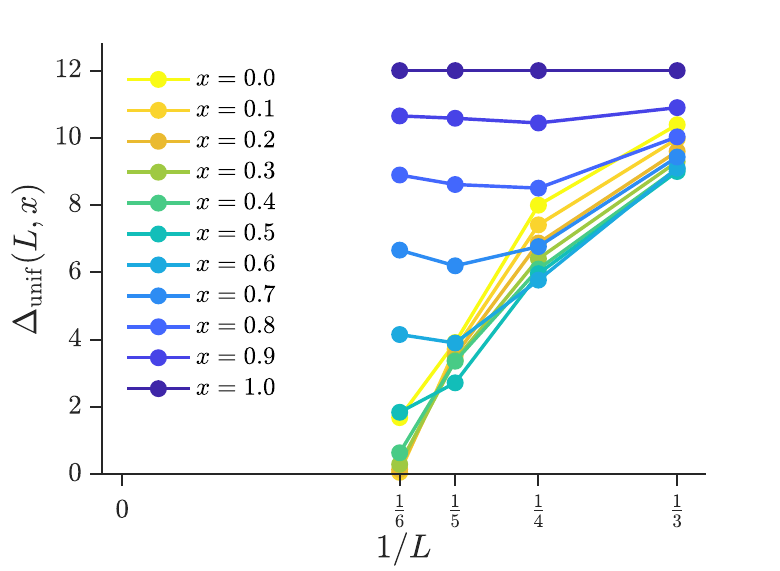}
		\caption{Uniform-sector gap $\Delta_{\rm unif}$ for the interpolation in
			Eq.~\eqref{eq:two-wp-H}, plotted versus $1/L$ on $L \times L$ tori. Lines
			connect data points belonging to a fixed $x$ and are guides to the eye.}
		\label{fig:two-wp-gap-scaling}
	\end{minipage}
	
	\vspace{1em}                                                                        
	
	\begin{minipage}{\linewidth}                                                        
		\centering
		\includegraphics[width=\linewidth]{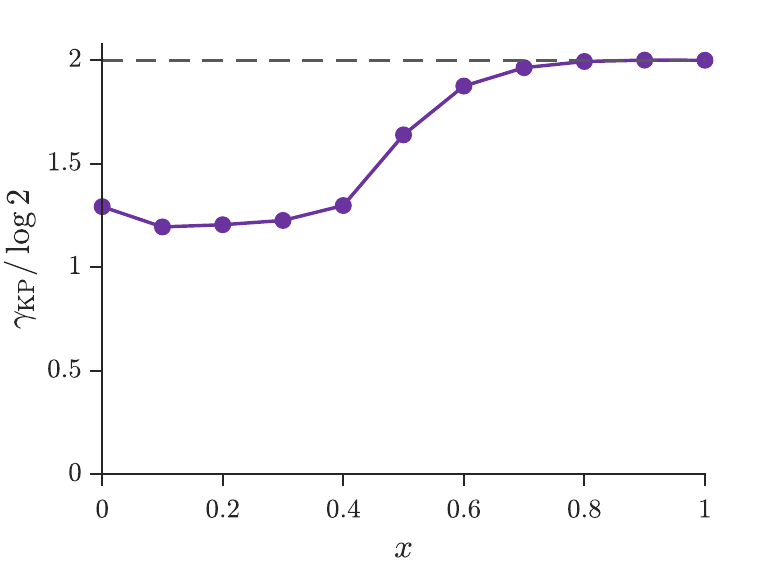}
		\caption{Kitaev--Preskill combination of entropies,                                                                                                                        
			$\gamma_{\rm KP}$ [Eq.~\eqref{eq:two-wp-kp}], for the interpolation in
			Eq.~\eqref{eq:two-wp-H} on a $6\times6$ torus. The dashed line is the exact                                                                                                
			$(\mathrm{Wen\!-\!plaquette})^{\otimes 2}$ value
			$\gamma_{\rm KP}=2\log(2)$.}
		\label{fig:two-wp-tee}
	\end{minipage}
\end{figure}


Figure~\ref{fig:two-wp-gap-scaling} shows the scaling of the gap with $1/L$ for various $x$. The data are consistent with a transition near
$x\simeq0.4$--$0.5$ between the nodal liquid and the gapped two-copy
Wen--plaquette phase.

To characterize the transition, we study the long-range tripartite entanglement in
the $6\times6$ ground state that corresponds to the  Kitaev--Preskill combination~\cite{Kitaev06_1}
\begin{align}
\gamma_{\rm KP}
=-\big(&S_A+S_B+S_C
\nonumber\\[-1mm]
&-S_{AB}-S_{AC}-S_{BC}+S_{ABC}\big).
\label{eq:two-wp-kp}
\end{align}
Regions $A, B, C$ are chosen as follows: we take a contractible $4\times4$ region, divide it into two stacked $2\times2$ regions $A,B$ and an adjacent
$2\times4$ region $C$.  Figure~\ref{fig:two-wp-tee} shows that
$\gamma_{\rm KP}/\log 2$ remains between approximately $1.2$ and $1.3$
through $x=0.4$, rises across the same interval in which the gap opens, and
rapidly approaches $2$.  Since the small-$x$ phase is
gapless, its finite-size value is not expected to take a value that corresponds to TEE~\cite{Kitaev06_1,Levin06} of any gapped topological order.  We leave a detailed study of this transition, including a field-theory description, to the future.

\subsection{Explicitly breaking subsystem symmetry}
\label{subsec:explicit-subsystem-breaking}
The subsystem symmetries protect the nodal lines found at large $N$.  We now
ask whether the nodal liquid remains stable when these symmetries are
explicitly broken.  We first consider the onsite and nearest-neighbor
perturbations
\begin{align}
 \delta H=-\sum_{\bm r}\big[&
 h_X\left(X_{\bm r}+X_{\bm r}^{\dagger}\right)
 \nonumber\\[-1mm]
 &+t_D\left(D^X_{\bm r}+(D^X_{\bm r})^{\dagger}\right)\big].
 \label{eq:main-subsystem-breaking-perturbations}
\end{align}
Here $D^X_{\bm r}=X_{{\bm r}+\hat y}X_{\bm r}^{\dagger}$.  The onsite
$Z$ and $\mathcal Y$ perturbations are related to $X$ by the
triangular-lattice symmetry.

The correlation functions at the unperturbed large-$N$ fixed point are
strongly anisotropic.  As detailed in Appendix~\ref{subsec:ssbrg-correlators},
that subsystem symmetry forces unequal-space correlators of $X$ to vanish,
while its equal-space correlator decays as $|\tau|^{-4/N}$.  The `dipole' $D^X_{\bm r}$ has the same power law both in time and along its symmetry-allowed spatial direction.

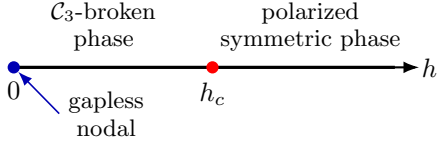
\begin{figure}[t]
	\centering
	\begin{tikzpicture}[x=0.72cm,y=0.72cm]
	\draw[very thick] (0,0)--(7,0);
	\draw[-{Latex[length=2.1mm]},thick] (0,0)--(7.45,0)
	node[right=-1mm] {$h$};
	\fill[blue!70!black] (0,0) circle[radius=2.3pt];
	\fill[red] (3.65,0) circle[radius=2.3pt];
	\node[align=center,font=\small] at (1.65,0.75)
	{$\mathcal C_3$-broken\\phase};
	\node[align=center,font=\small] at (5.45,0.75)
	{polarized\\symmetric phase};
	\node[below=2pt] at (0,0) {$0$};
	\node[below=2pt] at (3.65,0) {$h_c$};
	\node[align=center,font=\small] (nodalpoint) at (1.70,-0.85)
	{gapless\\nodal};
	\draw[-{Latex[length=1.9mm]},semithick,blue!70!black]
	(nodalpoint.west)--(0.08,-0.08);
	\end{tikzpicture}
	\caption{Schematic phase diagram of $H(h)$ in Eq.~\eqref{eq:onsite-field-H} suggested by mean-field theory and the
		finite-size spectra.  The blue
		point at $h=0$ is the gapless nodal XYZ model, while the red point marks
		the transition from the $\mathcal C_3$-broken phase to the polarized
		symmetric phase.}
	\label{fig:onsite-field-schematic}
\end{figure}

We use these correlators in Appendix~\ref{subsec:ssbrg-direct-RG} to carry
out an RG about the nodal lines.   The RG scheme is broadly similar to the patch RG used for Fermi surfaces
\cite{Polchinski1992FermiSurface,Shankar1994RG}; its application to a
bosonic theory with nodal lines is closely related to
Ref.~\cite{Lake2022EBLRG}.  In this scheme, we keep the tangential momentum fixed and
rescale frequency together with the momentum normal to the nodal lines. We find
\begin{align}
 \frac{d h_X}{d\ell}&=\left(2-\frac{2}{N}\right)h_X+\cdots,
 \nonumber\\
 \frac{d t_D}{d\ell}&=\left(2-\frac{2}{N}\right)t_D+\cdots .
 \label{eq:main-subsystem-breaking-flow}
\end{align}
Thus both perturbations are relevant at controlled large $N$.  As discussed in Appendix~\ref{subsec:ssbrg-general-gamma}, if one considers an anisotropic Hamiltonian where  the two Gaussian modes have unequal stiffnesses, $H_\gamma=\frac{\kappa}{2}\sum_{\bm r}
\left[
(\gamma+\gamma^{-1})(\theta_A^2+\theta_B^2)
+2(\gamma^{-1}-\gamma)\theta_A\theta_B
\right].
$, then the RG equations are 

\begin{align}
\frac{dh_X}{d\ell}
&=\left[2-\frac{3\gamma+\gamma^{-1}}{2N}\right]h_X+\cdots,
\nonumber\\
\frac{dt_D}{d\ell}
&=\left[2-\frac{\gamma+3\gamma^{-1}}{2N}\right]t_D+\cdots.
\end{align}

Therefore, a sufficiently large $\gamma$ may be able to stabilize the nodal phase against subsystem symmetry breaking perturbations. The anisotropy however breaks the dynamical split, since the coefficient of the $\theta_A \theta_B$ term in $H_\gamma$ is proportional to $\left(\gamma^{-1}-\gamma\right)$. The stiffness ratio $\gamma$ plays a role analogous to the ratio of the ring-exchange coupling to the charging energy in the exciton Bose liquid: in both cases, it controls
the scaling dimensions of perturbations that break subsystem symmetry~\cite{ParamekantiBalentsFisher2002,Lake2022EBLRG}.

\begin{figure*}[t]
	\centering
	\includegraphics[width=\textwidth]{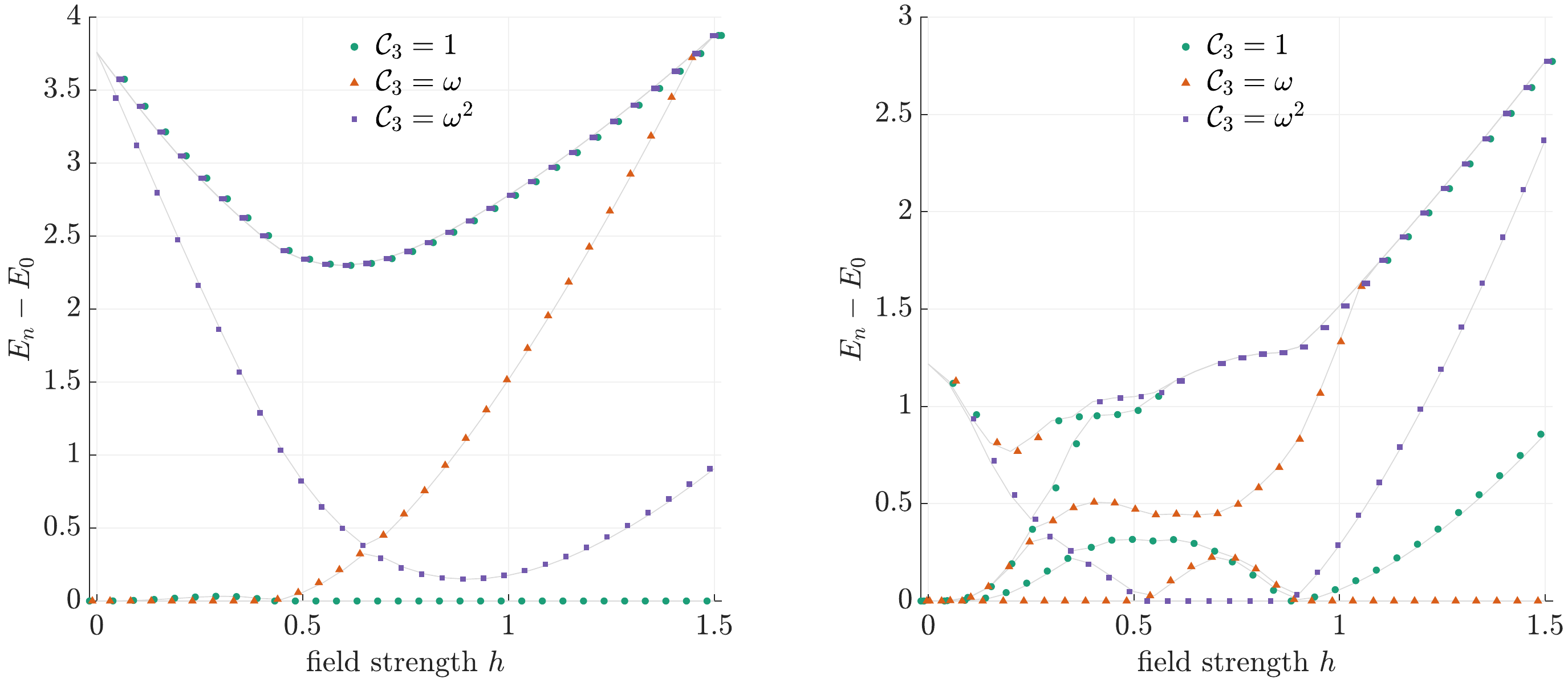}
	\par\vspace{1mm}
	\makebox[\textwidth]{%
		\makebox[0.5\textwidth]{\small (a) $L=3$}%
		\makebox[0.5\textwidth]{\small (b) $L=4$}}
	\caption{The six lowest excitation energies of
		$H(h)$ [Eq.~\eqref{eq:onsite-field-H}] and their $\mathcal C_3$ eigenvalues as functions of $h$.  The ground-state energy is set to zero.  Markers for degenerate states are displaced
		slightly in the horizontal direction for visibility.}
	\label{fig:onsite-field-c3-levels}
\end{figure*}
The RG result above
applies to the usual thermodynamic limit where the system size $L\to\infty$ at a fixed lattice spacing
$a$.  Following Refs.~\cite{SeibergShao,GorantlaLamSeibergShao}, one may
instead consider a different limit where  $a\to0$ and $L\to\infty$, while keeping the physical lengths $\ell=aL$ and the coefficient $v$ in
$\Omega({\bm k})\simeq v|k_xk_y(k_x+k_y)|$ fixed.  In this version of the continuum limit, acting with $X$ or $D^X$ on the ground state changes the subsystem-symmetry quantum numbers of the ground state, and the minimum excitation energy in the resulting sector scales as 
\begin{equation}
 E_{\rm charged}\sim\frac{v}{Na^2\ell}\longrightarrow\infty .
 \label{eq:main-continuum-charged-energy}
\end{equation}
They are therefore absent from the finite-energy operator content of this
continuum theory; the derivation is given in
Appendix~\ref{subsec:ssbrg-scope}. For the actual lattice model, the usual thermodynamic limit, $L \to \infty$ at fixed $a$, is more consequential, and as discussed above, in this limit, the RG equations show that subsystem symmetry breaking terms are relevant, at least when the anisotropy $\gamma$ is small.

\subsubsection{Breaking subsystem symmetry with a uniform onsite field}
\label{subsec:onsite-field-c3}

The large-$N$ calculation indicates that at least in the absence of the anisotropy $\gamma$, the ground state is unstable against perturbations that explicitly break the subsystem symmetry, but it does not determine the phase toward which it is unstable to.  Moreover, it is difficult to know if the conclusion holds at
$N=2$.  To study a concrete potential instability of the $N=2$ spin
model, we consider adding the simplest perturbation corresponding to the onsite field. Specifically, we consider the Hamiltonian
\begin{equation}
 H(h)=H_{\rm XYZ}
 +h\sum_{\bm r}\left(X_{\bm r}+Y_{\bm r}+Z_{\bm r}\right),
 \label{eq:onsite-field-H}
\end{equation}
where $h \ge 0$. This perturbation breaks both the subsystem symmetries and dynamical
splitting, but preserves the combined lattice--spin rotation
\begin{equation}
 \begin{aligned}
 \mathcal C_3:\quad (x,y)&\mapsto(-x+y,-x),\\
 X&\mapsto Y\mapsto Z\mapsto X ,
 \end{aligned}
 \label{eq:onsite-field-C3}
\end{equation}
on an $L\times L$ torus.  We choose $\mathcal C_3^3=\mathds{1}$, so its
eigenvalues are $1,\omega,\omega^2$, with
$\omega=e^{2\pi\mathrm{i}/3}$.

As a simple guess for the resulting state due to the perturbation $h$, we consider a translationally invariant mean-field ansatz with
\(\langle(X,Y,Z)\rangle=(n_x,n_y,n_z)\), with
\(\bm n^2=1\). The energy per site within this ansatz is
\begin{equation}
 e_{\rm MF}(\bm n)
 =-2J n_xn_yn_z+h(n_x+n_y+n_z).
 \label{eq:onsite-field-mean-field}
\end{equation}
At $h=0$, the interaction energy is minimized by four directions
\begin{equation}
 \begin{aligned}
 \bm n_1&=\frac{1}{\sqrt3}(1,-1,-1),&
 \bm n_2&=\frac{1}{\sqrt3}(-1,1,-1),\\
 \bm n_3&=\frac{1}{\sqrt3}(-1,-1,1),&
\bm n_4&=\frac{1}{\sqrt3}(1,1,1),\\
 \end{aligned}
 \label{eq:onsite-field-three-minima}
\end{equation}
At infinitesimal $h$, one selects $\bm n_1, \bm n_2, \bm n_3$
which are favored over the fourth direction $\bm n_4$.  Therefore, at small $h$, the mean field predicts a state where the aforementioned $\mathcal{C}_3$ symmetry is spontaneously broken. At large $h$, by contrast, the unique minimum approaches
$-(1,1,1)/\sqrt3$.  Mean-field theory therefore suggests a phase transition between a conventional
$\mathcal C_3$-broken and a polarized
symmetric phase as $h/J$ is increased.

We test this picture using ED on $3\times3$ and $4\times4$ tori. Due to the absence of subsystem symmetry as well as dynamical split, we are restricted to small sizes. We resolve the low-energy states by their $\mathcal C_3$ eigenvalues; whenever levels are exactly degenerate, we diagonalize $\mathcal C_3$ within the degenerate subspace. As shown in Fig.~\ref{fig:onsite-field-c3-levels}, the low-energy states
reorganize into branches carrying all three eigenvalues
$1,\omega,\omega^2$.  The detailed rearrangement is strongly size dependent. On the $3\times3$ torus, the two states that are degenerate at $h=0$
remain very close in energy up to $h\approx0.5$.  For $L=4$, the fourfold
ground-state manifold at $h=0$ acquires an appreciable splitting already
near $h\approx0.2$.  Near $h\simeq0.9$, the three lowest states, carrying
three distinct $\mathcal C_3$ eigenvalues, become nearly degenerate.  These states remain within the excitation-energy window
$E_n-E_0 \lesssim 0.3$ for $h\lesssim 0.9$.  This may be
a finite-size signature of $\mathcal C_3$ breaking.  Overall, the spectra
are suggestive of $\mathcal C_3$ breaking but do not establish it, given
the limited system sizes accessible to us.  The available sizes also do not determine whether an arbitrarily weak field immediately opens a gap in the nodal liquid. At larger field, the ground
state is nondegenerate and evolves toward the polarized state.

\section{Discussion}
\label{sec:discussion}
In this work, we studied the highly frustrated spin Hamiltonian in
Eq.~\eqref{eq:model-HXYZ}, which we call the chiral XYZ model.  The model was originally 
introduced as a special limit of the Majorana--Hubbard model in
Ref.~\cite{LiFranzMajoranaHubbard2018}.  A central feature of the model is dynamical splitting: the Hamiltonian is a sum of two terms whose operator algebras mutually commute. Together with its $O(L)$ subsystem symmetries, this structure approximately halves the active number of qubits in exact diagonalization (ED), thus making accessible lattices containing roughly twice as many spins as in conventional ED.  For the qubit ($N=2$) model, we carried out calculations up to
$9\times9$ lattice sites.  We also
constructed a $\mathbb{Z}_N$ generalization, and solved it at large $N$. We find that its excitation spectrum contains three
nodal lines, and the gapless ground state has bipartite entanglement scaling
as $L\log L$. Spectral gap and entanglement results provide evidence that these
features survive at $N=2$. We also
studied several deformations of the spin-$1/2$ model as well as the large-$N$ model. Among the deformations studied here, our numerical results suggest that the nodal phase survives a finite-strength perturbation that preserves the subsystem symmetries but breaks dynamical splitting, before giving way to a gapped phase described by two copies of the Wen--plaquette model.

We now briefly mention a few broader connections and directions.

Bosonic systems with extended manifolds of low-energy modes have been extensively studied in a variety of
settings~\cite{ParamekantiBalentsFisher2002,MotrunichSpinonFermiSurface2005,
MotrunichFisher2007,FisherMotrunichSheng2008,BlockShengMotrunichFisher2011}.  The existence of such manifolds need not rely on any subsystem symmetries.  The distinctive features of the present model are the unusually large-system ED enabled by dynamical splitting, which provides strong support for $L \log L$ entanglement scaling, a simple large-$N$ description, and the coexistence of nodal gaplessness with an exact topology-dependent degeneracy. In the present model, the large-$N$ analysis (Sec.~\ref{subsec:explicit-subsystem-breaking} and Appendix~\ref{app:general-stability}) shows that the leading perturbations which break the subsystem symmetries are relevant at the dynamically split point, although whether this conclusion holds at $N=2$ is not established. Even if the $N=2$ microscopic Hamiltonian is indeed fine-tuned with respect to generic local perturbations, this would not imply that the nodal liquid phase itself must be fine-tuned. More generally, as discussed in Sec.\ref{subsec:explicit-subsystem-breaking}, the large-$N$ scaling dimensions of subsystem-symmetry-breaking perturbations depend on the Gaussian stiffnesses, and anisotropy in these stiffnesses, which itself breaks dynamical splitting, can render such perturbations irrelevant, allowing the subsystem symmetries to emerge at low energies. Since the characteristic nodal structure persists within the subsystem-symmetric theory even after dynamical splitting is broken, the exactly split model provides a computationally tractable route to the same qualitative low-energy physics. A more interesting question is whether dynamical splitting itself can emerge at low energies in a generic microscopic model.

Motivated by the discussion in Sec.~\ref{subsec:pure-model}, where the terms in the Chiral XYZ model were interpreted as magnetic flux and Gauss-law operators, one can  extend this analogy by coupling the chiral XYZ model to matter while preserving dynamical splitting. Let us place a spinless fermion $c_{\bm r}$ on every vertex, with
$n_{\bm r}=c_{\bm r}^{\dagger}c_{\bm r}$, and couple it minimally to the
Pauli link variables.   With the square-lattice link convention of
Eq.~\eqref{eq:model-CS-link-assignment}, the resulting gauge--matter
Hamiltonian is
\begin{align}
H_{\rm GM}={}&-t\sum_{\bm r}\left(
c_{\bm r}^{\dagger}X_{\bm r}c_{{\bm r}+\hat x}
+\mathrm{h.c.}\right)
\nonumber\\
&-t\sum_{\bm r}\left(
c_{\bm r}^{\dagger}Z_{{\bm r}+\hat y}c_{{\bm r}+\hat y}
+\mathrm{h.c.}\right)
\nonumber\\
&-K\sum_{\bm r}a_{\bm r}
-g\sum_{\bm r}(-1)^{n_{\bm r}}b_{{\bm r}-\hat x}.
\label{eq:discussion-matter-gauge}
\end{align}
The first two lines describe fermion hopping, $a_{\bm r}$ is the magnetic-flux
term, and $(-1)^{n_{\bm r}}b_{{\bm r}-\hat x}$ is the Gauss-law operator,
the lattice analogue of
$(\bm\nabla\!\cdot\!\bm E)(-1)^{n_{\bm r}}$, included energetically in the
Hamiltonian.  As in the pure chiral XYZ model, each Gauss-law operator
commutes with every hopping and magnetic-flux term.  The Gauss-law operators
do not commute among themselves, however, and therefore cannot be imposed
simultaneously as constraints on the Hilbert space; this is why they enter
Eq.~\eqref{eq:discussion-matter-gauge} energetically.  The Hamiltonian therefore still possesses dynamical splitting. One can also construct the $\mathbb{Z}_N$ generalization of this theory and it will be interesting to study its large-$N$
limit.

At the level of operator algebras, Haah's invertible subalgebras provide an example of dynamical splitting with additional structure~\cite{HaahInvertibleSubalgebras2023}.
Let $\mathcal A$ be a local subalgebra of the physical operator algebra and
$\mathcal B = \mathcal A'$ its commutant.  Choosing the terms in $H_A$ from
$\mathcal A$ and those in $H_B$ from $\mathcal B$ immediately gives a
dynamically split Hamiltonian $H=H_A+H_B$.  Ref.~\cite{HaahInvertibleSubalgebras2023} calls
$\mathcal A$ invertible if every local physical operator $O$ can be written
as
\begin{equation}
 O=\sum_i A_iB_i,
 \qquad A_i\in\mathcal A,\quad B_i\in\mathcal B,
 \label{eq:discussion-local-reconstruction}
\end{equation}
with $A_i$ and $B_i$ supported within a fixed distance of the support of $O$, independent of the system size. 
More generally, a pair of commuting algebras can relate to the full physical operator algebra in three different ways. In the first case, $\mathcal A$ is an invertible subalgebra in the sense of Ref.~\cite{HaahInvertibleSubalgebras2023}. A second possibility is that the two algebras reconstruct the complete operator algebra on every finite system, but a local physical operator may require factor operators whose support grows with system size.  Finally, the
two algebras may jointly generate only a proper subalgebra, so that some
physical operators cannot be written at all in the form of
Eq.~\eqref{eq:discussion-local-reconstruction} even if nonlocal $A_i$ and
$B_i$ are allowed.  On the full Hilbert space, the two triangle algebras of
the chiral XYZ model belong to the last class: for example, an onsite Pauli
operator changes some subsystem charges and therefore cannot be generated by
products of $a$- and $b$-triangle operators. This does not pose a problem for our ED since for the numerical calculations performed in our work, it suffices to use the
tensor-product decomposition separately within each center sector of
interest; reconstructing the full operator algebra is unnecessary. 

Relatedly, one can define the chiral XYZ model with open boundary conditions with suitable boundary terms such that the dynamical split continues to exist. In this case, the triangle algebras $\mathcal A$ and $\mathcal B$ have no common center and become algebraically complementary, and therefore, every physical operator can be written using operators from the two algebras. This still does not make them invertible in the sense of Eq.~\eqref{eq:discussion-local-reconstruction}, however.  To see the obstruction, let's write $Y=\mathrm{i}XZ$ so that every $a_{\bm r}$
and $b_{\bm r}$ contains an even number of $X$ operators and an even number of $Z$ operators.  The same is therefore true of any product of bulk triangle
operators, which cannot reproduce a single onsite Pauli operator such as $X_{\bm r}$.  Appropriate boundary operators remove this obstruction on a finite open system, but representing $X_{\bm r}$ then requires a string extending from $\bm r$ to the boundary, thereby violating  Eq.~\eqref{eq:discussion-local-reconstruction}. Therefore, this situation realizes the second possibility mentioned right after Eq.~\eqref{eq:discussion-local-reconstruction}.

Another direction is to consider dynamical splitting into more than two active factors. Here is a simple illustrative example. Consider a chain of \(L=nM\) qubits and introduce the usual Jordan--Wigner Majoranas
\[
\gamma_{2j}=\left(\prod_{k<j}Y_k\right)X_j,\qquad
\gamma_{2j+1}=\left(\prod_{k<j}Y_k\right)Z_j.
\]
Define the local Pauli operators
\[
u_r=i\gamma_r\gamma_{r+n},
\]
and divide them into \(n\) families according to \(r=\alpha\pmod n\):
\[
H=\sum_{\alpha=0}^{n-1}H_\alpha,\qquad
H_\alpha=-J\sum_m u_{\alpha+mn}.
\]
Although the individual Majoranas contain Jordan--Wigner strings, these strings cancel in \(u_r\): each \(u_r\) is a geometrically local Pauli operator supported on at most \(\lceil n/2\rceil+1\) consecutive physical spins, consisting of \(X\) or \(Z\) operators at its endpoints and \(Y\)'s in between.  For example, for \(n=3\), the three factor Hamiltonians take the particularly simple form

  \begin{flalign*}
&H_0=-J\sum_{k=0}^{M-1}\big(
Z_{3k}Z_{3k+1}
+X_{3k+1}Y_{3k+2}X_{3k+3}
\big),
&&\displaybreak[3]\\
&H_1=-J\sum_{k=0}^{M-1}\big(
X_{3k}Y_{3k+1}X_{3k+2}
+Z_{3k+2}Z_{3k+3}
\big),
&&\displaybreak[3]\\
&H_2=-J\sum_{k=0}^{M-1}\big(
Z_{3k+1}Z_{3k+2}
+X_{3k+2}Y_{3k+3}X_{3k+4}
\big),
&&
\end{flalign*}
where the physical-site labels are understood modulo \(L=3M\). Thus each factor acts throughout the entire physical chain and contains mutually noncommuting terms.

By construction, for any $n$, every term in \(H_\alpha\) commutes with every term in \(H_\beta\) for \(\alpha\neq\beta\), since the two terms are even products of disjoint sets of Majoranas, whereas consecutive terms within a given \(H_\alpha\) anticommute because they share one Majorana. Importantly, similar to the splitting in the chiral XYZ model, each \(H_\alpha\) acts throughout the entire physical chain: as \(m\) is varied, the supports of its terms cover every physical qubit. On a periodic chain, each factor has one global \(\mathbb{Z}_2\) center charge. After fixing these \(n\) charges, each factor acts as the full operator algebra of \(M-1\) factor qubits, and hence
\[
\mathcal{H}_{\lambda}\simeq
\bigotimes_{\alpha=0}^{n-1}\mathcal{H}_{\alpha,\lambda},
\qquad
\dim\mathcal{H}_{\alpha,\lambda}=2^{M-1},
\]
with
\[
H\big|_{\lambda}
=\sum_{\alpha=0}^{n-1}
1\otimes\cdots\otimes H_\alpha(\lambda)\otimes\cdots\otimes 1.
\]
Thus, parametrically, a problem involving \(L=nM\) physical qubits is reduced to \(n\) problems involving approximately \(L/n\) factor qubits. Although, the example above is a free-fermion model, since the construction is based on algebras, one can make the factors non-integrable by multiplying terms that belong to the same algebra. For example, adding terms such as
\[
U\sum_m u_{\alpha+mn}u_{\alpha+(m+2)n}
\]
within every \(H_\alpha\) produces four-Majorana interactions while preserving the exact \(n\)-fold dynamical splitting. Following an argument similar to that for the chiral XYZ model, one can show that the subalgebras in this example are also non-invertible in the sense of Eq.~\eqref{eq:discussion-local-reconstruction}.

Another question about a system with dynamical splitting concerns the distinction between locality of a particular factor Hamiltonian and a local tensor-product realization of all its factor degrees of freedom. To illustrate this point, let's consider the example from Ref.~\cite{HaahInvertibleSubalgebras2023}
of two mutually commuting and invertible algebras, $\mathcal A$ and $\mathcal B=\mathcal A'$, with local commuting Hamiltonians that realize chiral $\mathbb Z_3$ anyon theories of opposite chirality; their sum is equivalent to the nonchiral $\mathbb Z_3$ toric-code Hamiltonian. In this example, $H_{\mathcal A}$ is a geometrically local commuting
Hamiltonian, but as discussed in Ref.~\cite{HaahInvertibleSubalgebras2023}, the chiral anyon content of  $\mathcal A$ provides evidence that its local-operator structure cannot be that of  an ordinary onsite qutrit algebra. Thus
locality of a particular factor Hamiltonian does not by itself imply a local onsite realization of the complete factor algebra.

This suggests an interesting problem.  Can one find a geometrically local, dynamically split Hamiltonian for which even the factor Hamiltonian itself cannot be made
geometrically local for any choice of factor variables or boundary conditions?  An especially interesting possibility would be an LDPC-like
factor Hamiltonian: each term would act on $O(1)$ factor qudits and each
factor qudit would occur in only $O(1)$ terms, while the resulting interaction
graph would have no geometrically local realization.  The kind of nontriviality mentioned above in the context of $\mathbb{Z}_3$ anyon theory does not lead to this stronger property because it constrains the
complete factor algebra rather than the factor Hamiltonian, which in fact is local in that example.  The chiral XYZ
model also does not yet provide such an example: the construction in
Appendix~\ref{app:tensor-factorization} is geometrically local with open boundaries, while on the torus only an $O(L)$ set of terms along the chosen
cuts has a range that grows with $L$.

Dynamical splitting may also be useful for studying decoherence in topological states.  As a potential application, consider a split $\mathbb Z_N$ topological model whose two factors carry opposite chiralities
and apply a local quantum channel to only one factor.  The other factor then
remains intact, while the decohered factor can undergo a mixed-state
transition, possibly leaving an intrinsically mixed topological phase with
chiral anyon content.  A framework for such phases was developed in
Refs.~\cite{SohalPremMixedState2025,EllisonChengMixedState2025}, while statistical-mechanics descriptions of a class of such transitions and related diagnostics of mixed-state chirality were subsequently developed in
Refs.~\cite{SunAldossariDevkotaLuo2026,EllisonManyBodyChirality2026}.
The advantage of a dynamically split model is that the channel preserves the
split even away from exactly solvable fixed points.  ED calculations can therefore be performed in a reduced Hilbert space, as in the chiral XYZ
model, making exact finite-size studies of mixed-state diagnostics possible
as functions of both a coherent chiral deformation and the chiral
decoherence strength.

\begin{acknowledgments}
	I thank Matthew Fisher, John McGreevy and Yi-Zhuang You for helpful discussions. I  acknowledge use of Codex/ChatGPT for help with writing exact-diagonalization codes and for useful discussions. T.G. is supported by the National Science Foundation under Grant No. DMR-2521369.
\end{acknowledgments}

\makeatletter

\appendix

\makeatother

\onecolumngrid

\addtocontents{toc}{\protect\hidesubsections}
\newpage
\section{Good quantum numbers of the chiral XYZ model}
\label{app:mixed-center-generator}

In this appendix, we identify the good quantum numbers associated with the subsystem symmetries of the $N = 2$ chiral XYZ model and explain how they depend on the lattice size. The convention described below is the same as the one in our ED study. Recall from the main text that
\begin{equation}
 n_D=\gcd(L_x,L_y),\qquad L_x=n_D a,\qquad L_y=n_D b,
 \label{eq:app-gcd-definitions}
\end{equation}
where $a$ and $b$ are coprime.  Since $a$ and $b$ cannot both be even,
\begin{equation}
 L_x\ \text{and}\ L_y\ \text{are both even}
 \quad\Longleftrightarrow\quad
 n_D\ \text{is even}.
 \label{eq:app-even-even}
\end{equation}
On an even-by-even lattice, the allowed parity classes of $(a,b)$ are therefore
(odd, odd), (odd, even), and (even, odd).

The ``relative line operators'' 
\begin{equation}
 c_x=C_xC_0,\qquad
 r_y=R_yR_0,\qquad
 d_s=D_sD_0
\label{eq:app-relative-lines}
\end{equation}
commute with the Hamiltonian and with every line symmetry. Following the
convention of the main text, the same symbols $c_x,r_y,d_s$ also denote
their $\pm1$ eigenvalues when a symmetry sector is specified. Their
eigenvalues therefore provide simultaneous quantum numbers for labeling the
energy eigenstates. Equivalently, these operators belong to the center of
the algebra generated by the line symmetries. There are
\begin{equation}
 (L_x-1)+(L_y-1)+(n_D-1)=L_x+L_y+n_D-3
 \label{eq:app-relative-count}
\end{equation}
of them, equal to the rank of the center.  They are independent unless both
$L_x$ and $L_y$ are even.  Indeed, modulo overall Pauli phases, the full line
operators obey the single relation
\begin{equation}
 \prod_x C_x\prod_y R_y\prod_s D_s\sim\id.
 \label{eq:app-binary-global-relation}
\end{equation}
A nontrivial product of relative strings can reproduce this relation only if
all relative strings are included and $L_x-1$, $L_y-1$, and $n_D-1$ are all
odd.  This occurs precisely when $L_x$, $L_y$, and $n_D$ are all even.
Keeping the phase in the ordering shown gives
\begin{equation}
 \left(\prod_x C_x\right)
 \left(\prod_y R_y\right)
 \left(\prod_s D_s\right)
 =i^{L_xL_y}\id.
 \label{eq:app-exact-global-relation}
\end{equation}
For an even-by-even lattice, $L_xL_y$ is divisible by four, and hence
\begin{equation}
 \prod_{x=1}^{L_x-1}c_x
 \prod_{y=1}^{L_y-1}r_y
 \prod_{s=1}^{n_D-1}d_s
 =\id.
 \label{eq:app-relative-relation}
\end{equation}
Thus the relative strings have one redundancy and span only
$L_x+L_y+n_D-4$ independent central generators.

The missing generator can be chosen as a product of the three reference
strings.  For
\begin{equation}
 C_0^\alpha R_0^\beta D_0,
 \qquad \alpha,\beta\in\mathbb F_2,
 \label{eq:app-reference-product}
\end{equation}
commutation with an arbitrary column and row gives the phases
$(-1)^{\beta+b}$ and $(-1)^{\alpha+a}$, respectively.  The product is
therefore central for
\begin{equation}
 \alpha=a\pmod 2,\qquad \beta=b\pmod 2.
 \label{eq:app-centrality-condition}
\end{equation}
Its commutation phase with any diagonal string is then
$(-1)^{b\alpha+a\beta}=1$.  Up to an overall phase, the required mixed central generator is consequently
\begin{equation}
 C_0^\alpha R_0^\beta D_0,
 \qquad
 \alpha=a\pmod 2,\quad \beta=b\pmod 2.
 \label{eq:app-mixed-generator-binary}
\end{equation}
On an even-by-even lattice this operator is not in the span of the relative
strings.  Every product of relative strings contains an even number of
operators from each of the column, row, and diagonal families, and the global
relation changes each of these numbers by an even amount.  By contrast,
Eq.~\eqref{eq:app-mixed-generator-binary} has odd parity in the diagonal
family and in at least one of the other two families.

For numerical sector labels, we choose the canonical Hermitian Pauli
representative: after reducing the operator to an onsite product of
$I,X,Y,Z$, its overall coefficient is positive.  First define
\begin{equation}
 \overline Q=
 \begin{cases}
  -iC_0R_0D_0,
    & a,b\ \text{odd},\\[2pt]
  C_0D_0,
    & a\ \text{odd},\ b\ \text{even},\\[2pt]
  R_0D_0,
    & a\ \text{even},\ b\ \text{odd}.
 \end{cases}
 \label{eq:app-Qbar}
\end{equation}
In the first case, $C_0$, $R_0$, and $D_0$ pairwise anticommute, so their
product is anti-Hermitian and the factor $-i$ makes $\overline Q$ Hermitian.
The remaining sign required by the canonical convention is
\begin{equation}
 Q_{\rm mix}=\sigma_{a,b}\overline Q,
 \qquad
 \sigma_{a,b}=
 \begin{cases}
  (-1)^{(a+b-2)/2},
    & a,b\ \text{odd},\\[2pt]
  (-1)^{b/2},
    & a\ \text{odd},\ b\ \text{even},\\[2pt]
  (-1)^{a/2},
    & a\ \text{even},\ b\ \text{odd}.
 \end{cases}
 \label{eq:app-Qmix}
\end{equation}
The signs in Eq.~\eqref{eq:app-Qmix} follow directly from the overlaps of the
reference strings.  The intersections $C_0\cap D_0$ and $R_0\cap D_0$ contain
$b$ and $a$ sites, respectively.  Using $ZY=-iX$ and $XY=iZ$, the onsite
phase of $C_0D_0$ is $(-i)^b=(-1)^{b/2}$ when $b$ is even, while that of
$R_0D_0$ is $i^a=(-1)^{a/2}$ when $a$ is even.  When $a$ and $b$ are both
odd, the onsite phase of $-iC_0R_0D_0$ is
$(-i)^{b-1}i^{a-1}=(-1)^{(a+b-2)/2}$.  Thus the factor $\sigma_{a,b}$ makes
the overall onsite coefficient positive.  With this convention,
$Q_{\rm mix}^2=\id$.

An even-by-even symmetry sector may therefore be labeled by the eigenvalue of
$Q_{\rm mix}$ together with an independent subset of the relative-string
eigenvalues, which satisfy
\begin{equation}
 \prod_{x>0}c_x\prod_{y>0}r_y\prod_{s>0}d_s=1.
 \label{eq:app-charge-constraint}
\end{equation}
When at least one of $L_x,L_y$ is odd, the relative strings are already
independent, and the mixed central operator lies in their span rather than
providing an additional label.  This distinction concerns only the labeling
of symmetry sectors; the exact twofold degeneracy follows for every periodic
geometry from
\begin{equation}
 [H,C_0]=[H,R_0]=0,\qquad C_0R_0=-R_0C_0.
 \label{eq:app-universal-doublet}
\end{equation}
\section{Algebraic details of the dynamical split symmetry and its numerical implementation in ED}
\label{app:tensor-factorization}

This appendix derives the tensor-product structure used in the exact
diagonalization, illustrates how it arises from the triangle operators, and
describes its numerical implementation.

\subsection{Binary representation of the triangle operators}
\label{subsec:appendix-binary-representation}

A Pauli string on $n$ qubits, modulo its overall phase
$\{\pm1,\pm i\}$, can be represented by a binary vector
\begin{equation}
v=(x\mid z)\in\mathbb F_2^{2n},
\qquad
P(v)=i^{x\cdot z}X^xZ^z.
\end{equation}
At each site, the pairs
$(x_j,z_j)=(0,0),(1,0),(0,1),(1,1)$ represent $I,X,Z,Y$,
respectively.  Multiplication of Pauli strings becomes addition of their
binary vectors, while their commutation relation is determined by
\begin{equation}
\langle v,w\rangle
=x\cdot z'+z\cdot x'
\pmod 2,
\qquad
P(v)P(w)=(-1)^{\langle v,w\rangle}P(w)P(v).
\label{eq:symp}
\end{equation}

Let $V_A$ and $V_B$ denote the binary spaces generated by the two families
of triangle operators:
\begin{equation}
V_A=\operatorname{span}_{\mathbb F_2}\{v(a_r)\},
\qquad
V_B=\operatorname{span}_{\mathbb F_2}\{v(b_r)\}.
\label{eq:binaryspaces}
\end{equation}
The corresponding operator algebras are
\begin{equation}
\mathcal A
=\operatorname{span}_{\mathbb C}\{P(v):v\in V_A\},
\qquad
\mathcal B
=\operatorname{span}_{\mathbb C}\{P(v):v\in V_B\}.
\label{eq:operatoralgebras}
\end{equation}
Since every $a$-triangle commutes with every $b$-triangle,
\begin{equation}
[a_r,b_{r'}]=0
\quad\text{for all }r,r',
\qquad\Longleftrightarrow\qquad
\langle V_A,V_B\rangle=0.
\label{eq:crosscommute}
\end{equation}

\subsection{Common center}
\label{subsec:appendix-common-center}

The operators in $V_A$ that commute with all of $V_A$ form its radical,
\begin{equation}
\operatorname{rad}(V_A)
=
\{v\in V_A:
\langle v,w\rangle=0
\text{ for every }w\in V_A\}.
\label{eq:radical}
\end{equation}
Their Pauli representatives generate the center of $\mathcal A$, and
similarly for $V_B$ and $\mathcal B$.

For the periodic chiral XYZ lattices considered here, binary row reduction
gives
\begin{equation}
Z\equiv
\operatorname{rad}(V_A)
=
\operatorname{rad}(V_B)
=
V_A\cap V_B.
\label{eq:commoncenter}
\end{equation}
Thus the two triangle algebras have the same center.  Let
$G_1,\ldots,G_c$ be independent Hermitian generators of $Z$, and denote
their eigenvalues by
\begin{equation}
G_j|\psi\rangle=\lambda_j|\psi\rangle,
\qquad
\lambda_j=\pm1.
\end{equation}
A fixed center sector is labeled by
$\lambda=(\lambda_1,\ldots,\lambda_c)$.

The equality in Eq.~\eqref{eq:commoncenter} concerns operators generated by
the triangle terms themselves.  Other operators may commute with a triangle
algebra without belonging to it; the logical row and column strings provide
examples.

\subsection{Tensor-product structure within a fixed center sector}
\label{subsec:appendix-sector-factorization}

After fixing the center sector $\lambda$, the remaining operators in each
triangle algebra can be arranged into independent anticommuting Pauli pairs.
If
\begin{equation}
\dim(V_A/Z)=2q_A,
\end{equation}
one can choose representatives
\begin{equation}
e_1,f_1,\ldots,e_{q_A},f_{q_A},
\qquad
\langle e_i,f_j\rangle=\delta_{ij},
\end{equation}
with all other pairings zero.  Their Pauli representatives act as the
$X$ and $Z$ operators of $q_A$ effective qubits.  Products of these
operators generate the full operator algebra on those qubits:
\begin{equation}
\left.\mathcal A\right|_\lambda
\simeq M_{2^{q_A}}(\mathbb C).
\label{eq:fullA}
\end{equation}
The same construction gives
\begin{equation}
\left.\mathcal B\right|_\lambda
\simeq M_{2^{q_B}}(\mathbb C).
\end{equation}
This is the symplectic analogue of the usual Gram--Schmidt procedure
\cite{WildeLogicalOperators}.  One begins with any nonzero vector $e_1$;
nondegeneracy guarantees a vector $f_1$ such that
$\langle e_1,f_1\rangle=1$.  Every remaining vector $g$ is then replaced by
\begin{equation}
g\longmapsto
g+\langle g,f_1\rangle e_1+\langle g,e_1\rangle f_1,
\end{equation}
which makes it commute with both $e_1$ and $f_1$ without changing the space
generated by all the vectors.  Setting this pair aside and repeating the
procedure produces the canonical pairs above.

Because $\mathcal A$ is the full operator algebra on $q_A$ effective qubits,
any operator commuting with $\mathcal A$ acts only on the remaining degrees
of freedom.  Since $\mathcal B$ commutes with $\mathcal A$ and is itself the
full operator algebra on $q_B$ effective qubits, the two algebras act on
separate tensor factors.  Any degrees of freedom on which neither algebra
acts form an additional multiplicity factor.  Thus, within each fixed center
sector,
\begin{equation}
\mathcal H_\lambda
\simeq
\mathcal H_{A,\lambda}
\otimes
\mathcal H_{B,\lambda}
\otimes
\mathcal H_{P,\lambda}.
\label{eq:ABfactor}
\end{equation}

For an $L_x\times L_y$ torus, let
\begin{equation}
n=L_xL_y,
\qquad
n_D=\gcd(L_x,L_y),
\qquad
c=L_x+L_y+n_D-3.
\end{equation}
The binary ranks are
\begin{equation}
\dim(V_A/Z)=\dim(V_B/Z)=2q,
\qquad
2q=n-L_x-L_y-n_D+2.
\end{equation}
Fixing the $c$ central eigenvalues leaves
\begin{equation}
\dim\mathcal H_\lambda
=2^{n-c}
=2^{2q+1}.
\end{equation}
The $A$ and $B$ factors each have dimension $2^q$, so the remaining
multiplicity factor is two-dimensional.  It is precisely the logical qubit
generated by the anticommuting row and column symmetries.  Therefore
\begin{equation}
\boxed{
	\mathcal H_\lambda
	\simeq
	\mathbb C_P^2
	\otimes
	\mathbb C_A^{2^q}
	\otimes
	\mathbb C_B^{2^q}.}
\label{eq:finalfactorization}
\end{equation}

The Hamiltonian correspondingly takes the form
\begin{equation}
H_\lambda
=
\mathds{1}_P\otimes
\left[
H_{A,\lambda}\otimes\mathds{1}_B
+
\mathds{1}_A\otimes H_{B,\lambda}
\right].
\label{eq:factorham}
\end{equation}
The two factor Hamiltonians can therefore be diagonalized independently, and
their eigenvalues are added to obtain the full spectrum.  The logical-qubit
factor supplies an exact twofold multiplicity to every energy level.

\subsection{Worked example: the chiral XYZ model on a $3\times3$ torus}
\label{subsec:appendix-3x3}

For the $3\times3$ torus, the Hamiltonian is
\begin{equation}
H=-J\sum_{x,y\in\mathbb Z_3}\left(a_{x,y}+b_{x,y}\right),
\end{equation}
with
\begin{align}
a_{x,y}&=X_{x,y}Y_{x,y+1}Z_{x+1,y+1},
\nonumber\\
b_{x,y}&=Z_{x,y}Y_{x+1,y}X_{x+1,y+1},
\end{align}
where all coordinates are understood modulo three.  Binary row reduction gives
\begin{align}
\dim V_A=\dim V_B&=8,
\nonumber\\
\dim\operatorname{rad}(V_A)
=\dim\operatorname{rad}(V_B)&=6.
\end{align}
The two radicals coincide with the common center.  A convenient basis is
$c_i=C_iC_0$, $r_i=R_iR_0$, and
$d_i=D_iD_0$, with $i=1,2$.
We denote their eigenvalues in a fixed center sector by
\begin{equation}
\lambda=(c_1,c_2,r_1,r_2,d_1,d_2),
\qquad
c_i,r_i,d_i=\pm1.
\end{equation}

Since each triangle space has dimension eight and a six-dimensional center,
its noncentral part is two-dimensional and therefore describes a single
factor qubit.  We choose
\begin{equation}
\tau_{A,1}^{x}=a_{0,0},
\qquad
\tau_{A,1}^{z}=a_{1,0},
\qquad
\tau_{A,1}^{y}=i\tau_{A,1}^{x}\tau_{A,1}^{z},
\end{equation}
and similarly
\begin{equation}
\tau_{B,1}^{x}=b_{0,0},
\qquad
\tau_{B,1}^{z}=b_{1,0},
\qquad
\tau_{B,1}^{y}=i\tau_{B,1}^{x}\tau_{B,1}^{z}.
\end{equation}
Every $A$ triangle is then a product of a central operator and one of
$\tau_{A,1}^{x},\tau_{A,1}^{y},\tau_{A,1}^{z}$, and every $B$ triangle has
the analogous form.  For example,
\begin{equation}
a_{2,0}=r_1\tau_{A,1}^{y},
\qquad
b_{2,0}=-r_1\tau_{B,1}^{y}.
\end{equation}
After fixing $\lambda$, the central operators become signs, and the two
triangle sums reduce to single-qubit Hamiltonians,
\begin{align}
H_A(\lambda)&=-J\left(A_x\tau_{A,1}^{x}
+A_y\tau_{A,1}^{y}+A_z\tau_{A,1}^{z}\right),
\label{eq:HA-3x3}\\
H_B(\lambda)&=-J\left(B_x\tau_{B,1}^{x}
+B_y\tau_{B,1}^{y}+B_z\tau_{B,1}^{z}\right),
\label{eq:HB-3x3}
\end{align}
where
\begin{align}
A_x={}&1+c_1c_2r_2d_1d_2+c_2r_1r_2d_1,
\nonumber\\
A_y={}&r_1+c_1r_1r_2d_1+c_1c_2r_2d_2,
\nonumber\\
A_z={}&1+c_2r_2d_2+c_1r_1r_2d_1d_2,
\label{eq:A-coeffs}
\end{align}
and
\begin{align}
B_x={}&1+c_1c_2r_2d_2+c_2r_1r_2d_1d_2,
\nonumber\\
B_y={}&-r_1-c_1r_1r_2d_1d_2-c_1c_2r_2d_1,
\nonumber\\
B_z={}&1+c_2r_2d_1+c_1r_1r_2d_2.
\label{eq:B-coeffs}
\end{align}

Fixing the six central signs leaves an eight-dimensional sector,
\begin{equation}
\dim\mathcal H_\lambda=2^{9-6}=8
=2_A\times2_B\times2_P,
\end{equation}
where the final factor is the logical qubit generated by the anticommuting
row and column symmetries.  Thus
\begin{equation}
\mathcal H_\lambda\simeq
\mathbb C_A^2\otimes\mathbb C_B^2\otimes\mathbb C_P^2,
\end{equation}
and the full Hamiltonian in this sector is
\begin{equation}
\boxed{
	H\big|_\lambda=
	\left[H_A(\lambda)\otimes I_B
	+I_A\otimes H_B(\lambda)\right]\otimes I_P.}
\label{eq:3x3-full-H}
\end{equation}
The two factor spectra are
\begin{equation}
\operatorname{spec}H_A=\{\pm J|\boldsymbol A|\},
\qquad
\operatorname{spec}H_B=\{\pm J|\boldsymbol B|\},
\end{equation}
so the energies are
\begin{equation}
E_{s,t}=sJ|\boldsymbol A|+tJ|\boldsymbol B|,
\qquad s,t=\pm1,
\end{equation}
with an exact twofold multiplicity from the $P$ qubit.  For the all-positive
center sector, $\boldsymbol A=(3,3,3)$ and
$\boldsymbol B=(3,-3,3)$, reproducing the direct diagonalization of the
original nine-qubit Hamiltonian.

\subsection{Explicit algebraic duality and almost local form of the factor Hamiltonians}

The choice of factor qubits is not unique and depends on the choices made during the symplectic  Gram-Schmidt procedure. Of course the physical properties of the actual model do not depend on this choice. We now use this freedom to ask a structural question: can the physical
triangle terms be represented by interactions among only a few nearby factor
qubits?  This choice has no numerical advantage for the diagonalization, but it is still conceptually interesting as a form of ``algebraic duality''.

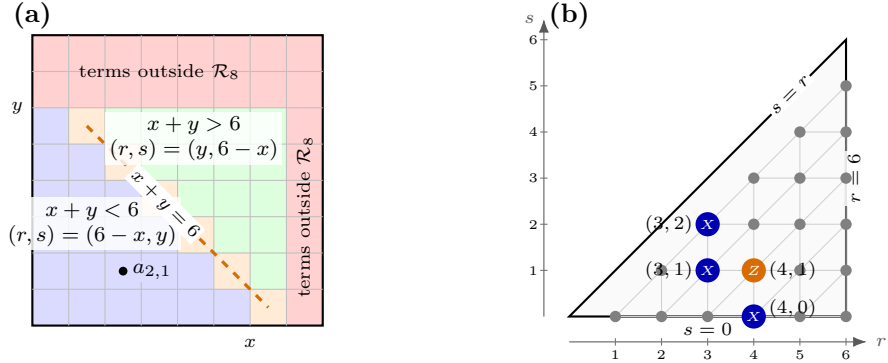
\begin{figure}[t]
	\centering
	\begin{tikzpicture}[font=\small]
	\begin{scope}[x=0.48cm,y=0.48cm]
	\foreach \x in {0,...,6}{
		\foreach \y in {0,...,5}{
			\pgfmathtruncatemacro{\xy}{\x+\y}
			\ifnum\xy<6
			\fill[blue!14] (\x,\y) rectangle ++(1,1);
			\else
			\ifnum\xy>6
			\fill[green!13] (\x,\y) rectangle ++(1,1);
			\else
			\fill[orange!16] (\x,\y) rectangle ++(1,1);
			\fi
			\fi
		}
	}
	\fill[red!18] (0,6) rectangle (8,8);
	\fill[red!18] (7,0) rectangle (8,6);
	\draw[black,thick] (0,0) rectangle (8,8);
	\foreach \x in {1,...,7}{\draw[black!25] (\x,0)--(\x,8);}
	\foreach \y in {1,...,7}{\draw[black!25] (0,\y)--(8,\y);}
	\draw[orange!85!black,very thick,dashed] (1.5,5.5)--(6.5,0.5);
	
	\node[font=\bfseries] at (0,8.55) {(a)};
	\node[align=center,font=\footnotesize,fill=white,fill opacity=0.82,
	text opacity=1,inner sep=1.5pt] at (1.65,2.8)
	{$x+y<6$\\$(r,s)=(6-x,y)$};
	\node[align=center,font=\footnotesize,fill=white,fill opacity=0.82,
	text opacity=1,inner sep=1.5pt] at (4.45,5.15)
	{$x+y>6$\\$(r,s)=(y,6-x)$};
	\node[rotate=-45,fill=white,inner sep=1pt,font=\scriptsize]
	at (3.65,3.35) {$x+y=6$};
	\node[align=center,rotate=90,font=\scriptsize] at (7.5,3)
	{terms outside $\mathcal R_8$};
	\node[align=center,font=\scriptsize] at (3.5,7)
	{terms outside $\mathcal R_8$};
	
	\fill[black] (2.5,1.5) circle (0.12);
	\node[anchor=west,fill=blue!14,inner sep=1pt,font=\footnotesize]
	at (2.68,1.5) {$a_{2,1}$};
	\node[font=\scriptsize] at (6,-0.48) {$x$};
	\node[font=\scriptsize] at (-0.42,6) {$y$};
	\end{scope}
	
	\begin{scope}[shift={(7.1cm,0.12cm)},x=0.61cm,y=0.61cm]
	\node[font=\bfseries] at (0,6.53) {(b)};
	
	\fill[gray!4] (0,0)--(6,0)--(6,6)--cycle;
	\draw[black,thick] (0,0)--(6,0)--(6,6)--cycle;
	\node[rotate=90,fill=white,inner sep=1pt,font=\scriptsize]
	at (6.18,3.0) {$r=6$};
	\node[rotate=45,fill=white,inner sep=1pt,font=\scriptsize]
	at (4.8,4.8) {$s=r$};
	\node[fill=white,inner sep=1pt,font=\scriptsize] at (3.0,-0.28) {$s=0$};
	
	\foreach \r/\n in {1/0,2/1,3/2,4/3,5/4}{
		\pgfmathtruncatemacro{\rp}{\r+1}
		\foreach \s in {0,...,\n}{
			\pgfmathtruncatemacro{\sp}{\s+1}
			\draw[black!18] (\r,\s)--(\rp,\s);
			\draw[black!18] (\r,\s)--(\rp,\sp);
		}
	}
	\foreach \r/\n in {2/0,3/1,4/2,5/3,6/4}{
		\foreach \s in {0,...,\n}{
			\pgfmathtruncatemacro{\sp}{\s+1}
			\draw[black!18] (\r,\s)--(\r,\sp);
		}
	}
	\foreach \r/\n in {1/0,2/1,3/2,4/3,5/4,6/5}{
		\foreach \s in {0,...,\n}{
			\fill[black!50] (\r,\s) circle (0.075cm);
		}
	}
	
	\foreach \r/\s in {3/1,3/2,4/0}{
		\fill[blue!70!black] (\r,\s) circle (0.16cm);
		\node[text=white,font=\bfseries\tiny] at (\r,\s) {$X$};
	}
	\fill[orange!85!black] (4,1) circle (0.16cm);
	\node[text=white,font=\bfseries\tiny] at (4,1) {$Z$};
	
	\node[anchor=east,font=\scriptsize] at (2.88,1) {$(3,1)$};
	\node[anchor=east,font=\scriptsize] at (2.88,2) {$(3,2)$};
	\node[anchor=west,font=\scriptsize] at (4.12,0.18) {$(4,0)$};
	\node[anchor=west,font=\scriptsize] at (4.12,1) {$(4,1)$};
	
	\draw[-{Latex[length=1.8mm]},black!65] (0,-0.55)--(6.45,-0.55)
	node[right,font=\scriptsize] {$r$};
	\draw[-{Latex[length=1.8mm]},black!65] (-0.55,0)--(-0.55,6.45)
	node[left,font=\scriptsize] {$s$};
	\foreach \r in {1,...,6}{
		\draw[black!55] (\r,-0.63)--(\r,-0.47);
		\node[font=\tiny] at (\r,-0.82) {\r};
	}
	\foreach \s in {1,...,6}{
		\draw[black!55] (-0.63,\s)--(-0.47,\s);
		\node[font=\tiny] at (-0.78,\s) {\s};
	}
	\end{scope}
	\end{tikzpicture}
	\caption{Folding construction for the factor qubits, illustrated here for $L=8$.
		(a) Each cell labels one physical triangle term $a_{x,y}$ in the chiral XYZ Hamiltonian. The factor qubits are labeled by $(r,s)$. The blue half
		of the rectangular core maps to $(r,s)=(6-x,y)$, while the green half maps to
		$(r,s)=(y,6-x)$, thus these two maps fold the rectangle onto the same triangular
		domain.  The orange cells lie on the fold, and the red $3L-2$ terms remain
		outside the core.  (b) The factor qubits are drawn in their actual
		Cartesian coordinates, within the boundaries $s=0$, $r=6$, and $s=r$. Following Eq.~\eqref{eq:factor-lower-fold-rule}, the marked core term has the representation $a_{2,1}|_\lambda=\eta^A_{2,1}(\lambda)
		\tau^z_{A;(4,1)}\tau^x_{A;(3,1)}
		\tau^x_{A;(3,2)}\tau^x_{A;(4,0)}$ where $\eta^A_{2,1}(\lambda)$ is a sign that depends on the center sector $\lambda$.}
	\label{fig:factor-local-core}
\end{figure}

We give an explicit construction for the $A$ factor; the $B$ factor follows
in the same way.  We start from the rectangular set of triangle terms
\begin{equation}
 \mathcal R_L=\{(x,y):0\leq x\leq L-2,\ 0\leq y\leq L-3\}.
 \label{eq:factor-rectangular-core}
\end{equation}
It contains $(L-1)(L-2)=2q$ terms.  The remaining $3L-2$ terms form one
column and two rows of the periodic lattice, as shown in
Fig.~\ref{fig:factor-local-core}(a).  This is only a convenient choice of
coordinates: the physical Hamiltonian still has periodic boundary
conditions.

Next, we place the $q=(L-1)(L-2)/2$ factor qubits on an auxiliary triangular
lattice and label them by
\begin{equation}
 (r,s),\qquad 1\leq r\leq L-2,\qquad 0\leq s<r.
 \label{eq:factor-triangular-coordinates}
\end{equation}
The coordinates $(x,y)$ and $(r,s)$ therefore label different objects:
$(x,y)$ labels a physical triangle term, whereas $(r,s)$ labels a factor
qubit.  The relation between them is most easily viewed as folding the
rectangle along the line $x+y=L-2$, see Fig.~\ref{fig:factor-local-core}.  Well inside the lower half, define
$(r,s)=(L-2-x,y)$.  The local rule is
\begin{equation}
 a_{x,y}\big|_\lambda=\eta^A_{x,y}(\lambda)
 \tau^z_{A;(r,s)}
 \tau^x_{A;(r-1,s)}
 \tau^x_{A;(r-1,s+1)}
 \tau^x_{A;(r,s-1)}.
 \label{eq:factor-lower-fold-rule}
\end{equation}
A key sanity check for this map is that it preserves the commutation relations for the triangle operators $\{a_{x,y}\}$.  For example, increasing $x$ by one maps
$(r,s)$ to $(r-1,s)$, and hence
\begin{equation}
a_{x+1,y}\big|_\lambda\propto
\tau^z_{A;(r-1,s)}
\tau^x_{A;(r-2,s)}
\tau^x_{A;(r-2,s+1)}
\tau^x_{A;(r-1,s-1)}.
\end{equation}
The representations of $a_{x,y}$ and $a_{x+1,y}$ therefore contain,
respectively, $\tau^x_{A;(r-1,s)}$ and
$\tau^z_{A;(r-1,s)}$.  These $\tau$ matrices are their only anticommuting overlap, so
\begin{equation}
\{a_{x,y},a_{x+1,y}\}=0.
\end{equation}
Similarly, the pairs separated by $(0,1)$ and $(1,1)$ have one
anticommuting overlap, on the factor qubits at $(r,s)$ and
$(r-1,s+1)$, respectively.
All other pairs have zero or an even number of such overlaps and therefore
commute.  Thus the factor-qubit representation reproduces the commutation
relations among the physical triangle terms.

Likewise, well inside the upper half, define $(r,s)=(y,L-2-x)$. The corresponding local rule is
\begin{equation}
 a_{x,y}\big|_\lambda=\eta^A_{x,y}(\lambda)
 \tau^z_{A;(r,s)}
 \tau^x_{A;(r,s+1)}
 \tau^x_{A;(r+1,s)}
 \tau^x_{A;(r+1,s-1)}.
 \label{eq:factor-upper-fold-rule}
\end{equation}

Thus the two halves of
the rectangle fold onto the same factor-qubit lattice, with each generic
triangle term acting on a small cluster.  Figure~\ref{fig:factor-local-core}
illustrates this map for $L=8$. The terms next to the fold and the outer edges require modifications, but their range and weight remain bounded, as we will  discuss shortly. Factor qubits whose coordinates lie outside the triangular domain will also be discussed below.

For completeness, we now specify the factor-qubit representation of every
term in $\mathcal R_L$, including the terms near the boundary of the
rectangle and along the fold. For the factor qubit at $(r,s)$, let $\mathcal X_{r,s}$ and
$\mathcal Z_{r,s}$ be sets of physical triangle coordinates $(x,y)$.
The representation of $a_{x,y}$ acts on this factor qubit as
$\tau^x_{A;(r,s)}$ if $(x,y)$ belongs only to $\mathcal X_{r,s}$,
as $\tau^z_{A;(r,s)}$ if it belongs only to $\mathcal Z_{r,s}$,
as $\tau^y_{A;(r,s)}$ if it belongs to both sets, and as
$\mathds{1}$ if it belongs to neither. For
$1\leq r\leq L-4$ and $0\leq s\leq r-1$, define
\begin{equation}
 u=L-2-r,\qquad v=L-2-s,
\end{equation}
and
\begin{equation}
 \mathcal Z_{r,s}=\{(u,s),(v,r)\}.
 \label{eq:factor-complete-Z-bulk}
\end{equation}
For $s\leq r-2$,
\begin{align}
 \mathcal X_{r,s}=\{&
 (u-1,s-1),(u-1,s),(u,s+1), (v-1,r-1),(v,r-1),(v+1,r)\},
 \label{eq:factor-complete-X-bulk}
\end{align}
while for $s=r-1$,
\begin{align}
 \mathcal X_{r,r-1}=\{&
 (u-1,r-2),(u-1,r-1),(u+1,r-1),
  (u,r),(u+1,r),(u+2,r)\}.
 \label{eq:factor-complete-X-fold}
\end{align}
The two outer rows of factor qubits are specified, for
$0\leq s\leq L-4$, by
\begin{align}
 \mathcal Z_{L-2,s}&=\{(1,s),(L-2-s,L-3)\},
 \nonumber\\
 \mathcal X_{L-2,s}&=\{(1,s-1),(L-3-s,L-4),
 (L-2-s,L-4),(L-3-s,L-3)\},
 \label{eq:factor-complete-outer-one}\\
 \mathcal Z_{L-3,s}&=\{(0,s)\},
 \nonumber\\
 \mathcal X_{L-3,s}&=\{(1,s),(0,s+1),(1,s+1)\}.
 \label{eq:factor-complete-outer-two}
\end{align}
Finally,
\begin{equation}
 \mathcal X_{L-2,L-3}=\{(0,L-3)\},\qquad
 \mathcal Z_{L-2,L-3}=\{(1,L-3)\}.
 \label{eq:factor-complete-corner}
\end{equation}
Any coordinate produced by these formulas that lies outside
$\mathcal R_L$ is discarded. The lists are understood modulo two: if the
same coordinate occurs twice, the two occurrences cancel because
$(\tau^\alpha)^2=\mathds{1}$. They give
\begin{equation}
 a_{x,y}\big|_\lambda=\eta^A_{x,y}(\lambda)
 \prod_{j\in\mathcal N_{x,y}}\tau_{A,j}^{\alpha_j},
 \qquad |\mathcal N_{x,y}|\leq6,
 \label{eq:factor-local-core-term}
\end{equation}
All the qubits in $\mathcal N_{x,y}$ lie within distance two on the
triangular lattice, and every factor qubit occurs in at most eight core
terms.  These bounds are independent of $L$.

As one may verify, similar to the maps  in Eqs.\eqref{eq:factor-lower-fold-rule} and  \eqref{eq:factor-upper-fold-rule}, the map in Eq.~\eqref{eq:factor-local-core-term} is again compatible with commutation relations of $\{a_{x,y}\}$ operators, as it should.  Moreover, no nontrivial product of the $2q$ Pauli strings representing the
terms in $\mathcal R_L$ commutes with every term in $\mathcal R_L$.
Their commutation matrix is therefore nondegenerate, so they span the full
Pauli algebra of the $q$ factor qubits. This also shows that the map can be inverted: each factor
Pauli can be expressed as a product of core triangle terms,
\begin{equation}
 \tau_{A,j}^{\alpha}
 \longleftrightarrow
 \prod_{(x,y)\in\mathcal R_L}
 a_{x,y}^{\,n^{\alpha}_{x,y;j}},
 \qquad n^{\alpha}_{x,y;j}=0,1.
 \label{eq:factor-inverse-physical-map}
\end{equation}
These products can extend across the microscopic lattice; just because $a_{x,y}$ for $(x,y)\in\mathcal R_L$ can be represented as a product of an
$O(1)$ number of Pauli operators $\tau_{A,j}^{\alpha}$, it does not imply that each $\tau_{A,j}^{\alpha}$  can be written as a product of an $O(1)$ number of $a_{x,y}$ terms.

 The complement of $\mathcal R_L$ consists of two rows and one column.
Because the core strings form a Pauli basis, the representation of each
remaining triangle term is uniquely fixed by its commutators with the core.
The resulting strings are supported on fixed-width neighborhoods of
one-dimensional paths in the triangular factor lattice.  Let $\operatorname{wt}_{\tau}(O)$ denote the number of factor qubits on which the factor-qubit representation of $O$ acts nontrivially. Direct counting gives, for $L\geq6$,
\begin{align}
\max_{0\leq x\leq L-2}
\operatorname{wt}_{\tau}(a_{x,L-2})&=4,
\nonumber\\
\max_{0\leq x\leq L-2}
\operatorname{wt}_{\tau}(a_{x,L-1})&=3L-11,
\nonumber\\
\max_{0\leq y\leq L-1}
\operatorname{wt}_{\tau}(a_{L-1,y})&=3L-12.
\end{align}
Thus the factor Hamiltonian is geometrically local away from the chosen
cuts, while the remaining $3L-2$ terms have weight at most
$3L-11=O(L)$.  The locations of these longer terms depend on the
factor-qubit basis, and we do not claim that this construction is optimal.

The same construction clarifies the role of boundary conditions.  With open boundary conditions, after fixing the boundary-center eigenvalues, all terms in the factor
Hamiltonians have weight at most six and bounded geometric range, and every
factor qubit occurs in at most eight terms.  The qubits close to the boundaries can be removed by adding suitable split-preserving one-site boundary terms,
without destroying geometric locality.

\subsubsection{Explicit factor Hamiltonians}
\label{subsubsec:explicit-factor-hamiltonians}

We now give two explicit factor-qubit encodings for $4\times4$ and
$6\times6$ tori.  Each row expresses one physical triangle operator as a
single Pauli string on the $q=(L-1)(L-2)/2$ factor qubits.  Identity factors
and the overall sign fixed by the center character are omitted.  The
$B$-triangle operators have the same Pauli strings with $A$ replaced by $B$,
although their sector-dependent signs can differ.
In the tables, we assign the factor qubit at $(r,s)$ the integer label
$j=r(r-1)/2+s+1$, with $1\leq r\leq L-2$ and $0\leq s<r$.

\subsubsection*{The \(4\times 4\) torus}

Each factor contains \(q=3\) factor qubits.

\begin{longtable}{@{}ccc@{\hspace{4em}}>{\raggedright\arraybackslash}p{0.68\textwidth}@{}}
\caption{An explicit factor-qubit representation of the \(A\)-triangle operators on the \(4\times 4\) torus. Identity factors and the overall sector-dependent signs are omitted. The corresponding \(B\)-triangle operators are obtained by replacing \(A\) with \(B\).}\label{tab:explicit-factor-terms-L4}\\
\toprule
\(x\) & \(y\) & weight & \multicolumn{1}{l}{\(a_{x,y}\) in terms of factor-qubit Pauli operators} \\
\midrule
\endfirsthead
\toprule
\(x\) & \(y\) & weight & \multicolumn{1}{l}{\(a_{x,y}\) in terms of factor-qubit Pauli operators} \\
\midrule
\endhead
\midrule
\multicolumn{4}{r}{\emph{Continued on the next page}}\\
\endfoot
\bottomrule
\endlastfoot
0 & 0 & 1 & \(\tau_{A,1}^{z}\) \\
1 & 0 & 2 & \(\tau_{A,1}^{x}\,\allowbreak \tau_{A,2}^{y}\) \\
2 & 0 & 1 & \(\tau_{A,2}^{x}\) \\
3 & 0 & 2 & \(\tau_{A,1}^{y}\,\allowbreak \tau_{A,2}^{z}\) \\
0 & 1 & 2 & \(\tau_{A,1}^{x}\,\allowbreak \tau_{A,3}^{x}\) \\
1 & 1 & 3 & \(\tau_{A,1}^{x}\,\allowbreak \tau_{A,2}^{x}\,\allowbreak \tau_{A,3}^{z}\) \\
2 & 1 & 1 & \(\tau_{A,2}^{z}\) \\
3 & 1 & 2 & \(\tau_{A,2}^{y}\,\allowbreak \tau_{A,3}^{y}\) \\
0 & 2 & 1 & \(\tau_{A,3}^{z}\) \\
1 & 2 & 1 & \(\tau_{A,3}^{y}\) \\
2 & 2 & 3 & \(\tau_{A,1}^{z}\,\allowbreak \tau_{A,2}^{x}\,\allowbreak \tau_{A,3}^{z}\) \\
3 & 2 & 3 & \(\tau_{A,1}^{z}\,\allowbreak \tau_{A,2}^{x}\,\allowbreak \tau_{A,3}^{y}\) \\
0 & 3 & 2 & \(\tau_{A,1}^{y}\,\allowbreak \tau_{A,3}^{y}\) \\
1 & 3 & 2 & \(\tau_{A,2}^{z}\,\allowbreak \tau_{A,3}^{x}\) \\
2 & 3 & 3 & \(\tau_{A,1}^{z}\,\allowbreak \tau_{A,2}^{z}\,\allowbreak \tau_{A,3}^{z}\) \\
3 & 3 & 1 & \(\tau_{A,1}^{x}\) \\
\end{longtable}

\subsubsection*{The \(6\times 6\) torus}

Each factor contains \(q=10\) factor qubits.

\begin{longtable}{@{}ccc@{\hspace{4em}}>{\raggedright\arraybackslash}p{0.68\textwidth}@{}}
\caption{An explicit factor-qubit representation of the \(A\)-triangle operators on the \(6\times 6\) torus. Identity factors and the overall sector-dependent signs are omitted. The corresponding \(B\)-triangle operators are obtained by replacing \(A\) with \(B\).}\label{tab:explicit-factor-terms-L6}\\
\toprule
\(x\) & \(y\) & weight & \multicolumn{1}{l}{\(a_{x,y}\) in terms of factor-qubit Pauli operators} \\
\midrule
\endfirsthead
\toprule
\(x\) & \(y\) & weight & \multicolumn{1}{l}{\(a_{x,y}\) in terms of factor-qubit Pauli operators} \\
\midrule
\endhead
\midrule
\multicolumn{4}{r}{\emph{Continued on the next page}}\\
\endfoot
\bottomrule
\endlastfoot
0 & 0 & 1 & \(\tau_{A,4}^{z}\) \\
1 & 0 & 5 & \(\tau_{A,2}^{x}\,\allowbreak \tau_{A,3}^{x}\,\allowbreak \tau_{A,4}^{x}\,\allowbreak \tau_{A,7}^{z}\,\allowbreak \tau_{A,8}^{x}\) \\
2 & 0 & 2 & \(\tau_{A,1}^{x}\,\allowbreak \tau_{A,2}^{z}\) \\
3 & 0 & 1 & \(\tau_{A,1}^{z}\) \\
4 & 0 & 1 & \(\tau_{A,1}^{x}\) \\
5 & 0 & 6 & \(\tau_{A,1}^{z}\,\allowbreak \tau_{A,2}^{y}\,\allowbreak \tau_{A,3}^{x}\,\allowbreak \tau_{A,4}^{y}\,\allowbreak \tau_{A,7}^{z}\,\allowbreak \tau_{A,8}^{x}\) \\
0 & 1 & 2 & \(\tau_{A,4}^{x}\,\allowbreak \tau_{A,5}^{z}\) \\
1 & 1 & 5 & \(\tau_{A,3}^{x}\,\allowbreak \tau_{A,4}^{x}\,\allowbreak \tau_{A,5}^{x}\,\allowbreak \tau_{A,8}^{z}\,\allowbreak \tau_{A,9}^{x}\) \\
2 & 1 & 2 & \(\tau_{A,2}^{x}\,\allowbreak \tau_{A,3}^{z}\) \\
3 & 1 & 3 & \(\tau_{A,1}^{x}\,\allowbreak \tau_{A,2}^{x}\,\allowbreak \tau_{A,3}^{x}\) \\
4 & 1 & 2 & \(\tau_{A,1}^{y}\,\allowbreak \tau_{A,2}^{x}\) \\
5 & 1 & 6 & \(\tau_{A,1}^{z}\,\allowbreak \tau_{A,2}^{x}\,\allowbreak \tau_{A,3}^{z}\,\allowbreak \tau_{A,5}^{y}\,\allowbreak \tau_{A,8}^{z}\,\allowbreak \tau_{A,9}^{x}\) \\
0 & 2 & 2 & \(\tau_{A,5}^{x}\,\allowbreak \tau_{A,6}^{z}\) \\
1 & 2 & 3 & \(\tau_{A,5}^{x}\,\allowbreak \tau_{A,6}^{x}\,\allowbreak \tau_{A,9}^{y}\) \\
2 & 2 & 3 & \(\tau_{A,3}^{x}\,\allowbreak \tau_{A,8}^{x}\,\allowbreak \tau_{A,9}^{x}\) \\
3 & 2 & 3 & \(\tau_{A,3}^{y}\,\allowbreak \tau_{A,7}^{x}\,\allowbreak \tau_{A,8}^{x}\) \\
4 & 2 & 3 & \(\tau_{A,2}^{z}\,\allowbreak \tau_{A,3}^{x}\,\allowbreak \tau_{A,7}^{x}\) \\
5 & 2 & 4 & \(\tau_{A,2}^{z}\,\allowbreak \tau_{A,3}^{y}\,\allowbreak \tau_{A,6}^{y}\,\allowbreak \tau_{A,9}^{z}\) \\
0 & 3 & 2 & \(\tau_{A,6}^{x}\,\allowbreak \tau_{A,10}^{x}\) \\
1 & 3 & 3 & \(\tau_{A,6}^{x}\,\allowbreak \tau_{A,9}^{x}\,\allowbreak \tau_{A,10}^{z}\) \\
2 & 3 & 2 & \(\tau_{A,8}^{x}\,\allowbreak \tau_{A,9}^{z}\) \\
3 & 3 & 2 & \(\tau_{A,7}^{x}\,\allowbreak \tau_{A,8}^{z}\) \\
4 & 3 & 1 & \(\tau_{A,7}^{z}\) \\
5 & 3 & 4 & \(\tau_{A,7}^{y}\,\allowbreak \tau_{A,8}^{y}\,\allowbreak \tau_{A,9}^{y}\,\allowbreak \tau_{A,10}^{y}\) \\
0 & 4 & 1 & \(\tau_{A,10}^{z}\) \\
1 & 4 & 1 & \(\tau_{A,10}^{y}\) \\
2 & 4 & 3 & \(\tau_{A,6}^{z}\,\allowbreak \tau_{A,9}^{x}\,\allowbreak \tau_{A,10}^{z}\) \\
3 & 4 & 4 & \(\tau_{A,5}^{z}\,\allowbreak \tau_{A,6}^{x}\,\allowbreak \tau_{A,8}^{x}\,\allowbreak \tau_{A,9}^{x}\) \\
4 & 4 & 4 & \(\tau_{A,4}^{z}\,\allowbreak \tau_{A,5}^{x}\,\allowbreak \tau_{A,7}^{x}\,\allowbreak \tau_{A,8}^{x}\) \\
5 & 4 & 5 & \(\tau_{A,4}^{z}\,\allowbreak \tau_{A,5}^{y}\,\allowbreak \tau_{A,6}^{y}\,\allowbreak \tau_{A,7}^{x}\,\allowbreak \tau_{A,10}^{y}\) \\
0 & 5 & 4 & \(\tau_{A,4}^{y}\,\allowbreak \tau_{A,5}^{y}\,\allowbreak \tau_{A,6}^{y}\,\allowbreak \tau_{A,10}^{y}\) \\
1 & 5 & 5 & \(\tau_{A,2}^{x}\,\allowbreak \tau_{A,7}^{z}\,\allowbreak \tau_{A,8}^{y}\,\allowbreak \tau_{A,9}^{y}\,\allowbreak \tau_{A,10}^{x}\) \\
2 & 5 & 6 & \(\tau_{A,1}^{x}\,\allowbreak \tau_{A,2}^{y}\,\allowbreak \tau_{A,3}^{y}\,\allowbreak \tau_{A,6}^{z}\,\allowbreak \tau_{A,9}^{z}\,\allowbreak \tau_{A,10}^{z}\) \\
3 & 5 & 7 & \(\tau_{A,1}^{y}\,\allowbreak \tau_{A,2}^{x}\,\allowbreak \tau_{A,3}^{z}\,\allowbreak \tau_{A,5}^{z}\,\allowbreak \tau_{A,6}^{x}\,\allowbreak \tau_{A,8}^{z}\,\allowbreak \tau_{A,9}^{x}\) \\
4 & 5 & 7 & \(\tau_{A,1}^{z}\,\allowbreak \tau_{A,2}^{y}\,\allowbreak \tau_{A,3}^{x}\,\allowbreak \tau_{A,4}^{z}\,\allowbreak \tau_{A,5}^{x}\,\allowbreak \tau_{A,7}^{z}\,\allowbreak \tau_{A,8}^{x}\) \\
5 & 5 & 1 & \(\tau_{A,4}^{x}\) \\
\end{longtable}

\FloatBarrier

\subsection{Numerical implementation}
\label{subsec:appendix-ed-algorithm}

The numerical implementation has two stages.  First, each $n$-qubit Pauli
string is encoded by two binary strings $(x\,|\,z)\in\mathbb F_2^{2n}$ that
record its $X$ and $Z$ content.  Pauli multiplication then becomes binary
addition, while commutation is determined by the symplectic pairing in
Eq.~\eqref{eq:symp}.  For each lattice geometry, we use this representation
to construct the factor qubits and encode the Hamiltonian terms once.  The
resulting factor Hamiltonians are then diagonalized independently in each
center sector.

Fixing the $c$ common-center eigenvalues leaves a symmetry block of dimension
\begin{equation}
D_{\rm center}=2^{n-c}=2^{2q+1},
\end{equation}
where
\begin{equation}
2q=n-L_x-L_y-n_D+2.
\label{eq:factor-active-qubits}
\end{equation}
The tensor decomposition reduces this block to two independent $q$-qubit
problems,
\begin{equation}
H_\lambda=
\mathds{1}_P\otimes\left[
H_A(\lambda)\otimes\mathds{1}_B+
\mathds{1}_A\otimes H_B(\lambda)
\right].
\label{eq:ED-factor-H}
\end{equation}
After fixing the charge corresponding to the logical qubit $P$, the active degrees of freedom correspond to $A\otimes B$ which has a dimension $2^{2q}$.  Dynamical splitting avoids constructing this matrix:                                                                                                                            
the two factor Hamiltonians are diagonalized separately, and each matrix has                                                                                                                            
dimension $2^q$. 
Thus the largest matrix that must be diagonalized has dimension $2^q$, and
the full $2^{2q}$ tensor-product matrix is never constructed.  At fixed
aspect ratio, $q=n/2-O(\sqrt n)$, so the extensive part of the active-qubit
count is reduced by a factor of two.  For $L \times L$ tori,
$q=(L-1)(L-2)/2$; for example, the reductions for $L=4,6,8$ are
$2^7\to2^3$, $2^{21}\to2^{10}$, and $2^{43}\to2^{21}$, respectively.

For each triangle family $F=A,B$, binary row reduction first removes
redundant generators, while the null space of their commutation matrix gives
the common center.  After fixing its eigenvalues, symplectic Gram--Schmidt
produces $q$ factor-Pauli pairs
$(\tau_{F,i}^{x},\tau_{F,i}^{z})$.  We use
\begin{equation}
\tau_{F,i}^{y}=i\tau_{F,i}^{x}\tau_{F,i}^{z}.
\label{eq:factor-Pauli-y-definition}
\end{equation}
Every physical triangle term is then represented by a single Pauli string on
the factor qubits, multiplied by a sector-dependent sign:
\begin{equation}
T_r^F=\eta_r^F P_F(x_r^F,z_r^F)
\prod_{j=1}^{c}G_j^{m_{rj}^F},
\qquad \eta_r^F=\pm1,
\label{eq:encoded-term-short}
\end{equation}
where
$P_F(x,z)=i^{x\cdot z}\prod_{i=1}^q
(\tau_{F,i}^{x})^{x_i}(\tau_{F,i}^{z})^{z_i}$ is a Pauli string acting on the
$q$ factor qubits.  In a center sector
$\lambda=(\lambda_1,\ldots,\lambda_c)$, the central factor is replaced by
its eigenvalue, giving
\begin{equation}
H_F(\lambda)=
\sum_r h_r^F\eta_r^F
\left(\prod_j\lambda_j^{m_{rj}^F}\right)
P_F(x_r^F,z_r^F).
\label{eq:encoded-factor-H-short}
\end{equation}
This preprocessing requires only binary linear algebra and is reused when the
couplings are changed (assuming the coupling preserves the dynamical split).

For small $q$ we diagonalize $H_F(\lambda)$ densely; for larger $q$ we use a
matrix-free Hermitian Krylov method.  With $N_t$ encoded terms, a matrix-vector
multiplication costs $O(N_t2^q)$ operations and each Krylov vector requires
$O(2^q)$ memory.  If
\begin{align}
H_A(\lambda)|a_i;\lambda\rangle
&=e_i^A(\lambda)|a_i;\lambda\rangle,
\nonumber\\
H_B(\lambda)|b_j;\lambda\rangle
&=e_j^B(\lambda)|b_j;\lambda\rangle,
\end{align}
then the complete sector spectrum is obtained without further diagonalization:
\begin{equation}
E_{ij}(\lambda)=e_i^A(\lambda)+e_j^B(\lambda),
\label{eq:numerical-energy-sum}
\end{equation}
with an exact twofold multiplicity from the logical qubit $P$.  The global
ground-state energy is therefore
\begin{equation}
E_0=\min_\lambda
\left[e_0^A(\lambda)+e_0^B(\lambda)\right].
\label{eq:global-ground-sector-scan}
\end{equation}
Translations and point-group symmetries may further reduce the number of
center sectors that must be examined, but this is separate from the tensor
factorization itself.

For energies and factor-space observables, the calculation ends at this
stage. When a physical wavefunction or a real-space observable is required, we
additionally choose an eigenstate of one logical charge, such as $C_0$,
and map the resulting $P\otimes A\otimes B$ state back to the physical spin
basis using the Clifford encoding determined during preprocessing.  This inverse map is essential
for real-space entanglement, but constructing the full $2^n$-component
wavefunction is unnecessary for the spectral calculations reported here.

\section{Split-preserving deformation: analytical and numerical details}
\label{app:split-preserving-current}

This appendix provides the details of the split-preserving deformation in
Sec.~\ref{subsec:current-deformation-ed}.

As discussed in the main text, we first define the following Hermitian operators
\begin{align}
J_A&=\mathrm{i}\sum_{\bm r}\left(
a_{\bm r}a_{{\bm r}+\hat x}
+a_{\bm r}a_{{\bm r}+\hat y}
+a_{\bm r}a_{{\bm r}-\hat x-\hat y}\right),
\nonumber\\
J_B&=\mathrm{i}\sum_{\bm r}\left(
b_{\bm r}b_{{\bm r}+\hat x}
+b_{\bm r}b_{{\bm r}+\hat y}
+b_{\bm r}b_{{\bm r}-\hat x-\hat y}\right).
\label{eq:model-current}
\end{align}

To maintain the symmetries of the triangular lattice, we combine the above
two operators as
\begin{equation}
H_p=J_A-J_B.
\label{eq:model-Hp}
\end{equation}
The twofold rotation $C_2$ exchanges the triangle families and reverses the
orientation of the directed bonds:
\begin{equation}
C_2J_AC_2^{-1}=-J_B,
\qquad
C_2J_BC_2^{-1}=-J_A.
\label{eq:model-current-C2}
\end{equation}
Consequently, $H_p$ is invariant under $C_2$.

We study the one-parameter family
\begin{equation}
H_s(x)=xH_{\rm XYZ}+s(1-x)H_p,
\qquad 0\leq x\leq1,\qquad s=\pm1.
\label{eq:model-deformation}
\end{equation}
The XYZ model is recovered at $x=1$; below we focus mainly on the branch
$s=-1$.   The checkerboard antiunitary symmetry, Eq. \eqref{eq:model-checkerboard-antiunitary},  sends
$H_s(x)\mapsto H_{-s}(x)$, so it relates the two branches (which is why it suffices to study just one of them). Since $J_A$ and $J_B$ belong entirely to the $A$ and $B$ triangle
algebras, respectively, the deformation preserves all unitary subsystem
symmetries and the tensor-product structure in
Eqs.~\eqref{eq:model-factorization} and \eqref{eq:model-factor-H}.  In
particular, the exact logical doublet survives for every $x$ and either sign
$s$.  The deformation can
therefore potentially change the phase diagram without sacrificing either the 
subsystem or the computational reduction associated with dynamical splitting.

\subsection{Large-$N$ treatment}
\label{subsec:large-N-deformation}

At $N>2$, the clock-triangle operators $a_{\bm r}$ and $b_{\bm r}$ of
Eq.~\eqref{eq:pure-largeN-ab} are unitary rather than Hermitian.  We must
therefore specify how the Pauli products in $J_A$ and $J_B$ are continued
away from $N=2$.  We use the Hermitian continuation
\begin{align}
J_A&=\frac{\mathrm{i}}{2}\sum_{\bm r}\sum_{{\bm\delta}\in\mathcal D_\triangle}
\left(a_{\bm r}^{\dagger}a_{{\bm r}+{\bm\delta}}
-a_{{\bm r}+{\bm\delta}}^{\dagger}a_{\bm r}\right),
\nonumber\\
J_B&=\frac{\mathrm{i}}{2}\sum_{\bm r}\sum_{{\bm\delta}\in\mathcal D_\triangle}
\left(b_{\bm r}^{\dagger}b_{{\bm r}+{\bm\delta}}
-b_{{\bm r}+{\bm\delta}}^{\dagger}b_{\bm r}\right),
\label{eq:largeN-clock-deformation}
\end{align}
where $\mathcal D_\triangle=\{\hat x,\hat y,-\hat x-\hat y\}$.  At
$N=2$, $a_{\bm r}=a_{\bm r}^{\dagger}$ and neighboring $a$ triangles
anticommute, and similarly for $b$.  Each term in
Eq.~\eqref{eq:largeN-clock-deformation} then reduces to
$\mathrm{i}a_{\bm r}a_{{\bm r}+{\bm\delta}}$ or
$\mathrm{i}b_{\bm r}b_{{\bm r}+{\bm\delta}}$, as required.  The clock
Hamiltonian studied below is therefore
\begin{equation}
H_s(x)=xH_N+s(1-x)(J_A-J_B),
\qquad 0\leq x\leq1,\qquad s=\pm1,
\label{eq:largeN-clock-H}
\end{equation}
with $H_N$ defined in Eq.~\eqref{eq:pure-largeN-H}.

We now derive the leading large-$N$ energy functional rather than simply
postulating it.  As in Sec.~\ref{subsec:large-N-solution}, write
\begin{equation}
X_{\bm r}=e^{\mathrm{i}A_x({\bm r})},\qquad
Z_{\bm r}=e^{\mathrm{i}A_y({\bm r})},\qquad
[A_x({\bm r}),A_y({\bm r}')]=\frac{2\pi\mathrm{i}}{N}\delta_{{\bm r},{\bm r}'}.
\label{eq:largeN-deformation-phase-space}
\end{equation}
Thus $2\pi/N$ plays the role of an effective Planck constant, and the two
phase fields may be treated as commuting variables at leading order in
$1/N$.  The clock triangles have the corresponding phase-space representation
\begin{equation}
a_{\bm r}\longrightarrow e^{\mathrm{i}\theta_A({\bm r})},
\qquad b_{\bm r}\longrightarrow e^{\mathrm{i}\theta_B({\bm r})},
\label{eq:largeN-deformation-triangle-symbols}
\end{equation}
where
\begin{align}
\theta_A({\bm r})&=A_x({\bm r})-A_x({\bm r}+\hat y)-A_y({\bm r}+\hat y)+A_y({\bm r}+\hat x+\hat y),
\nonumber\\
\theta_B({\bm r})&=A_y({\bm r})-A_y({\bm r}+\hat x)-A_x({\bm r}+\hat x)+A_x({\bm r}+\hat x+\hat y).
\label{eq:largeN-deformation-theta}
\end{align}
In particular,
\begin{equation}
\frac{\mathrm{i}}{2}\left(a_{\bm r}^{\dagger}a_{{\bm r}+{\bm\delta}}
-a_{{\bm r}+{\bm\delta}}^{\dagger}a_{\bm r}\right)
\longrightarrow
\sin\!\left[\theta_A({\bm r})-\theta_A({\bm r}+{\bm\delta})\right],
\label{eq:largeN-deformation-bond-symbol}
\end{equation}
and the same relation holds with $A$ replaced by $B$.  Substituting these
expressions into Eq.~\eqref{eq:largeN-clock-H} gives the leading semiclassical
energy functional
\begin{equation}
\mathcal E_s=-x\sum_{F=A,B}\sum_{\bm r}\cos\theta_F({\bm r})
+\sum_{F=A,B}\sum_{\bm r}\sum_{{\bm\delta}\in\mathcal D_\triangle}\gamma_F
\sin\!\left[\theta_F({\bm r})-\theta_F({\bm r}+{\bm\delta})\right],
\qquad (\gamma_A,\gamma_B)=s(1-x)(1,-1).
\label{eq:largeN-classical-H}
\end{equation}
Here ``semiclassical energy functional'' means the ordinary function of
$A_x$ and $A_y$ obtained from the quantum Hamiltonian at leading order in
$1/N$.  Corrections from the
noncommutativity in Eq.~\eqref{eq:largeN-deformation-phase-space} and from
operator-ordering phases are higher order in $1/N$.

Changing $\gamma_F\to-\gamma_F$ together with
$\theta_F\to-\theta_F$ leaves the factor energy unchanged.  It is therefore
enough to minimize one factor with $\gamma_F=1-x$.  Its energy is
\begin{equation}
\mathcal E_F[\theta]=-x\sum_{\bm r}\cos\theta_{\bm r}
+(1-x)\sum_{\bm r}\sum_{{\bm\delta}\in\mathcal D_\triangle}
\sin(\theta_{\bm r}-\theta_{{\bm r}+{\bm\delta}}).
\label{eq:largeN-one-factor-energy}
\end{equation}
The oriented bonds form a triangular lattice.  Since every site belongs to
six elementary triangles and every directed bond belongs to two, the energy
can be written as a sum of triangle energies
\begin{equation}
h_\triangle=-\frac{x}{6}\sum_{j=0}^2\cos\theta_j
+\frac{1-x}{2}\sum_{j=0}^2\sin(\theta_j-\theta_{j+1}),
\qquad \theta_3\equiv\theta_0.
\label{eq:largeN-triangle-energy}
\end{equation}
For three phases, define
\begin{equation}
m=\left|e^{\mathrm{i}\theta_0}+e^{\mathrm{i}\theta_1}
+e^{\mathrm{i}\theta_2}\right|,
\qquad
\mathcal C_\triangle=\sum_{j=0}^2\sin(\theta_j-\theta_{j+1}).
\label{eq:largeN-triangle-invariants}
\end{equation}
A common shift of the three phases can always make their sum real and
positive, so that $\sum_j\cos\theta_j=m$.  At fixed $m$, the oriented sum
obeys the bound
\begin{equation}
\mathcal C_\triangle^2\leq\frac14(3-m)^3(1+m).
\label{eq:largeN-triangle-bound}
\end{equation}
The negative extremum, which minimizes Eq.~\eqref{eq:largeN-triangle-energy},
is attained by $(\theta_0,\theta_1,\theta_2)=(0,\alpha,-\alpha)$, up to a
common shift and cyclic permutation.  Since the lattice is three-colorable
by $n({\bm r})=r_x+r_y\pmod 3$, and every displacement in
$\mathcal D_\triangle$ sends $n\to n+1$, this local bound can be saturated
on every triangle simultaneously.  The exact minimum of this semiclassical
one-factor energy density per site is consequently
\begin{equation}
e_{\rm fac}(x)=\min_{0\leq m\leq3}\left[-\frac{x}{3}m
-\frac{1-x}{2}\sqrt{(3-m)^3(1+m)}\right].
\label{eq:largeN-energy-density}
\end{equation}

For the three-sublattice branch, $m=1+2\cos\alpha$.  Writing
$c=\cos\alpha$, stationarity of Eq.~\eqref{eq:largeN-energy-density} gives
\begin{equation}
\frac{x}{3(1-x)}=(1+2c)\sqrt{\frac{1-c}{1+c}},
\qquad -\frac12\leq c\leq0.
\label{eq:largeN-stationary}
\end{equation}
The uniform solution has $m=3$ and energy density $-x$.  It crosses the
three-sublattice branch at $c=0$, where the latter has $m=1$.  The order
parameter therefore jumps, establishing a first-order transition at
\begin{equation}
x_c=\frac34,
\qquad {\bm Q}=\left(\frac{2\pi}{3},\frac{2\pi}{3}\right).
\label{eq:largeN-transition}
\end{equation}
For $x<x_c$, the two factors have opposite orientations,
\begin{equation}
(\theta_{F,0}^{(0)},\theta_{F,1}^{(0)},\theta_{F,2}^{(0)})
=\nu_F(0,\alpha,-\alpha),
\qquad \nu_A=s,\qquad \nu_B=-s.
\label{eq:largeN-period-three-saddle}
\end{equation}
The pattern is unfrustrated on a finite torus only when both circumferences
are divisible by three, anticipating the strong commensuration effects seen
in ED.

Although $\theta_A$ and $\theta_B$ arise from the same two phase-space
fields, the two minimizing patterns above are compatible.  In momentum
space,
\begin{equation}
\begin{pmatrix}\theta_A({\bm k})\\ \theta_B({\bm k})\end{pmatrix}
=U({\bm k})\begin{pmatrix}A_x({\bm k})\\ A_y({\bm k})\end{pmatrix},
\qquad
U({\bm k})=\begin{pmatrix}1-w&w(z-1)\\-z(1-w)&1-z\end{pmatrix},
\label{eq:largeN-theta-map}
\end{equation}
where $z=e^{\mathrm{i}k_x}$ and $w=e^{\mathrm{i}k_y}$.  Since
$\det U=(1-z)(1-w)(1-zw)$ is nonzero at ${\bm k}=\pm{\bm Q}$, the two
oppositely oriented period-three patterns can be realized simultaneously by
$A_x$ and $A_y$.

We next determine the harmonic spectrum.  In the uniform phase, the sine
term in Eq.~\eqref{eq:largeN-classical-H} has a telescoping linear term and
no quadratic term.  Therefore the Gaussian approximation for the Hamiltonian in the uniform phase is
\begin{equation}
\mathcal E_{\rm u}^{(2)}=\frac{x}{2}\sum_{\bm r}
\left[\theta_A({\bm r})^2+\theta_B({\bm r})^2\right],
\label{eq:largeN-uniform-quadratic}
\end{equation}
where the subscript $u$ denotes `uniform'. Commutation relations in Eq.~\eqref{eq:largeN-deformation-phase-space} imply
\begin{equation}
[\theta_A({\bm k}),\theta_A(-{\bm k})]=-\frac{16\pi}{N}f({\bm k}),
\qquad
[\theta_B({\bm k}),\theta_B(-{\bm k})]=+\frac{16\pi}{N}f({\bm k}),
\label{eq:largeN-deformation-commutators}
\end{equation}
where $f({\bm k})=\sin\frac{k_x}{2}\sin\frac{k_y}{2}
\sin\frac{k_x+k_y}{2}$. This leads to the dispersion in the uniform phase:
\begin{equation}
\Omega_{\rm u}({\bm k};x)=\frac{16\pi x}{N}|f({\bm k})|.
\label{eq:largeN-uniform-deformed-dispersion}
\end{equation}
The uniform phase therefore retains all three nodal lines that existed in the pure chiral XYZ model.

For $0<x<x_c$, as discussed above, the ground state is non-uniform. To isolate the low-energy fluctuations, we write
$\theta_F({\bm r})=\theta_{F,n({\bm r})}^{(0)}+\varphi_F({\bm r})$.
After reversing the sign of $\varphi_F$ in the factor with the opposite
orientation, both factors have the same quadratic energy.  Suppressing the
factor label, it is
\begin{equation}
\mathcal E_3^{(2)}=\frac12\sum_{\bm r}\mu_{n({\bm r})}\varphi_{\bm r}^2
+\frac12\sum_{\bm r}\sum_{{\bm\delta}\in\mathcal D_\triangle}
\kappa_{n({\bm r})}
\left(\varphi_{\bm r}-\varphi_{{\bm r}+{\bm\delta}}\right)^2,
\label{eq:largeN-ordered-quadratic}
\end{equation}
where
\begin{equation}
(\mu_0,\mu_1,\mu_2)=x(1,c,c),
\qquad
(\kappa_0,\kappa_1,\kappa_2)=(1-x)\sin\alpha\,(1,-2c,1).
\label{eq:largeN-ordered-coefficients}
\end{equation}

Let ${\bm q}_a={\bm k}+a{\bm Q}$, $a=0,1,2$, with ${\bm k}$ in the
reduced Brillouin zone.  For either three-periodic coefficient
$v_n=\mu_n,\kappa_n$, define
\begin{equation}
\widehat v_{\ell}=\frac13\sum_{n=0}^2v_n e^{-2\pi\mathrm{i}\ell n/3}.
\label{eq:largeN-period-three-transform}
\end{equation}
In the folded basis $({\bm q}_0,{\bm q}_1,{\bm q}_2)$, the Hessian of
Eq.~\eqref{eq:largeN-ordered-quadratic} is the Hermitian matrix
\begin{equation}
M_{ab}({\bm k})=\widehat\mu_{a-b}
+\widehat\kappa_{a-b}\sum_{{\bm\delta}\in\mathcal D_\triangle}
\left(1-e^{-\mathrm{i}{\bm q}_a\cdot{\bm\delta}}\right)
\left(1-e^{+\mathrm{i}{\bm q}_b\cdot{\bm\delta}}\right),
\qquad a,b=0,1,2,
\label{eq:largeN-folded-Hessian}
\end{equation}
where the indices are understood modulo three.  Apart from the overall
factor $\pm16\pi/N$, whose sign is opposite for the two factors, the
commutator matrix in the same folded basis is
\begin{equation}
D_f({\bm k})=\operatorname{diag}
\left[f({\bm q}_0),f({\bm q}_1),f({\bm q}_2)\right].
\label{eq:largeN-folded-commutator}
\end{equation}
The linearized Heisenberg equation therefore has dynamical matrix
$(16\pi/N)D_fM$, and the three excitation frequencies are
\begin{equation}
\Omega_n^{(3)}({\bm k};x)=\frac{16\pi}{N}
\left|\lambda_n\!\left[D_f({\bm k})M({\bm k})\right]\right|,
\qquad n=1,2,3.
\label{eq:largeN-ordered-frequencies}
\end{equation}
Here $n=1,2,3$ labels the three folded bands. On the support of $M$, $D_fM$ is similar to the Hermitian matrix
$M^{1/2}D_fM^{1/2}$, so these eigenvalues are real.  Moreover,
\begin{equation}
\det[D_fM]=f({\bm k})f({\bm k}+{\bm Q})f({\bm k}-{\bm Q})\det M.
\label{eq:largeN-folded-determinant}
\end{equation}
Thus at least one band vanishes whenever any displayed $f$ factor vanishes.
Near the origin of the reduced Brillouin zone, the lowest band has
\begin{equation}
\Omega_{\rm lowest}^{(3)}({\bm k};x)\propto
\frac1N|k_xk_y(k_x+k_y)|,
\qquad 0<x<x_c,
\label{eq:largeN-ordered-cubic}
\end{equation}
while the other two bands are generically nonzero there.

At the pure-deformation endpoint, $x=0$, one has $c=-1/2$, and all three
bond stiffnesses in Eq.~\eqref{eq:largeN-ordered-coefficients} become
$\sqrt3/2$.  The Hessian is then translation invariant.  Defining
\begin{equation}
d_\triangle({\bm k})=6-2\left[\cos k_x+\cos k_y+\cos(k_x+k_y)\right],
\end{equation}
one obtains
\begin{equation}
\Omega_{x=0}({\bm k})=\frac{8\sqrt3\pi}{N}
|f({\bm k})|d_\triangle({\bm k}).
\label{eq:largeN-quintic-lattice}
\end{equation}
Since $d_\triangle({\bm k})=O(k^2)$, the common nodal intersection softens
from cubic to quintic.  Therefore, at large $N$, we obtain a
first-order transition at $x=3/4$ between two gapless phases.

\subsection{Exact diagonalization}
\label{subsec:deformed-transition}

We now use exact diagonalization to test the two main predictions of the
large-$N$ analysis: the competition between uniform and nonuniform states,
and the persistence of a gapless uniform branch away from the chiral XYZ
point.

\paragraph{Reconstruction of the ground-state sector.}
We first exhaustively scan all center sectors on $L \times L$ tori with
$L=3,\ldots,7$.  The ground state remains in the uniform sector for
$L=3,4,5$, and $7$.  On the $6\times6$ torus, however, a nonuniform sector
has the lowest energy at small $x$ and crosses the uniform sector near
$x\simeq0.3$.  A representative of this nonuniform sector is
\begin{equation}
 \begin{gathered}
 c=r=(+,-,+,-,+,-),\\
 d=(+,+,-,+,+,-),\qquad Q_{\rm mix}=+1,
 \end{gathered}
 \label{eq:deformed-L6-texture-sector}
\end{equation}
which combines period-two and period-three line-charge patterns.

\begin{figure*}[t]
 \centering
 \includegraphics[width=0.82\textwidth]{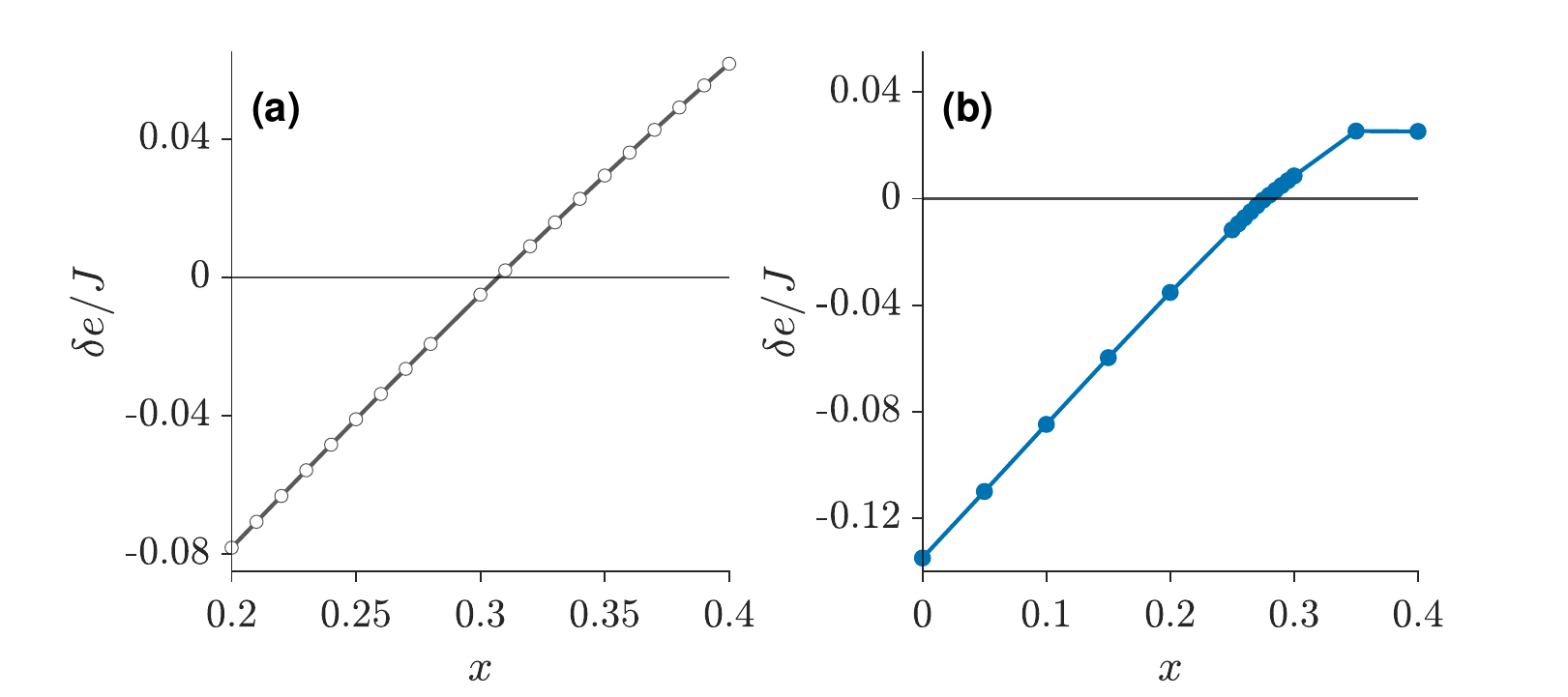}
 \caption{Numerical test for translational symmetry breaking in the ground state of the split-preserving deformation, Eq.~\eqref{eq:model-deformation}.
 Energy-density difference
 $\delta e=(E_{\rm nonuniform}-E_{\rm uniform})/(L_xL_y)$ on
 (a) $6\times6$ and (b) $6\times9$ tori.  Panel (a) follows the
 nonuniform sector in Eq.~\eqref{eq:deformed-L6-texture-sector}, while
 panel (b) shows the lowest nonuniform sector at each $x$.  Negative values
 mean that a nonuniform sector lies below the uniform sector.}
 \label{fig:deformed-sector-reconstruction}
\end{figure*}

Figure~\ref{fig:deformed-sector-reconstruction}(a) compares its energy
density directly with that of the uniform sector.  Panel (b) shows the
corresponding exhaustive comparison on a $6\times9$ torus, where we plot
the lowest nonuniform sector at each value of $x$.  The identity of this
sector changes briefly near the crossing, but this finite-size detail does
not affect the main result: the nonuniform sector is favored at small $x$
and the uniform sector at larger $x$ on both geometries.  The location of
the crossing and the detailed line-charge pattern depend strongly on the
shape and size of the torus.  These data therefore support the competition
between uniform and nonuniform states found at large $N$, but do not by
themselves establish the precise pattern of the translation-breaking in the thermodynamic limit.

\begin{figure*}[t]
 \centering
 \includegraphics[width=\textwidth]{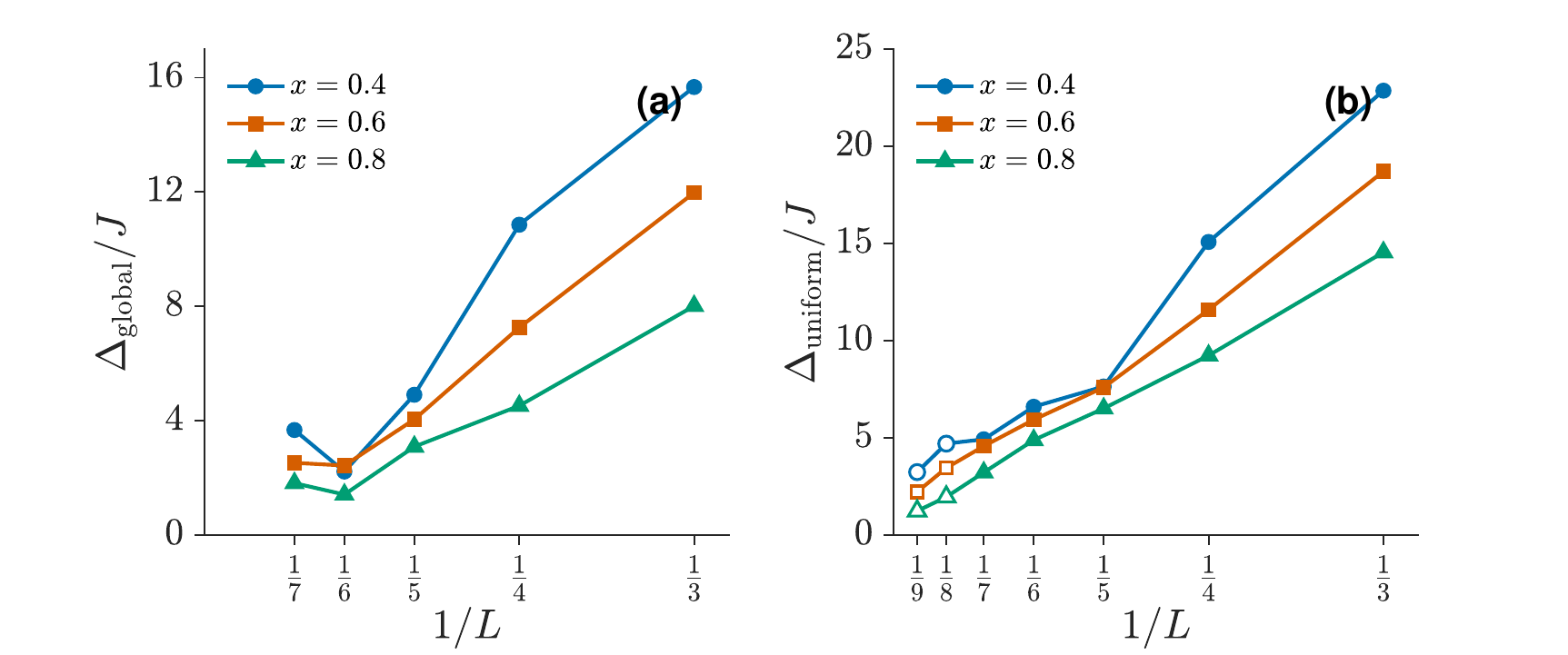}
 \caption{Raw finite-size gaps at \(x=0.4,0.6\), and \(0.8\).
 (a) The global gap \(\Delta_{\rm global}\), obtained by minimizing over
 all center sectors for \(L=3,\ldots,7\).  (b) The first distinct excitation above the logical doublet
 \(\Delta_{\rm uniform}\) in the fixed uniform center sector through
 \(L=9\).  Open symbols denote the \(L=8,9\) calculations, which are
 certified within the specified uniform sector rather than by an
 exhaustive scan.  Lines are guides to the eye.}
 \label{fig:deformed-gap-scaling}
\end{figure*}

\paragraph{Gapless uniform branch.}
We next focus on $x=0.4,0.6$, and $0.8$, where the uniform sector is the
global winner on every exhaustively scanned size through $L=7$.
Figure~\ref{fig:deformed-gap-scaling}(a) shows the global gap
$\Delta_{\rm global}$, defined as the first distinct energy above the
complete ground-state manifold after minimizing over all center sectors.
Panel (b) shows the first distinct excitation above the logical doublet $\Delta_{\rm uniform}$ after fixing
all center charges to their uniform values.  Both gaps decrease strongly
with system size at all three couplings.  The available sizes are not
sufficient for a controlled extrapolation, but give no indication that the
uniform branch develops a conventional gap.

\begin{figure*}[t]
 \centering
 \includegraphics[width=\textwidth]{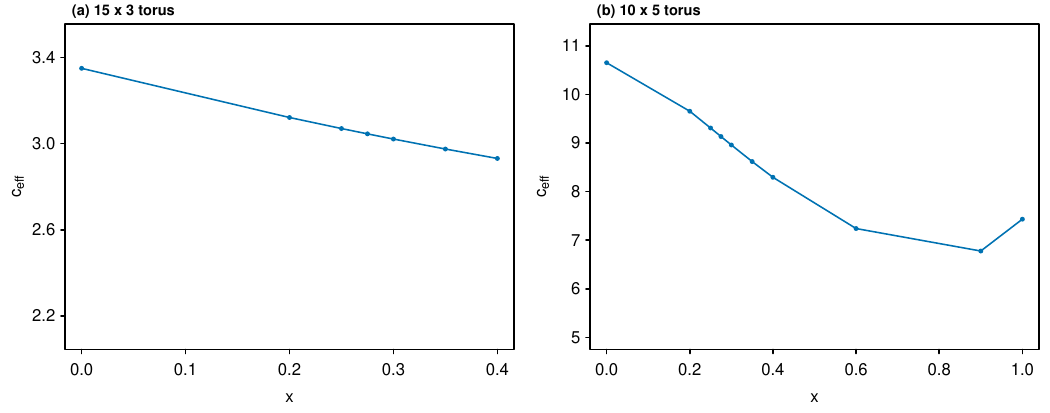}
 \caption{Effective von Neumann coefficient \(c_{\rm eff}=3b_1\) obtained
 from the chord fits on (a) \(15\times3\) and (b) \(10\times5\) tori.
 Every state is a definite representative of an exhaustively identified
 global ground sector.  Lines connect the sampled couplings and are guides
 to the eye.}
 \label{fig:deformed-chord-ceff}
\end{figure*}

The ground-state entanglement gives an independent indication of gaplessness. In particular, half-torus entropies retain a positive $L\log L$
contribution throughout the interpolation.  We also move the cut within
$15\times3$ and $10\times5$ tori and fit the chord form in
Eq.~\eqref{eq:pure-line-node-entropy-scaling}.  Denoting the fitted slope by
$b_1$, Fig.~\ref{fig:deformed-chord-ceff} shows the corresponding coefficient
$c_{\rm eff}=3b_1$, which remains positive throughout the sampled range. Overall, the positive coefficient shows that the logarithmic chord dependence
survives away from the pure XYZ point.

Together, the gap and entanglement results support an extended gapless
uniform branch.  The long rectangles remain in this branch even at the
smallest sampled $x$, whereas the $6\times6$ and $6\times9$ tori favor
nonuniform sectors there.  Recall that the large-$N$ analysis discussed above predicts that the nonuniform phase remains gapless; although the present finite-size ED data cannot test this directly at $N=2$, we suspect that the nonuniform phase remains gapless.
\section{Large-$N$ correlation functions and RG analysis for instability against explicit subsystem-symmetry breaking}
\label{app:general-stability}

In this appendix we study the stability of the nodal liquid for the large-$N$ version of the pure chiral XYZ model against local perturbations that explicitly break its subsystem symmetries.  The leading perturbations take the form:
\begin{equation}
\delta H=-\sum_{\bm r}\big[
h_X\big(X_{\bm r}+X_{\bm r}^{\dagger}\big)
+t_D\big(D^X_{\bm r}+(D^X_{\bm r})^{\dagger}\big)\big],
\label{eq:ssbrg-perturbations}
\end{equation}
where $D^X_{\bm r}=X_{{\bm r}+\hat y}X_{\bm r}^{\dagger}$.  The other
onsite clock fields, $Z+Z^{\dagger}$ and
$\mathcal Y+\mathcal Y^{\dagger}$, are related to
$X+X^{\dagger}$ by triangular-lattice symmetry and need not be considered
separately. 

Our goal is to find the RG $\beta$-functions for $h_X$ and $t_D$. As usual, one of the first steps is to find the correlation functions $\left\langle T_\tau\mathcal O_{\bm r}(\tau)
\mathcal O_{\bm r'}^{\dagger}(0)\right\rangle$ where $\mathcal O_{\bm r} = X_r, D^X_r$ at the unperturbed fixed point (i.e. pure Chiral XYZ). A conceptual difference between onsite operators such as $\big(X_{\bm r}+X_{\bm r}^{\dagger}\big)$ and the dipole type operators $\big(D^X_{\bm r}+(D^X_{\bm r})^{\dagger}\big)$ is that the subsystem symmetry forces the two-point  unequal-space correlations of the former type objects to vanish identically while for the latter, such correlations will be non-zero along certain directions in space. Let's proceed to calculate such correlators.

\subsection{Two-point correlations in the pure Chiral XYZ model}
\label{subsec:ssbrg-correlators}

As in the main text, at large $N$, we write
\begin{equation}
X_{\bm r}=e^{\mathrm{i}A_x({\bm r})},\qquad
Z_{\bm r}=e^{\mathrm{i}A_y({\bm r})},\qquad
[A_x({\bm r}),A_y({\bm r}')]=\frac{2\pi\mathrm{i}}{N}
\delta_{{\bm r},{\bm r}'}.
\label{eq:ssbrg-canonical}
\end{equation}
Within Gaussian approximation, which is justified at large-$N$, the Hamiltonian of the pure chiral XYZ model is
\begin{equation}
H_0=\frac12\sum_{\bm r}
\left[\theta_A({\bm r})^2+\theta_B({\bm r})^2\right],
\label{eq:ssbrg-pure-H}
\end{equation}
where $\theta_A$ and $\theta_B$ are given in
Eq.~\eqref{eq:pure-largeN-ab-symbols}.  

The corresponding imaginary-time action is

\begin{equation}
S_E=\int d\tau\left\{
H_0-\frac{\mathrm{i}}{2\hbar_N}\sum_{\bm r}
\left[A_y({\bm r})\partial_\tau A_x({\bm r})
-A_x({\bm r})\partial_\tau A_y({\bm r})\right]\right\},
\label{eq:ssbrg-realspace-action}
\end{equation}

which, after Fourier transformation, becomes
\begin{align}
S_E&=\frac12\int_{\omega,{\bm k}}
{\bm A}(-\omega,-{\bm k})^T
\mathcal D(\omega,{\bm k})
{\bm A}(\omega,{\bm k}),
\nonumber\\
\mathcal D(\omega,{\bm k})
&=K({\bm k})+\frac{\omega}{\hbar_N}
\begin{pmatrix}0&1\\-1&0\end{pmatrix},
\label{eq:ssbrg-Euclidean-action}
\end{align}
where 

\begin{equation}
K({\bm k})=8
\begin{pmatrix}
\sin^2(k_y/2)&
-\sin(k_x/2)\sin(k_y/2)\cos[(k_x+k_y)/2]\\
-\sin(k_x/2)\sin(k_y/2)\cos[(k_x+k_y)/2]&
\sin^2(k_x/2)
\end{pmatrix},
\label{eq:ssbrg-K-matrix}
\end{equation}

$\hbar_N=2\pi/N$, $\bm A=(A_x,A_y)^T$ and
$\int_{\omega,{\bm k}}=\int d\omega/(2\pi)
\int_{\rm BZ}d^2k/(2\pi)^2$. 

The determinant of $K$ is
\begin{equation}
\det K({\bm k})=64
\sin^2\frac{k_x}{2}\sin^2\frac{k_y}{2}
\sin^2\frac{k_x+k_y}{2}.
\label{eq:ssbrg-det-K}
\end{equation}

We define the imaginary-time-ordered two-point function by
\begin{equation}
G_{ij}(\tau,{\bm k})
\equiv\left\langle T_\tau A_i({\bm k},\tau)
A_j(-{\bm k},0)\right\rangle,
\qquad i,j\in\{x,y\}.
\label{eq:ssbrg-field-correlator-definition}
\end{equation}
The frequency-space correlator is $\mathcal D^{-1}(\omega,{\bm k})$.
Direct inversion gives
\begin{equation}
\mathcal D^{-1}(\omega,{\bm k})
=\frac{1}{\det K+(\omega/\hbar_N)^2}
\begin{pmatrix}
K_{yy}&-K_{xy}-\omega/\hbar_N\\
-K_{xy}+\omega/\hbar_N&K_{xx}
\end{pmatrix}.
\label{eq:ssbrg-inverse-kernel}
\end{equation}
For $\tau\neq0$, Fourier transforming from frequency to imaginary time gives
\begin{equation}
G_{ij}(\tau,{\bm k})
=\frac{\hbar_N}{2}e^{-\Omega({\bm k})|\tau|}
\left[
\frac{\operatorname{adj}K({\bm k})}{\sqrt{\det K({\bm k})}}
+\mathrm{i}\,\operatorname{sgn}(\tau)
\begin{pmatrix}0&1\\-1&0\end{pmatrix}
\right]_{ij}.
\label{eq:ssbrg-time-propagator}
\end{equation}
Here $\Omega({\bm k})=\hbar_N\sqrt{\det K({\bm k})} = \frac{16\pi}{N}
\left|\sin\frac{k_x}{2}\sin\frac{k_y}{2}
\sin\frac{k_x+k_y}{2}\right|$ is the energy of excitations above the ground state.
This is a somewhat complicated expression, so let us consider a simple example that will be relevant below. Consider $\left\langle T_\tau A_x(k_x,k_y,\tau)A_x(-k_x,-k_y,0)\right\rangle$ at small $k_y = q$ at generic non-zero $k_x = p$. At such momenta,
\begin{align}
\sqrt{\det K(p,q)}
&=4\sin^2\frac p2\,|q|+O(q^2),
\nonumber\\
\Omega(p,q)&=v(p)|q|+O(q^2),
\qquad
v(p)=\frac{8\pi}{N}\sin^2\frac p2.\nonumber\\
[\operatorname{adj}K]_{xx}&=K_{yy}=8\sin^2(p/2)
\label{eq:ssbrg-column-expansion} 
\end{align}
Therefore, Eq.~\eqref{eq:ssbrg-time-propagator} gives the most singular part of  $\left\langle T_\tau A_x(k_x,k_y,\tau)A_x(-k_x,-k_y,0)\right\rangle$ (which will determine its long-time correlations),
\begin{equation}
\left\langle T_\tau A_x(p,q,\tau)A_x(-p,-q,0)\right\rangle_{\rm sing}
=\frac{2\pi/N}{|q|}e^{-v(p)|q||\tau|}.
\label{eq:ssbrg-column-soft-propagator}
\end{equation}
This shows that the $A_x$ field is soft close to the $k_y = 0$ nodal line. Similarly, $A_y$ is the soft mode close to the $k_x = 0$ nodal line and $(A_x - A_y)/2$ is soft close to the $k_x + k_y = 0$ nodal line. 
With these normalizations, each soft mode has the singular propagator
shown in Eq.~\eqref{eq:ssbrg-column-soft-propagator}; only $p$, $q$, and
$v(p)$ change from one line to another.

\paragraph{Onsite $X$.} We first notice that unequal-space correlators $\langle X_{\bm r} X^{\dagger}_{\bm 0}\rangle$ vanish identically due to subsystem symmetry (assuming subsystem symmetry is unbroken), because one can find a subsystem symmetry under which $X_{\bm r} \to \omega X_r, X_{\bm 0} \to X_{\bm 0}$. Therefore, only unequal-time, equal-space correlator,  
\begin{equation}
G_X(\tau)=\left\langle T_\tau X_{\bm r}(\tau)
X_{\bm r}^{\dagger}(0)\right\rangle,
\label{eq:ssbrg-X-definition}
\end{equation}
is non-trivial. Since $X_{\bm r}=e^{\mathrm{i}A_x({\bm r})}$ and $A_x$ is Gaussian,
\begin{align}
-\log|G_X(\tau)|
&=\frac12\left\langle T_\tau
\left[A_x({\bm r},\tau)-A_x({\bm r},0)\right]^2\right\rangle
\nonumber\\
&=\int_{\rm BZ}\frac{d^2k}{(2\pi)^2}
\left[G_{xx}(0,{\bm k})-G_{xx}(\tau,{\bm k})\right].
\label{eq:ssbrg-X-variance}
\end{align}
The $k_y=0$ nodal line contributes through the $A_x$ propagator in
Eq.~\eqref{eq:ssbrg-column-soft-propagator}.  The diagonal nodal line gives
an identical singular contribution.  Indeed, writing
$k_x=p$ and $k_y=-p+q$, Eq.~\eqref{eq:ssbrg-time-propagator} gives
\begin{equation}
G_{xx}(\tau;p,-p+q)_{\rm sing}
=\frac{2\pi/N}{|q|}e^{-v(p)|q||\tau|},
\qquad
v(p)=\frac{8\pi}{N}\sin^2\frac p2.
\label{eq:ssbrg-diagonal-Ax-propagator}
\end{equation}
On the remaining nodal line, $k_x=0$, the soft field is $A_y$ and
$G_{xx}$ is nonsingular.  Keeping the singular neighborhoods of the two
contributing lines in Eq.~\eqref{eq:ssbrg-X-variance} therefore gives
\begin{align}
-\log|G_X(\tau)|
&=2\int_{-\pi}^{\pi}\frac{dp}{2\pi}
\int_{-\Lambda}^{\Lambda}\frac{dq}{2\pi}
\frac{2\pi/N}{|q|}
\left[1-e^{-v(p)|q||\tau|}\right]+O(1)
\nonumber\\
&=\frac{4}{N}\int_{-\pi}^{\pi}\frac{dp}{2\pi}
\int_0^\Lambda\frac{dq}{q}
\left[1-e^{-v(p)q|\tau|}\right]+O(1)
\nonumber\\
&=\frac{4}{N}\log|\tau|+O(1).
\label{eq:ssbrg-X-log}
\end{align}
It follows that
\begin{equation}
|G_X(\tau)|\sim|\tau|^{-4/N} \Rightarrow 
\Delta_X=\frac{2}{N}+O(N^{-2}).
\label{eq:ssbrg-X-correlation}
\end{equation}

\paragraph{Dipole $D^X$.}
We next consider the ``dipole operator'', defined as
\begin{equation}
D^X_{\bm r}=e^{\mathrm{i}Q_D({\bm r})},\qquad
Q_D({\bm r})=A_x({\bm r}+\hat y)-A_x({\bm r}).
\label{eq:ssbrg-DX-phase}
\end{equation}
Its momentum-space form factor is
$f_D({\bm k})=e^{\mathrm{i}k_y}-1$, i.e., $Q_D({\bm k})=f_D({\bm k})A_x({\bm k})$.  We first calculate its unequal-time, equal-space correlator,
\begin{equation}
G_{D^X}(0,\tau)=\left\langle T_\tau D^X_{\bm r}(\tau)
\big[D^X_{\bm r}(0)\big]^{\dagger}\right\rangle.
\label{eq:ssbrg-DX-time-definition}
\end{equation}
As before, the Gaussian average gives
\begin{equation}
-\log|G_{D^X}(0,\tau)|
=\int_{\rm BZ}\frac{d^2k}{(2\pi)^2}|f_D({\bm k})|^2
\left[G_{xx}(0,{\bm k})-G_{xx}(\tau,{\bm k})\right].
\label{eq:ssbrg-DX-time-variance}
\end{equation}
The form factor vanishes on $k_y=0$, while on $k_x=0$ the soft field is
$A_y$.  Thus only the neighborhood of the diagonal nodal line contributes
to the logarithm.  We parametrize this neighborhood by
$k_x=p$ and $k_y=-p+q$, where $q=k_x+k_y$ is the momentum perpendicular to
the nodal line.  The form factor is then
\begin{equation}
|f_D(p,-p+q)|^2
=\left|e^{-\mathrm{i}p+\mathrm{i}q}-1\right|^2
=4\sin^2\frac{p-q}{2}
=4\sin^2\frac p2+O(q).
\label{eq:ssbrg-DX-weight}
\end{equation}
Retaining the leading term in
Eq.~\eqref{eq:ssbrg-DX-weight} and substituting
Eq.~\eqref{eq:ssbrg-diagonal-Ax-propagator} into
Eq.~\eqref{eq:ssbrg-DX-time-variance} gives
\begin{align}
-\log|G_{D^X}(0,\tau)|
&=\int_{-\pi}^{\pi}\frac{dp}{2\pi}4\sin^2\frac p2
\int_{-\Lambda}^{\Lambda}\frac{dq}{2\pi}
\frac{2\pi/N}{|q|}
\left[1-e^{-v(p)|q||\tau|}\right]+O(1)
\nonumber\\
&=\frac{2}{N}\int_{-\pi}^{\pi}\frac{dp}{2\pi}4\sin^2\frac p2
\int_0^\Lambda\frac{dq}{q}
\left[1-e^{-v(p)q|\tau|}\right]+O(1)
\nonumber\\
&=\frac{4}{N}\log|\tau|+O(1),
\label{eq:ssbrg-DX-time-log}
\end{align}
where we used
$\int_{-\pi}^{\pi}dp\,4\sin^2(p/2)/(2\pi)=2$.  Hence
\begin{equation}
|G_{D^X}(0,\tau)|\sim|\tau|^{-4/N},\qquad
\Delta_{D^X}=\frac{2}{N}+O(N^{-2}).
\label{eq:ssbrg-DX-dimension}
\end{equation}

The subsystem charges also permit an unequal-space dipole correlator when
the two dipoles are separated along $(1,1)$.  At equal time, define
\begin{equation}
G_{D^X}(s,0)=
\left\langle D^X_{{\bm r}+s(1,1)}(0)
\big[D^X_{\bm r}(0)\big]^{\dagger}\right\rangle,
\label{eq:ssbrg-DX-spacetime}
\end{equation}
where $s$ is an integer.  Its Gaussian average is
\begin{align}
-\log|G_{D^X}(s,0)|
&=\int_{\rm BZ}\frac{d^2k}{(2\pi)^2}|f_D({\bm k})|^2
G_{xx}(0,{\bm k})
\left[1-\cos\big(s(k_x+k_y)\big)\right]
\nonumber\\
&=\int_{-\pi}^{\pi}\frac{dp}{2\pi}4\sin^2\frac p2
\int_{-\Lambda}^{\Lambda}\frac{dq}{2\pi}
\frac{2\pi/N}{|q|}\left[1-\cos(sq)\right]+O(1)
\nonumber\\
&=\frac{2}{N}\int_{-\pi}^{\pi}\frac{dp}{2\pi}4\sin^2\frac p2
\int_0^\Lambda\frac{dq}{q}\left[1-\cos(sq)\right]+O(1)
\nonumber\\
&=\frac{4}{N}\log|s|+O(1).
\label{eq:ssbrg-DX-space-log}
\end{align}
Therefore,
\begin{equation}
|G_{D^X}(s,0)|\sim|s|^{-4/N},\qquad
|G_{D^X}(0,\tau)|\sim|\tau|^{-4/N}.
\label{eq:ssbrg-DX-limits}
\end{equation}
At generic spatial separations the dipole correlation vanishes by the same
subsystem-charge selection rule.

\subsection{Renormalization Group Flow}
\label{subsec:ssbrg-direct-RG}

We now use the exponents found above to determine the RG flow of leading relevant perturbations.  An isotropic rescaling of both $k_x$ and $k_y$ would retain contribution only
from the common intersection of the three nodal lines and discard their generic
segments.  Instead, as in the RG for the exciton Bose
liquid~\cite{ParamekantiBalentsFisher2002,Lake2022EBLRG}, we treat each
generic segment as a $1+1$-dimensional mode.

The corresponding one-component action follows directly from the
two-component action in Eq.~\eqref{eq:ssbrg-Euclidean-action}.  Consider
again the line $k_y=0$, with $k_x=p$ and $k_y=q$.  At a generic point $p$
on this line, $A_x$ is the low-energy field: its quadratic coefficient
vanishes as $q,\omega\to0$, whereas the coefficient
$K_{yy}=8\sin^2(p/2)$ of $|A_y|^2$ remains nonzero.  Because the action is
quadratic, $A_y$ can be integrated out exactly.  The resulting action for
$A_x$ has the form
\begin{equation*}
S_{\rm eff}[A_x]=\frac12\int\frac{dp\,d\omega\,dq}{(2\pi)^3}
\mathcal D_{\rm eff}(p,q,\omega)|A_x(p,q,\omega)|^2,
\end{equation*}
where
\begin{align}
\mathcal D_{\rm eff}
&=K_{xx}
-\frac{(K_{xy}+\omega/\hbar_N)(K_{xy}-\omega/\hbar_N)}
{K_{yy}}
\nonumber\\
&=\frac{\det K+(\omega/\hbar_N)^2}{K_{yy}}.
\label{eq:ssbrg-effective-soft-kernel}
\end{align}
Using Eqs.~\eqref{eq:ssbrg-K-matrix} and \eqref{eq:ssbrg-det-K} and retaining
the leading terms in $q$ therefore gives
\begin{equation}
S_{\rm eff}[A_x]=\frac12\int\frac{dp\,d\omega\,dq}{(2\pi)^3}
\left[
\frac{\omega^2}{8\hbar_N^2\sin^2(p/2)}
+2\sin^2\frac p2\,q^2
\right]|A_x(p,q,\omega)|^2.
\label{eq:ssbrg-explicit-column-patch}
\end{equation}
The other two nodal lines give the same result after the triangular-lattice
rotation that permutes the three normalized low-energy fields described above.
Thus, near any generic segment, the
quadratic action may be written as
\begin{equation}
S_0=\frac12\int\frac{dp\,d\omega\,dq}{(2\pi)^3}
\left[Z(p)\omega^2+\rho(p)q^2\right]
|\phi_p(q,\omega)|^2.
\label{eq:ssbrg-patch-action}
\end{equation}
For the explicit segment above,
\begin{equation}
Z(p)=\frac{1}{8\hbar_N^2\sin^2(p/2)},\qquad
\rho(p)=2\sin^2\frac p2.
\label{eq:ssbrg-patch-coefficients}
\end{equation}
Consequently,
\begin{equation}
\sqrt{\frac{\rho(p)}{Z(p)}}=v(p),\qquad
\frac{1}{2\sqrt{Z(p)\rho(p)}}=\hbar_N=\frac{2\pi}{N}.
\label{eq:ssbrg-patch-consistency}
\end{equation}
As a sanity check, the propagator obtained from Eq.~\eqref{eq:ssbrg-patch-action} is therefore
precisely Eq.~\eqref{eq:ssbrg-column-soft-propagator}.  

In the RG below, $p$ is left unscaled while $\omega$ and $q$ are rescaled.
Thus, one is essentially doing RG for a collection of 1+1-D field theories labelled by distinct values of $p$.

We first define the RG transformation for the unperturbed action.  At fixed
$p$, separate $\phi_p$ into slow and fast modes, with the fast modes lying in
the shell
\begin{equation}
\frac{\Lambda}{b}<
\sqrt{\omega^2+v(p)^2q^2}<\Lambda,
\qquad b=e^{d\ell}>1.
\label{eq:ssbrg-fast-shell}
\end{equation}
After integrating out this shell, the remaining cutoff $\Lambda/b$ is
restored to $\Lambda$ by
\begin{equation}
\omega'=b\omega,\qquad q'=bq,\qquad p'=p.
\label{eq:ssbrg-rescaling}
\end{equation}
Since $d\omega\,dq=b^{-2}d\omega'\,dq'$ and
\begin{equation}
Z(p)\omega^2+\rho(p)q^2
=b^{-2}\left[Z(p')\omega'^2+\rho(p')q'^2\right],
\end{equation}
the slow part of Eq.~\eqref{eq:ssbrg-patch-action} becomes
\begin{align}
S_{0,<}
&=\frac12\int\frac{dp'\,d\omega'\,dq'}{(2\pi)^3}
b^{-4}\left[Z(p')\omega'^2+\rho(p')q'^2\right]
\left|\phi_{p',<}\left(\frac{q'}b,\frac{\omega'}b\right)\right|^2.
\end{align}
Defining
\begin{equation}
\phi'_{p'}(q',\omega')
=b^{-2}\phi_{p',<}\left(\frac{q'}b,\frac{\omega'}b\right)
\label{eq:ssbrg-field-rescaling}
\end{equation}
therefore restores the original action,
\begin{equation}
S_{0,<}=\frac12\int\frac{dp'\,d\omega'\,dq'}{(2\pi)^3}
\left[Z(p')\omega'^2+\rho(p')q'^2\right]
|\phi'_{p'}(q',\omega')|^2.
\label{eq:ssbrg-rescaled-patch-action}
\end{equation}
In real space, Eq.~\eqref{eq:ssbrg-field-rescaling} reads
\begin{equation}
\phi_{p,<}(x_\perp,\tau)
=\phi'_p(x_\perp',\tau'),
\qquad
x_\perp'=\frac{x_\perp}{b},\quad
\tau'=\frac{\tau}{b}.
\label{eq:ssbrg-realspace-field-rescaling}
\end{equation}
The coordinate $x_\parallel$ conjugate to $p$ is not rescaled.

We now apply this transformation to
\begin{equation}
\delta S_{\mathcal O}
=-2h_{\mathcal O}\int d\tau\sum_{\bm r}
\cos Q_{\mathcal O}({\bm r},\tau).
\label{eq:ssbrg-uniform-perturbation}
\end{equation}
Let $\lambda$ label a nodal line and write a nearby momentum as
${\bm k}={\bm k}_\lambda(p)+q\hat{\bm n}_\lambda$.  In this patch, the
soft part of the phase appearing in Eq.~\eqref{eq:ssbrg-uniform-perturbation}
is
\begin{equation}
Q^{(\lambda)}_{\mathcal O,<}(p,q,\omega)=u_{\mathcal O,\lambda}(p)\,
\phi_{\lambda,p,<}(q,\omega)+O(q).
\label{eq:ssbrg-slow-phases}
\end{equation}
Here $u_{\mathcal O,\lambda}(p)$ is the corresponding form factor.
Because $p'=p$, this form factor is unchanged by the RG transformation.
Writing $Q_{\mathcal O}=Q_<+Q_>$, the term linear in
$h_{\mathcal O}$ is renormalized according to
\begin{equation}
\left\langle
\cos\left[Q_<({\bm r},\tau)+Q_>({\bm r},\tau)\right]
\right\rangle_>
=\cos Q_<({\bm r},\tau)
\exp\left[-\frac12
\left\langle Q_>({\bm r},\tau)^2\right\rangle_>\right].
\label{eq:ssbrg-local-shell-average}
\end{equation}
The variance in Eq.~\eqref{eq:ssbrg-local-shell-average} is
\begin{align}
\left\langle Q_>({\bm r},\tau)^2\right\rangle_>
&=\sum_\lambda\int_{-\pi}^{\pi}\frac{dp}{2\pi}
|u_{\mathcal O,\lambda}(p)|^2
\int_{\rm shell}\frac{d\omega\,dq}{(2\pi)^2}
\frac{1}{Z(p)\omega^2+\rho(p)q^2}.
\label{eq:ssbrg-local-shell-variance}
\end{align}
The inner integral can be evaluated by setting
$\widetilde q=v(p)q$:
\begin{align}
\int_{\rm shell}\frac{d\omega\,dq}{(2\pi)^2}
\frac{1}{Z(p)\omega^2+\rho(p)q^2}
&=\frac{1}{\sqrt{Z(p)\rho(p)}}
\int_{\Lambda/b<\sqrt{\omega^2+\widetilde q^2}<\Lambda}
\frac{d\omega\,d\widetilde q}{(2\pi)^2}
\frac{1}{\omega^2+\widetilde q^2}
\nonumber\\
&=\frac{\log b}{2\pi\sqrt{Z(p)\rho(p)}}
=\frac{2}{N}\log b.
\label{eq:ssbrg-explicit-shell-integral}
\end{align}
In the last equality we used Eq.~\eqref{eq:ssbrg-patch-consistency}.  It
follows that
\begin{equation}
\frac12\left\langle Q_>({\bm r},\tau)^2\right\rangle_>
=\Delta_{\mathcal O}\log b,
\qquad
\Delta_{\mathcal O}
=\frac1N\sum_\lambda\int_{-\pi}^{\pi}\frac{dp}{2\pi}
|u_{\mathcal O,\lambda}(p)|^2.
\label{eq:ssbrg-shell-dimension}
\end{equation}
For $X$, two nodal lines contribute with unit weight.  For $D^X$, only the
diagonal line contributes and
$\int dp\,4\sin^2(p/2)/(2\pi)=2$.  Thus both operators have
$\Delta_{\mathcal O}=2/N$, in agreement with their two-point functions.
Equation~\eqref{eq:ssbrg-local-shell-average} consequently becomes
\begin{equation}
\left\langle\cos(Q_<+Q_>)\right\rangle_>
=b^{-\Delta_{\mathcal O}}\cos Q_<.
\label{eq:ssbrg-shell-average}
\end{equation}

Finally,
$d\tau\,dx_\perp\,dx_\parallel
=b^2d\tau'\,dx_\perp'\,dx_\parallel'$.  Combining this rescaling of the
measure with Eq.~\eqref{eq:ssbrg-shell-average} gives
\begin{equation}
h_{\mathcal O}'=b^{2-\Delta_{\mathcal O}}h_{\mathcal O}.
\label{eq:ssbrg-coupling-rescaling}
\end{equation}
Expanding to first order in $d\ell=\log b$ therefore yields
\begin{equation}
\frac{dh_{\mathcal O}}{d\ell}
=\left(2-\Delta_{\mathcal O}\right)h_{\mathcal O}
+ \cdots
\label{eq:ssbrg-full-flow}
\end{equation}

Both operators have nonzero weight on a generic segment of at least one
nodal line.  Substituting the dimensions calculated above gives
\begin{equation}
\frac{dh_X}{d\ell}=\left(2-\frac2N\right)h_X+\cdots,\qquad
\frac{dt_D}{d\ell}=\left(2-\frac2N\right)t_D+\cdots.
\label{eq:ssbrg-isotropic-flows}
\end{equation}
The zero of the dipole form factor at the isolated intersection $p=0$ does
not change its flow, which is controlled by generic points of the diagonal
nodal line.  Both subsystem-symmetry-breaking perturbations are
therefore relevant at controlled large $N$.

\subsection{Dependence on the Gaussian stiffness ratio}
\label{subsec:ssbrg-general-gamma}

The dynamically split theory studied above has equal stiffnesses for the two
combinations
\begin{equation}
\Phi_F=\theta_A-\theta_B,\qquad
\Phi_C=\theta_A+\theta_B.
\label{eq:ssbrg-Phi-FC}
\end{equation}
It is useful to study how the result changes in the more general
subsystem-symmetric Gaussian theory
\begin{equation}
H_\gamma=\frac{\kappa}{2}\sum_{\bm r}
\left[\gamma\Phi_F({\bm r})^2
+\gamma^{-1}\Phi_C({\bm r})^2\right],
\qquad \gamma>0.
\label{eq:ssbrg-general-gamma-H}
\end{equation}
In the original triangle variables, this Hamiltonian is
\begin{equation}
H_\gamma=\frac{\kappa}{2}\sum_{\bm r}
\left[
(\gamma+\gamma^{-1})(\theta_A^2+\theta_B^2)
+2(\gamma^{-1}-\gamma)\theta_A\theta_B
\right].
\label{eq:ssbrg-general-gamma-theta}
\end{equation}
Thus the subsystem symmetries alone do not fix $\gamma$, but dynamical
splitting does: the absence of a mixed $\theta_A\theta_B$ term requires
$\gamma=1$ (unequal coefficients of $\theta_A^2$ and $\theta_B^2$ are
still compatible with splitting).

Repeating the kernel inversion in
Eq.~\eqref{eq:ssbrg-inverse-kernel}, the singular propagators on the first
two nodal lines acquire the common residue $\gamma$:
\begin{align}
k_y\longrightarrow0:\quad
G^{\rm s}(\tau,{\bm k})&\simeq
\frac{\hbar_N\gamma}{|k_y|}
\begin{pmatrix}1&0\\0&0\end{pmatrix}
e^{-\Omega|\,\tau|},
\nonumber\\
k_x\longrightarrow0:\quad
G^{\rm s}(\tau,{\bm k})&\simeq
\frac{\hbar_N\gamma}{|k_x|}
\begin{pmatrix}0&0\\0&1\end{pmatrix}
e^{-\Omega|\,\tau|}.
\label{eq:ssbrg-general-axis-residues}
\end{align}

On the diagonal line, write $(k_x,k_y)=(p,-p+q)$ and define the
normalized soft field $\phi_d=(A_x-A_y)/2$.  Its singular propagator is
\begin{align}
\left\langle T_\tau
\phi_d(p,-p+q,\tau)\phi_d(-p,p-q,0)
\right\rangle_{\rm sing}
&=
\frac{\hbar_N r_\gamma(p)}{|q|}
e^{-v(p)|q||\tau|},
\nonumber\\
r_\gamma(p)
&=\gamma\cos^2\frac p2+\gamma^{-1}\sin^2\frac p2.
\label{eq:ssbrg-general-diagonal-residue}
\end{align}

For example, $X=e^{\mathrm{i}A_x}$ couples with unit weight to the
$k_y=0$ and diagonal lines.  Its dimension is therefore
\begin{align}
\Delta_X(\gamma)
&=\frac1N\left[
\gamma+\int_{-\pi}^{\pi}\frac{dp}{2\pi}r_\gamma(p)\right]
+O(N^{-2})
\nonumber\\
&=\frac{3\gamma+\gamma^{-1}}{2N}+O(N^{-2}).
\label{eq:ssbrg-general-X-dimension}
\end{align}
Applying the same calculation to the dipole gives
\begin{equation}
\Delta_{D^X}(\gamma)
=\frac{\gamma+3\gamma^{-1}}{2N}+O(N^{-2}).
\label{eq:ssbrg-general-dimensions}
\end{equation}
This result uses the diagonal form factor
$4\sin^2(p/2)$ from Eq.~\eqref{eq:ssbrg-DX-weight}.  Both dimensions
reduce to $2/N$ at the dynamically split point $\gamma=1$.

The RG rescaling in Eq.~\eqref{eq:ssbrg-rescaling} is unchanged by
$\gamma$.  Substitution into Eq.~\eqref{eq:ssbrg-full-flow} therefore yields
\begin{align}
\frac{dh_X}{d\ell}
&=\left[2-\frac{3\gamma+\gamma^{-1}}{2N}\right]h_X+\cdots,
\nonumber\\
\frac{dt_D}{d\ell}
&=\left[2-\frac{\gamma+3\gamma^{-1}}{2N}\right]t_D+\cdots.
\label{eq:ssbrg-general-gamma-flows}
\end{align}

For any fixed $\gamma=O(1)$, both perturbations remain relevant at
large $N$.  At large $N$, changing this conclusion requires extreme anisotropy $\gamma \sim N$. However, when $N = O(1)$, an $O(1)$ anisotropy may be enough to stabilize the nodal phase against explicit breaking of subsystem symmetry.

For completeness, for a general onsite perturbation $\delta H_{m,n}=-h_{m,n}\sum_{\bm r}\big[\mathcal O_{m,n}({\bm r})+\mathcal O_{m,n}
^{\dagger}({\bm r})\big]$, where $\mathcal O_{m,n}=e^{\mathrm{i}(mA_x+nA_y)}$ with $m,n\in\mathbb Z$, the same calculation gives $
\frac{dh_{m,n}}{d\ell}=\left\{2-\frac{1}{N}\left[\gamma(m^2+n^2)+\frac{\gamma+\gamma^{-1}}{2}(m-n)^2\right]+O(N^{-2})\right\}h_{m,n}+\cdots$. Therefore, for $\gamma > 1$, $X$ and $Z$ are the most relevant operators, while for $\gamma < 1$, $Y$ is the most relevant operator.

As an aside, following
Ref.~\cite{Lake2022EBLRG}, one way to improve upon our RG scheme is to use a single energy shell surrounding the complete three-line nodal manifold and retain the most general momentum-dependent marginal stiffnesses along it.   This refinement is unnecessary for the linear
stability analysis studied here, but would be
needed to classify more general perturbations.

\subsection{An alternative continuum limit}
\label{subsec:ssbrg-scope}

The calculation above keeps the lattice spacing fixed and then takes the thermodynamic limit. Following the analysis of subsystem-symmetric lattice theories in
Refs.~\cite{SeibergShao,GorantlaLamSeibergShao}, one may instead take
$a\to0$ and $L\to\infty$ while holding the physical lengths
$\ell\equiv aL$ fixed. Let us first restore the lattice spacing $a$ and write
the cubic dispersion, with ${\bm k}$ now denoting physical momentum, as
$\Omega({\bm k})\simeq v|k_xk_y(k_x+k_y)|$, where for
Eq.~\eqref{eq:ssbrg-general-gamma-H}
\begin{equation}
v=\frac{4\pi a^3\kappa}{N}.
\label{eq:ssbrg-continuum-v3}
\end{equation}
Thus, at fixed $N$ and $\gamma$, keeping $v$ fixed as $a\to0$ requires
$\kappa=Nv/(4\pi a^3)$.
\par
A local $X_{x_0,y_0}$ operator changes the eigenvalue of the column subsystem
symmetry $C_{x_0}$, as well as that of a diagonal subsystem symmetry.  It
therefore maps the ground state to a different subsystem-symmetry sector.
Within the Gaussian approximation, the lowest-energy configuration carrying one unit of the changed
column eigenvalue spreads the twist uniformly along that column:
\begin{equation}
A_y(x,y)=\frac{2\pi}{NL_y}\delta_{x,x_0},\qquad A_x(x,y)=0,
\qquad
E_{\rm col}=\frac{16\pi^2\kappa\gamma}{N^2L_y}.
\label{eq:ssbrg-column-sector-energy}
\end{equation}
Since $X_{x_0,y_0}$ also changes a diagonal eigenvalue, $E_{\rm col}$ is a
lower bound on the energy of the complete subsystem-symmetry sector that it
creates.  At fixed lattice spacing this bound vanishes as
$L_y^{-1}$ in the thermodynamic limit $L_y\to\infty$, which is consistent with our RG treatment above, where a single insertion of $X_{x_0,y_0}$ is  part of the low-energy theory. In the alternative limit, however, $\ell_y$ is fixed while $L_y=\ell_y/a\to\infty$. Using $\kappa\propto a^{-3}$, one finds
\begin{equation}
E_{\rm col}=\frac{4\pi\gamma v}{Na^2\ell_y}
\longrightarrow\infty.
\label{eq:ssbrg-continuum-sector-energy}
\end{equation}
Similarly, the dipole $D^X$ operator preserves the column-subsystem symmetry eigenvalue but changes diagonal subsystem-symmetry eigenvalues; the corresponding minimum energy has the
same $a^{-2}$ divergence.  Consequently, excitations associated with the insertion of $X$ and $D^X$ are absent from the finite-energy operator content of the continuum-first theory. 
This does not contradict our RG result in the last section, where the lattice
spacing is held fixed while the thermodynamic limit is taken. We believe this
fixed-lattice-spacing thermodynamic limit is the one relevant to the microscopic
spin model.
\ %

%

\end{document}